\documentclass[a4paper,fleqn]{article}
\usepackage[authoryear,longnamesfirst]{natbib}
\usepackage{empheq}
\usepackage{booktabs}
\usepackage{caption}
\usepackage{tikz}
\usepackage{authblk}
\usepackage{hyperref}
\usepackage{comment}
\usepackage{amssymb}
\usepackage{geometry}

\usepackage{subcaption}
\newcommand{\roundedcolorbox}[4]{%
    \begin{tikzpicture}
        \node[draw=#1, fill=#2, rounded corners=#3, inner sep=10pt] (box) {%
            \begin{minipage}{\dimexpr\linewidth-40pt\relax}
                #4
            \end{minipage}
        };
    \end{tikzpicture}%
}

\begin{document}


\title{Open-source MRI-informed computational model of human cortical folding}


\author[1]{Anne Kerachni}

\author[2,3,4]{Thomas Lavigne}

\author[5]{Stéphane Urcun}

\author[6]{Mireia Alenyà}

\author[6]{Oscar Camara}

\author[7]{Julien Lefèvre}

\author[1]{François Rousseau}


\affil[1]{LaTIM U1101 INSERM, IMT Atlantique, Plouzane, France}

\affil[2]{Institute of Computational Engineering, Department of Engineering, University of Luxembourg, Esch-sur-Alzette, Luxembourg}

\affil[3]{Institut de Biomecanique Humaine Georges Charpak, Arts et Metiers Institute of Technology, Paris, France}

\affil[4]{CNRS, Bordeaux INP, I2M, UMR 5295, I2M Bordeaux, Arts et Metiers Institute of Technology, Univ. of Bordeaux, Talence, France}

\affil[5]{Inria Research Center, Rennes University, Rennes, France}

\affil[6]{BCN MedTech, Department of Engineering, Universitat Pompeu Fabra, Barcelona, Spain}

\affil[7]{Institut de Neurosciences de la Timone, CNRS UMR 7289, Aix Marseille Universite, Marseille, France}

\maketitle

\begin{abstract}
The human cerebral cortex, initially smooth, progressively folds during fetal brain development \textit{in utero}, giving rise to cortical convolutions. Atypical cortical folding patterns can be associated with neurodevelopmental and neurological disorders. To better understand these conditions, it is crucial to first examine the factors governing healthy cortical folding. Computational modeling provides a powerful way for this purpose and has already helped understanding the influence of key biomechanical parameters on the folding pattern. However, most existing models use simplified geometries, limiting calibration and validation with fetal and neonatal brain Magnetic Resonance Imaging (MRI) and neglecting the influence of initial geometry on fold development. On the other hand, simulations on realistic brain geometries introduce additional challenges, including collision handling, fold characterization, and additional computational cost. Furthermore, model parameters are often difficult to interpret, complicating comparison, clinical translation, and calibration. Finally, computational models of cortical folding also remain rarely accessible.
In this work, we introduce a novel computational model of cortical folding, developed using the open-source code FEniCS to simulate folding on a whole-brain geometry generated from fetal MRI data. We also propose a modular, interpretable, and scalable simulation framework built around this computational model and openly available to the community. It uses fetal MRI data to generate realistic input brain meshes and estimate key biomechanical parameters such as cortical growth rate. The framework also integrates a spectral metric for cortical surface analysis to optimize folding pattern predictions from an healthy fetal MRI dataset.    
\end{abstract}






\section{Introduction}\label{sec1}


The mechanisms governing the formation and development of cortical convolutions - gyri and sulci - remain incompletely elucidated. Nevertheless, specific cortical folds have been shown to correlate with cognitive functions in both humans and animal models \cite{HdeVareilles2023}. Furthermore, the variability observed in certain folds is thought to underlie individual differences. In some cases, abnormal cortical folding has been associated with neurodevelopmental and neurological disorders, including autism or epilepsy \cite{VFernandez2016, HdeVareilles2023}. 
Consequently, investigating brain development, and in particular the processes underlying cortical folding, represents both a scientific challenge for understanding human cognitive functions and a clinical challenge for diagnosing neurodevelopmental disorders and predicting the trajectory of associated symptoms.

\vskip0.2cm

Computational modeling of brain cortical folding - also known as gyrification - offers a means to describe and predict the emergence of cortical folds, investigating the role of different biomechanical and biological factors underlying the phenomenon.

\noindent Moreover, advances in medical imaging, and in particular the availability of fetal magnetic resonance imaging (MRI) data, now enable the simulation of cortical folding on anatomically accurate brain geometries \cite{TTallinen2016, XWang2019, ZWang2021, MAlenya2022}. This progress opens the door to personalized approaches for studying and predicting cortical gyrification in both healthy and pathological cases \cite{MAlenya2022}. 

\vskip0.2cm

While gyrification models implemented on 2D or simplified geometries enable the testing of more complex hypotheses and multi-scale models (e.g. \cite{RToro2005, Sbudday2015bis}), simulating cortical folding on anatomically accurate brain meshes - relatively rare in the literature - offers the potential for inherently more relevant simulations. Such simulations also facilitate a more direct interpretation of the predictions, as they can be readily compared to fetal or neonatal brain medical images, thereby enabling potential validation of computational gyrification models. Moreover, this approach enables to perform patient-specific folding simulations \cite{MAlenya2022}, which could ultimately allow researchers to infer the factors underlying neurodevelopmental pathologies in young children based on MRI data.

\noindent However, MRI-informed gyrification models often lack realistic biophysical parameters, making direct comparison with medical data challenging. Furthermore, the brain computational models are not always readily accessible, either because they are implemented in isolated code bases that require substantial effort to interpret, are not publicly available, or have been developed using proprietary software that require paying a license fee, limiting the replicability of the simulations.

\noindent This work aims at developing a computational model of cortical folding that is interpretable and directly comparable to both other models and empirical data, meaning fetal and neonatal medical imaging data. Recent trends in gyrification modeling increasingly favor open-source approaches \cite{TTallinen2016, VernerGarikipati2018, ZWang2021, MAlenya2022, MSZarzor2023}. This shift is particularly important given that understanding cortical folding lies at the intersection of numerous disciplines - including genetics, biology, physics, mathematics, computer science, medical imaging, and medicine. Open-source models facilitate interdisciplinary collaboration, enabling diverse scientific communities to investigate cortical folding and its underlying mechanisms.

\vskip0.2cm

The primary objective, therefore, is to enhance the biomechanical model of human brain growth proposed by \cite{TTallinen2016, XWang2019}. 
This model, grounded on the hypothesis of differential tangential growth, allows to simulate fetal cortical folding on anatomically realistic brain geometries. A secondary objective of this work is to provide a computational model of gyrification within an interpretable and open-source framework. This involves, first, ensuring that the geometry and input parameters are defined using International System (SI) units, and second, employing parameters whose values have clear biophysical significance. 

\noindent Previous studies \cite{RToro2005, XWang2021,MAlenya2022} have demonstrated the influence of key model parameters - including cortical thickness, cortical growth, and brain stiffness - on the shape and pattern of simulated folds. Building on these findings, the present work aims to examine how these parameters affect quantitative metrics of cortical surface, some of which are employed here for the first time to calibrate the model against fetal brain MRI data.  
\vskip0.2cm

In this work, we introduce \textit{FetalFoldSim}, a novel FEniCS-based purely biomechanical computational model to simulate cortical folding during human gestation from a realistic brain mesh and MRI-derived parameters, such as cortical thickness. The simulated folding pattern is quantitatively compared with real fetal brain MRI data.

\noindent We first describe
the MRI fetal brain data used for the simulations (Section~\ref{subsec_mri_atlas_data}).
Then, we detail the biomechanical model (Section~\ref{subsec_computational_model_simulation_framework}), which has the following characteristics: cortical folding is driven by differential growth; the brain is represented as a nonlinear bilayer material; and mechanical collisions between the two hemispheres are penalized. The model is expressed in its variational form.
Next, we present the complete simulation pipeline built upon the \textit{FetalFoldSim} computational model in Section~\ref{subsec_computational_model_simulation_framework}, providing justification for the choice of model parameters, in particular the cortical growth rate. We also describe the spectral quantitative metric \cite{DGermanaud2012} employed to evaluate the folding patterns produced by our simulations (Section~\ref{subsec_biometrics}).
In the second step, cortical folding is simulated under healthy developmental conditions (see Section~\ref{sec3}). The validity of the model is assessed using healthy fetal MRI data, and parameter values that maximize similarity to the empirical data are identified using the spectral metric (Section \ref{sec:spectral_opt}).

\section{Materials and methods}\label{sec2}

\subsection{MRI atlas data of human healthy fetal brains}
\label{subsec_mri_atlas_data}

\begin{figure}[htbp]
  \centering
  \hspace*{-1cm}
  \includegraphics[width=6cm]{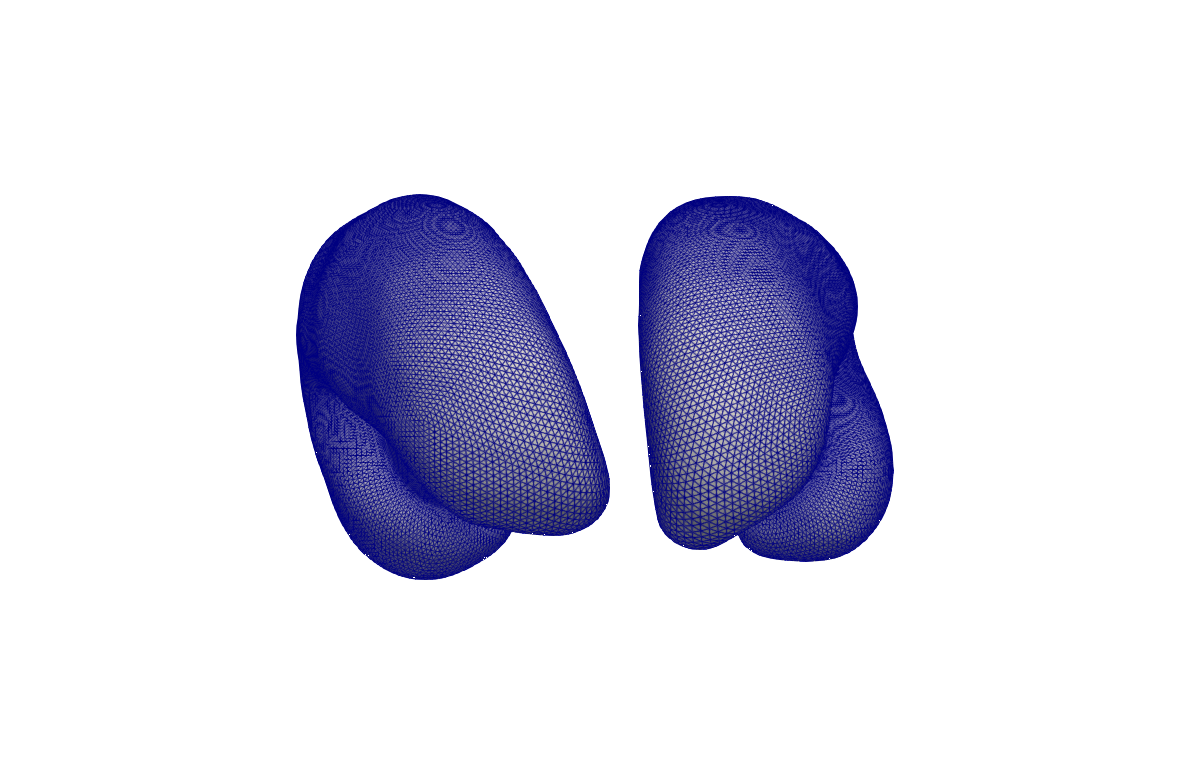}\label{fig:dHCP_fetal_brain_surface_atlas_2hemispheres_split_21GW}%
  \includegraphics[width=6cm]{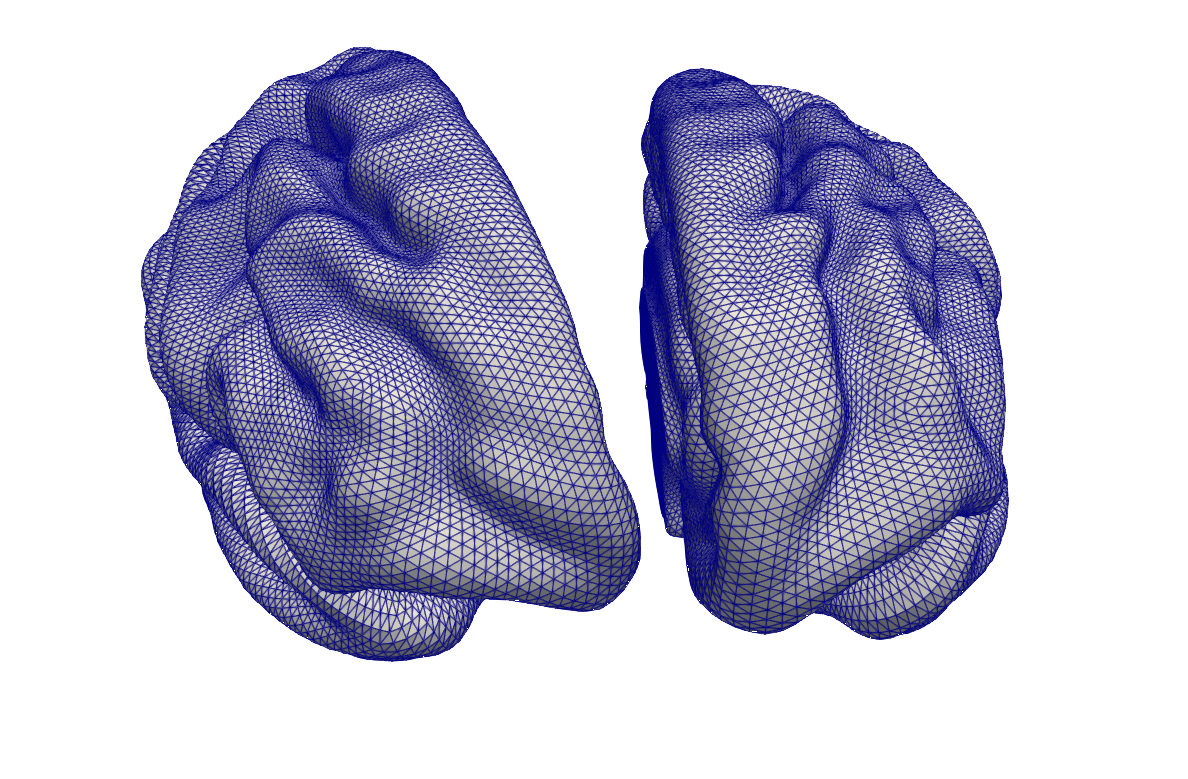}\label{fig:dHCP_fetal_brain_surface_atlas_2hemispheres_split_36GW}%
  \caption{The two surface atlas hemisphere meshes as provided by the dHCP project at 21 GW (a) and 36 GW (b)}
  \label{fig:dHCP_surface_data_21_36GW}
\end{figure}

The developing Human Connectome Project\footnote{\footnotesize\url{https://www.developingconnectome.org}} (dHCP) notably provides, for each gestational week from 21 to 36, the two \textit{hemispheric surface atlas} meshes\footnote{\footnotesize\url{https://gin.g-node.org/kcl_cdb/dhcp_fetal_brain_surface_atlas/src/master/atlas}} (see Fig. \ref{fig:dHCP_surface_data_21_36GW}). These 2D meshes have been generated from the MRI data of 242 healthy fetal scans and are chosen to simulate gyrification because of the high quality of the surface elements and the one-week temporal sampling to validate the model. At each gestational week (GW) a template mesh is computed from a weighted average of all fetal MRIs, taking into account their respective gestational age. The \textit{pial} surface of the brain corresponds to the external surface of the cortex, which is the surface of interest to analyze our simulations. Since longitudinal data for individual fetal brains on a large interval from 21 to 36 weeks are irregularly and sparsely sampled, the use of atlases is an alternative approach that allows a fine and regular sampling each week of development.

\vskip0.3cm

\begin{figure}[htbp]
  \centering
  \hspace*{-1cm}
  \includegraphics[width=4cm]{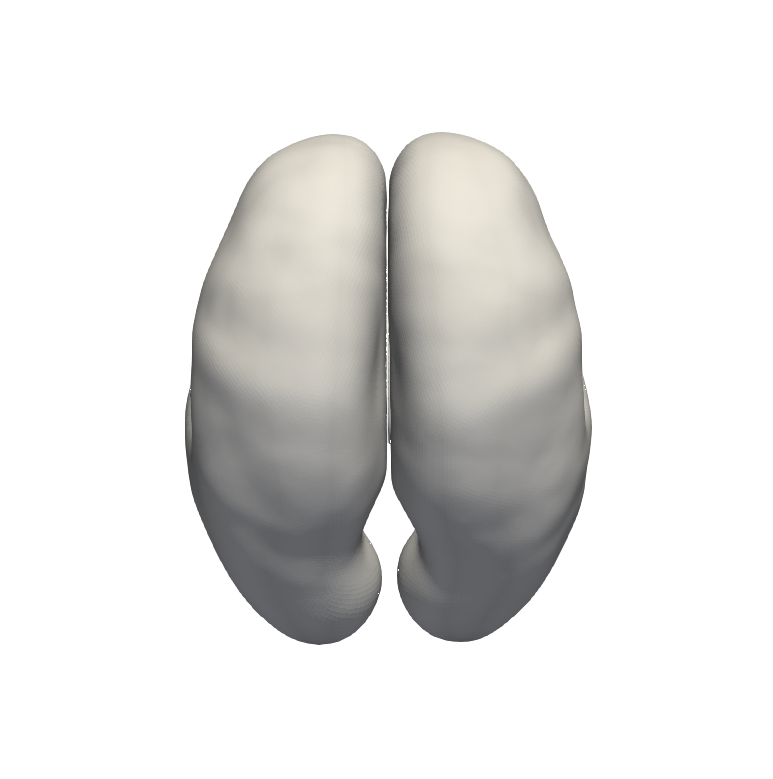} 
  \includegraphics[width=4cm]{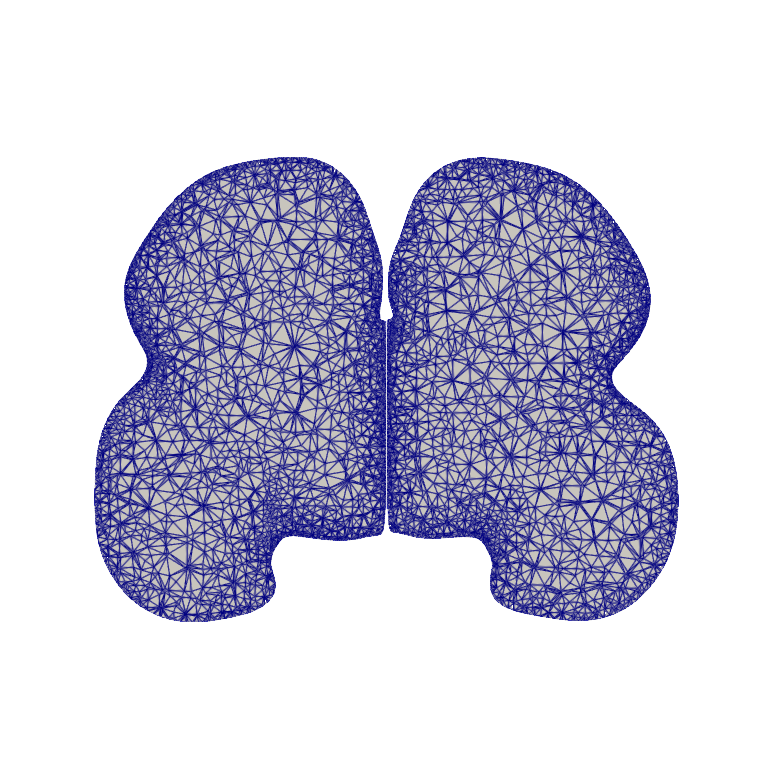}
  \caption{Initial brain geometry at 21GW (visualization with \textit{Paraview}). (a) Cortical surface mesh post-processed from the \textit{hemispheric surface atlas meshes} provided by the dHCP project. (b) 3D geometry generated with \textit{Netgen} from the cortical surface mesh}
  \label{fig:3D_input_brain_geometry_21GW_from_dHCP_Hemispheric_meshes} 
\end{figure}

\noindent For each gestational age, the two surface meshes are initially misoriented. Therefore, from 21 to 28, the two meshes are realigned symmetrically with respect to the interhemispheric fissure, using manual rigid-body transformations and surface normal inversion (\textit{Meshlab}\footnote{\url{www.meshlab.net}} \cite{cignoni2008meshlab}, \textit{Paraview}\footnote{\url{https://www.paraview.org/}} \cite{ahrens2005paraview}). The RAS+ coordinate system convention, widely employed in neuro-anatomical imaging, is chosen for the final orientation of the brain mesh. 
The two hemispheres are merged in the same file and the registered whole brain cortical surface mesh is then exported into .stl format (Fig. \ref{fig:3D_input_brain_geometry_21GW_from_dHCP_Hemispheric_meshes} (a)). Finally, the input 3D brain mesh associated to the cortical surface mesh at 21 GW is generated with \textit{Netgen}\footnote{\url{https://ngsolve.org/}} \cite{schoberl1997netgen} (Fig. \ref{fig:3D_input_brain_geometry_21GW_from_dHCP_Hemispheric_meshes} (b)). 

\vskip0.1cm

\noindent The brain mesh obtained, used for simulations, is composed by 129,960 faces on the external domain (pial surface of the brain) and 457,128 tetrahedrons as shown in Fig. \ref{fig:3D_input_brain_geometry_21GW_from_dHCP_Hemispheric_meshes}. It will be referred as to \textit{dHCP surface} brain mesh, below.

\vskip0.3cm

\subsection{Biomechanical model of brain cortical folding}
\label{subsec_computational_model_simulation_framework}

The biological process of cortical fold formation during human gestation can be mathematically formulated as a problem of growth-induced mechanical deformation \cite{PVBayly2013, SBudday2015, TTallinen2014}. Consequently, 
our
modeling approach is grounded in the theoretical frameworks of continuum and growth mechanics.  


\subsubsection{Kinematics of finite growth}

\noindent The \textit{multiplicative decomposition} of the geometric deformation gradient $\mathbf{F}$ into a growth-induced component $\mathbf{F}^\mathrm{g}$ and an elastic component $\mathbf{F}^\mathrm{e}$ (tensors) is considered   \cite{MBenAmar2005, JDervaux2008}, as proposed by \cite{ERodriguez1994}:
\begin{equation}\label{eqn:growth_multiplicative_decomposition}
\mathbf{F} = \mathbf{F}^\mathrm{e} \cdot \mathbf{F}^\mathrm{g}
\end{equation} 
The growth tensor $\mathbf{F}^\mathrm{g}$ reflects the change in mass and the irreversible deformation of the brain, 
while the elastic one $\mathbf{F}^\mathrm{e}$ traduces remodeling of the tissue, required to have \textit{compatibility} and \textit{integrity} of the body \cite{JDervaux2008} and is the reversible part of the deformation.

\subsubsection{Constitutive equations: non-linear elasticity}



\noindent The brain tissue is modeled as a bilayer solid material, with two distinct materials, referred as to the cortex and the inner layers of the brain, with distinct mechanical properties.

\noindent Cortical and subcortical layers are modeled as slightly compressible Neo-Hookean materials, as in \cite{TTallinen2016}. 
The strain energy density function of the brain material only depends on the elastic deformation gradient and reads: 

\begin{equation}\label{eqn:Neo_Hookean_strain_energy_density_function_I}
\psi = \psi(\mathbf{F}^\mathrm{e}) = \frac{\mu}{2} (\mathrm{tr}(\mathbf{C}^\mathrm{e})J^{\mathrm{e}-2/3} - 3) + \frac{K}{2}(J^\mathrm{e} - 1)^2
\end{equation}

\vskip0.1cm 

\noindent where $\mathbf{C}^\mathrm{e}$ is the right Cauchy-Green deformation tensor, with $\mathbf{C}^\mathrm{e} = (\mathbf{F}^\mathrm{e})^{\intercal} \mathbf{F}^\mathrm{e}$; $J^\mathrm{e}$ the Jacobian of the (elastic) deformation, with $J^\mathrm{e} = \det(\mathbf{F}^\mathrm{e})$; $\mu$ the local shear modulus of the brain, also referred to as rigidity; and $K$ the local bulk modulus of the brain.

\vskip0.2cm 

\noindent The first Piola-Kirchhoff stress $\mathbf{P}$, given by Eq. \eqref{eqn:third_governing_equation}, and its conjugate $\mathbf{F}$, the deformation gradient, are chosen to describe the internal efforts in our brain growth model. $\mathbf{P}$ stands for the \textit{total} stress, including growth- and elastic-stresses; $\mathbf{P}^\mathrm{e}$ stands for the first Piola-Kirchhoff stress tensor associated with the elastic deformation gradient $F^\mathrm{e}$ and depends on $\mathbf{C}^\mathrm{e}$ and $J^\mathrm{e}$.

\begin{equation}\label{eqn:third_governing_equation}
    \mathbf{P} = \dfrac {\partial {\psi(\mathbf{F}^\mathrm{e})}} {\partial \mathbf{F^\mathrm{e}}}:\dfrac {\partial {\mathbf{F}^\mathrm{e}}} {\partial \mathbf{F}} = \mathbf{P}^\mathrm{e}(\mathbf{F}^\mathrm{g})^{-\intercal}
\end{equation}

\subsubsection{Differential growth is driving folding}\label{section_differential_growth}

\noindent The \textit{growth-driven hypothesis} is used, which posits that cortical folding arises from a higher growth intensity in the cortical layer than in the inner layers. As in \cite{TTallinen2016}, we assume that only the cortical layer is growing.

\subsubsection{Growth tensor: tangential and adaptive growth in the cortex}\label{subsection_growth_law_cortex}

\noindent The growth tensor is often built using the reference configuration $\Omega_0$ as the exclusive reference \cite{AGoriely2007, AJavili2013, SiBudday2015, TTallinen2016, SNVerner2018}. \cite{ZWang2021} propose to use a "\textit{continuously evolving reference configuration}" instead (referred hereafter as the \textit{updated} reference configuration) to prevent introducing undesired elastic and growth-induced distortions in the geometry, at advanced stages of the simulation but also to enable incorporation of material changes, such as mass addition or local variations in material properties, into the model.

\noindent We use this approach in our model. The normal vectors used to build growth tensor orientation are updated on each updated reference configuration; the growth ratios are defined between two successive steps. 

\vskip0.2cm 

\noindent We primarly model cortical growth as tangential. The growth tensor, defined for any point of the domain, is given by: 

\begin{equation}
\mathbf{F}_\mathrm{g}({t}, \mathbf{X}) = (1 + \mathrm{d}g_{\mathrm{TAN}}(t, \mathbf{X}) ) \cdot (\mathbf{I} - \mathbf{N}_{\mathrm{t}} \otimes \mathbf{N}_\mathrm{t}) 
+ \hspace{1mm} \mathbf{N}_{\mathrm{t}} \otimes \mathbf{N}_{\mathrm{t}}
\label{eqn:growth_tensor_Cortex} 
\end{equation}

\vskip0.2cm 

\noindent where $\mathbf{X}$ is the position of any particle of the domain and $\mathbf{N}_\mathrm{t}$ the normal vector at the closest cortex surface location, in the updated reference configuration; $\mathrm{d}g_{\mathrm{TAN}}$ denotes the tangential growth stretch increment between two successive configurations, is non-zero in the cortex but equal to zero in the sublayers.

\vskip0.2cm 

\noindent In Eq. \eqref{eqn:growth_tensor_Cortex}, $\mathrm{d}g_{\mathrm{TAN}}$ equals $\alpha_{\mathrm{TAN}} \cdot \mathrm{d}t$ in the cortex, where the tangential growth rate in the cortex $\alpha_{\mathrm{TAN}} $ is assumed constant.

\vskip0.2cm 

\subsubsection{Relating Piola-Kirchhoff stress $\mathbf{P}$ to the displacement field $\mathbf{u}$ (unknown)} 

\noindent The objective of the biomechanical model is to obtain a differential equation that enables to deduce the displacement field between one undeformed state of the brain and the deformed one. To do so, continuum mechanics defines the coordinates $\mathbf{X}$ for any particle in the undeformed space $\Omega_0$ (\textit{reference} configuration) and its coordinates $\mathbf{x}$ in the deformed space $\Omega_\mathrm{t}$ (\textit{current} configuration).

\noindent The two configurations are related by the displacement field $\mathbf{u}$, through the following equation: 

\begin{equation}\label{eqn:u_x_X}
    \mathbf{u}(\mathbf{X}, t)= \mathbf{x} (\mathbf{X}, t) - \mathbf{X}
\end{equation}

\vskip0.2cm 

\noindent The total deformation gradient $\mathbf{F}$, can be defined as follows:
\begin{equation}\label{eqn:deformation_gradient}
    \mathbf{F}(\mathbf{X}, t)= \dfrac{\partial \mathbf{x}(t)}{\partial \mathbf{X}} = \mathbf{I} + \nabla_{\mathrm{X}}\mathbf{u}
\end{equation}

\vskip0.2cm 

\noindent Then, using Eqs. \eqref{eqn:growth_multiplicative_decomposition} and \eqref{eqn:deformation_gradient}, the elastic deformation gradient reads:

\begin{equation}\label{eqn:elastic_deformation_tensor_from_growth_tensor}
\mathbf{F}^\mathrm{e} = \mathbf{F} \cdot (\mathbf{F}^\mathrm{g})^{-1} = (\mathbf{I} + \nabla_{\mathrm{X}}\mathbf{u}) \cdot (\mathbf{F}^\mathrm{g})^{-1}
\end{equation}

\vskip0.2cm

\noindent The right Cauchy-Green deformation tensor $\mathbf{C}^\mathrm{e}$ defined previously and the Jacobian of the elastic part of the deformation $J^\mathrm{e}$ are therefore functions of the displacement field $\mathbf{u}$, and so is the mechanical stress $\mathbf{P}$.

\subsubsection{Governing equation}
\label{subsubsec_governing_equation}

In our mechanical model, neither feedback of the mechanical forces nor coupling with cellular processes are incorporated within the problem equations.\\ 

\vskip0.2cm 
\noindent As the phenomenon of cortical folding is a slow process, it is modeled by the quasistatic approximation of the linear momentum \cite{AJavili2013, PVBayly2013, MAHolland2013, SBudday2014, SiBudday2015, SNVerner2018, ZWang2021, SWang2021}. We assume no body forces are exerted on the brain volume.\\

\noindent $\forall (t, \mathbf{X}) \in (\mathbb{R}^{+}, \Omega_\mathrm{t})$:  
\begin{equation}\label{eqn:ODE}
    \nabla \cdot \mathbf{P}(\hspace{0.5mm} \mathbf{u}(\mathbf{X}, t) \hspace{0.5mm} ) = \mathbf{0} 
\end{equation}

\noindent where  $\mathbf{P}$ is the first Piola-Kirchhoff stress given by Eq. \eqref{eqn:third_governing_equation} and $\Omega_\mathrm{t}$ the spatial 3D brain domain. All quantities are defined at each new updated reference configuration time $t$.\\ 

\vskip0.2cm 

The variational problem Eq.~\eqref{eqn:weakform_IPP} corresponding to Eq.~\eqref{eqn:ODE} traduces the equilibrium between internal virtual efforts and external virtual efforts applying at the boundary of the domain $\partial \Omega_\mathrm{t}$ (cortical surface). It is a requirement of Finite Element formulation.
It reads:\\
\vskip0.1cm
\noindent Let $\mathbf{v}$ belong to $V=H_\mathrm{t}^{1}(\Omega_\mathrm{t})$, an Hilbert space. 
\vskip0.2cm
\noindent Find $\mathbf{u}\in{V}$ such as for all $\mathbf{v}\in{V}$:
\begin{equation} \label{eqn:weakform_IPP}
    \hspace{-7mm} \int_{\Omega_\mathrm{t}}^{} \mathbf{P}(\mathbf{u}):\nabla{\mathbf{v}} \, dX = \int_{\delta\Omega_\mathrm{t}}^{} \mathbf{N}_\mathrm{t} \cdot \mathbf{P}(\mathbf{u}) \cdot \mathbf{v} \, dS  
\end{equation}

\noindent with $\mathbf{v}$ the virtual field, chosen as the arbitrary test displacement field ${\partial \mathbf{u}}$ (test function) \cite{VAYastrebov2013}; $\mathbf{N}_\mathrm{t}$ the unit normal in the updated reference configuration, at the external surface of the brain (pial surface).

\subsubsection{Boundary conditions (BCs)}
\label{subsubsec_boundary_conditions}
\begin{figure}
  \centering
      \centering
      \includegraphics[width=0.33\textwidth]{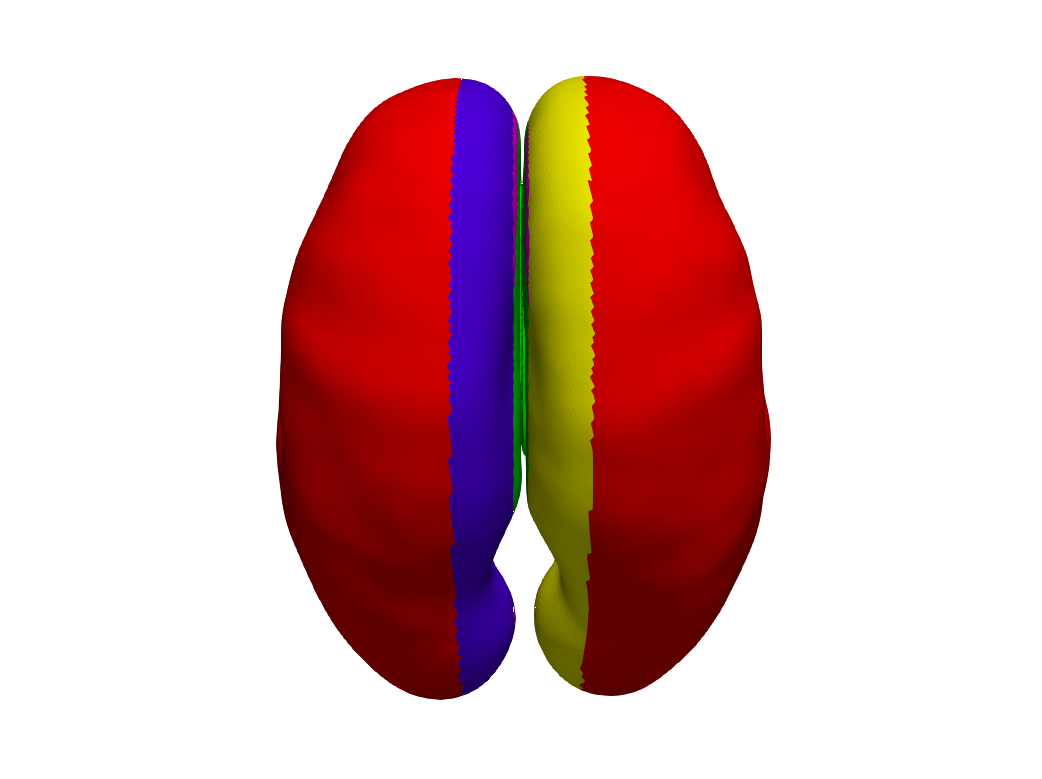}%
      \includegraphics[width=0.33\textwidth]{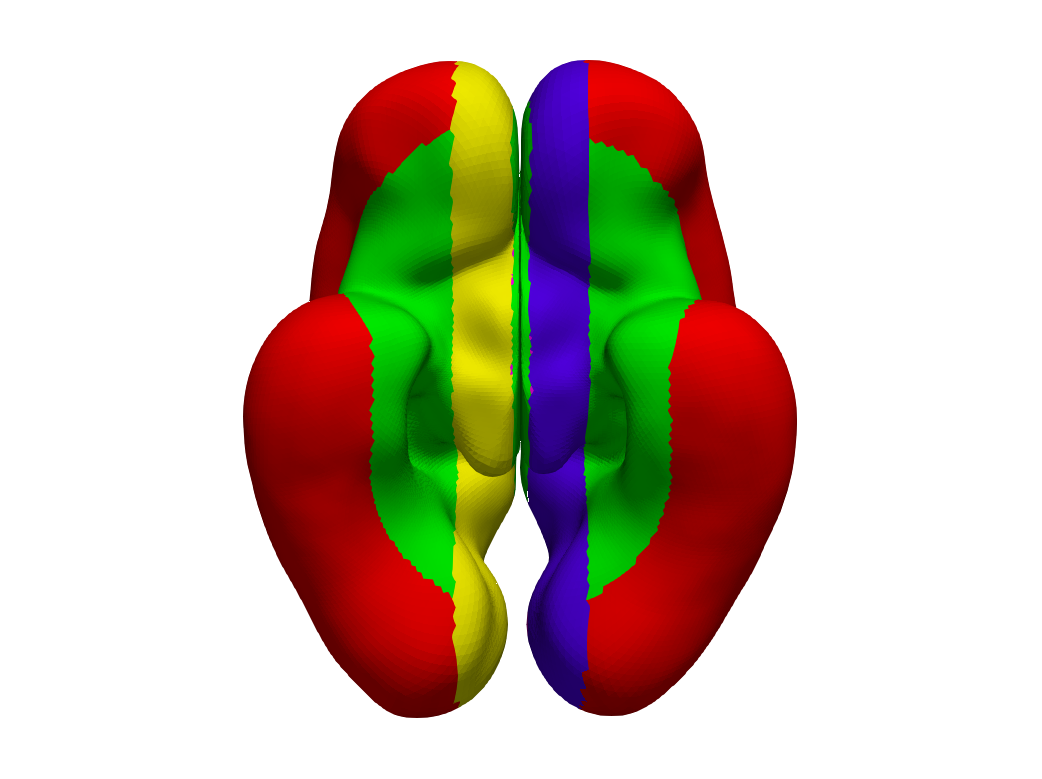}%
\caption{Distinct tags mark the different boundaries of the brain mesh: traction free cortical surface (red), fixed surface (green), boundaries with contact pressure (blue and yellow)}  
\label{fig:brain932K_boundaries}
\end{figure}

The brain boundaries, on which the external virtual efforts will apply, are defined as in Fig. \ref{fig:brain932K_boundaries}.
We assume the red cortical surface is traction-free (Neumann BCs: Eq. \eqref{eqn:weakform_neumann}) \cite{SWang2021}, and that the green boundary is fully fixed (Dirichlet BCs: Eq. \eqref{eqn:weakform_dirichlet}) \cite{TTallinen2016}. Such fixed boundary are chosen for computational convenience. 
Finally, the contact pressure generated by the collisions between the two hemispheres (through the blue and yellow boundaries) is the only external virtual effort considered on the surface of the brain. 

\noindent Variational problem Eq. \eqref{eqn:weakform_IPP} then reads:
\vskip0.2cm
\noindent Find $\mathbf{u}\in{V}$ such as:

\begin{subequations}\label{eqn:weakform_all}
\begin{empheq}[left=\empheqlbrace]{align}
    &\int_{\Omega_\mathrm{t}} \mathbf{P}(\mathbf{u}) : \partial \nabla{\mathbf{u}} \, dX = \int_{\Gamma_\mathrm{c}^1 \cup \Gamma_\mathrm{c}^2} \mathbf{N}_\mathrm{t} \cdot \mathbf{P}(\mathbf{u}) \cdot \partial \mathbf{u} \, dS \label{eqn:weakform_main} \\[8pt]
    &\int_{\Gamma_\mathrm{f}} \mathbf{P}_\mathrm{n}(\mathbf{u}) \cdot \partial \mathbf{u} \, dS = 0 \quad \text{(Neumann BCs)} \label{eqn:weakform_neumann} \\[8pt]
    &\int_{\Gamma_\mathrm{u}} \mathbf{N}_\mathrm{t} \cdot \mathbf{P}(\mathbf{u}) \cdot \partial \mathbf{u} \, dS = 0 \quad \text{(Dirichlet BCs)} \label{eqn:weakform_dirichlet}
\end{empheq}
\end{subequations}

\noindent where $\Gamma_\mathrm{c}^1$ and $\Gamma_\mathrm{c}^2$ are the contact boundaries on the left and right hemispheres (blue and yellow boundaries); $\Gamma_\mathrm{f}$ is the boundary where the Neumann boundary conditions apply (red boundary), and $\Gamma_\mathrm{u}$ is th boundary where the Dirichlet boundary conditions apply (green boundary).

\subsubsection{Penalizing contact between the two hemispheres}
\label{subsubsec_contact}
Collisions are rarely considered in simulations of cortical folding, likely because the use of simplified geometries (such as spherical or paved domains) does not result in significant contact. However, when simulations are conducted on realistic brain geometries, handling collisions becomes crucial. This is necessary to prevent these simulations from being limited to early gestational stages and to accurately reproduce realistic deformations. \cite{TTallinen2016} proposed an algorithm for correcting global collisions between surface nodes in brain folding simulations, but their approach is not formulated within a finite element framework. 

\noindent In our finite element model of cortical growth, we observed that the earliest collisions occur between the two hemispheres. We therefore propose a penalty-based approach integrated into the variational formulation of the cortical folding problem to avoid collisions between the two hemispheric surfaces.

\vskip0.2cm

\begin{figure}[htbp]
  \centering
  \includegraphics[width=4cm]{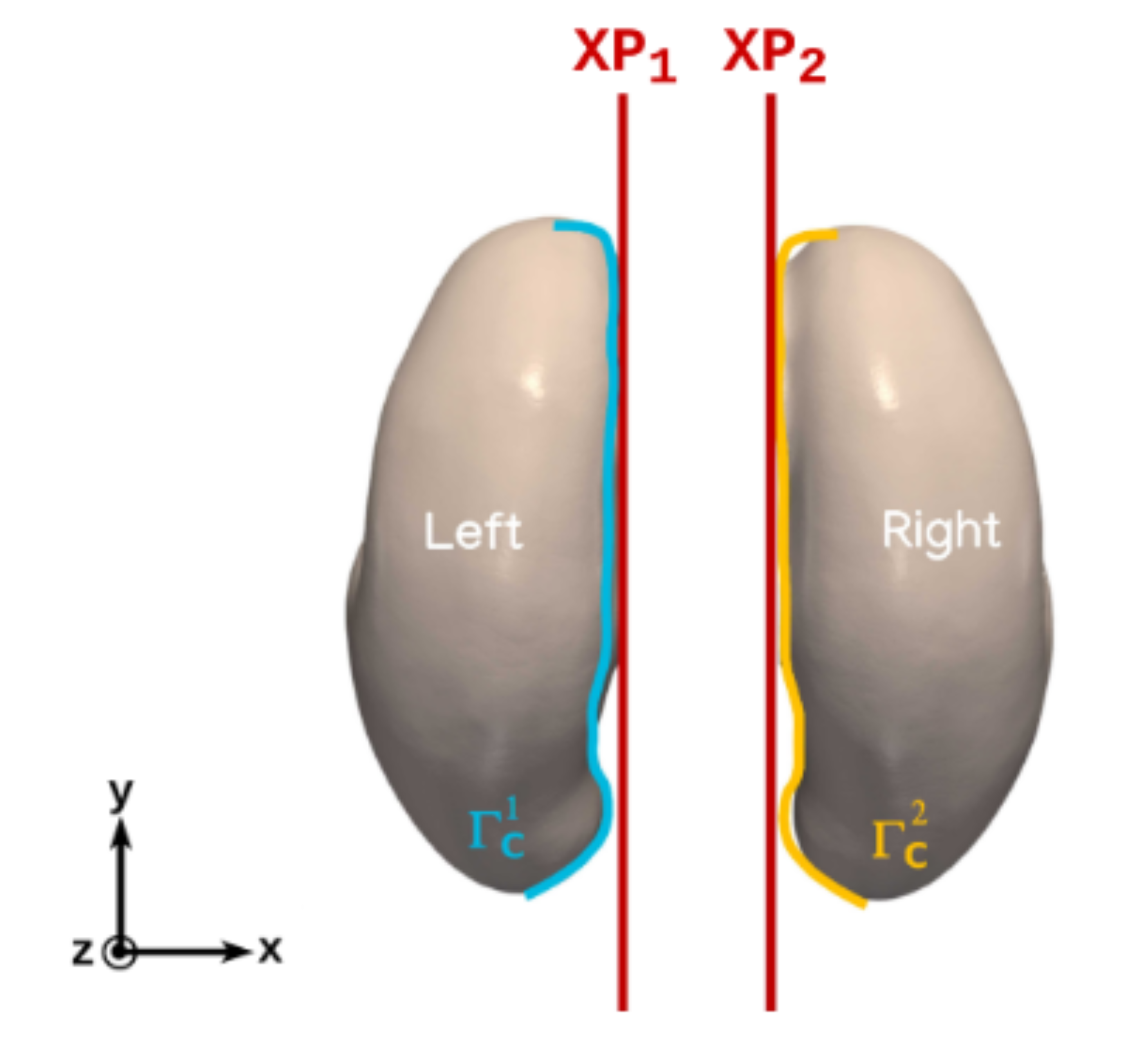} 
  \includegraphics[width=4cm]{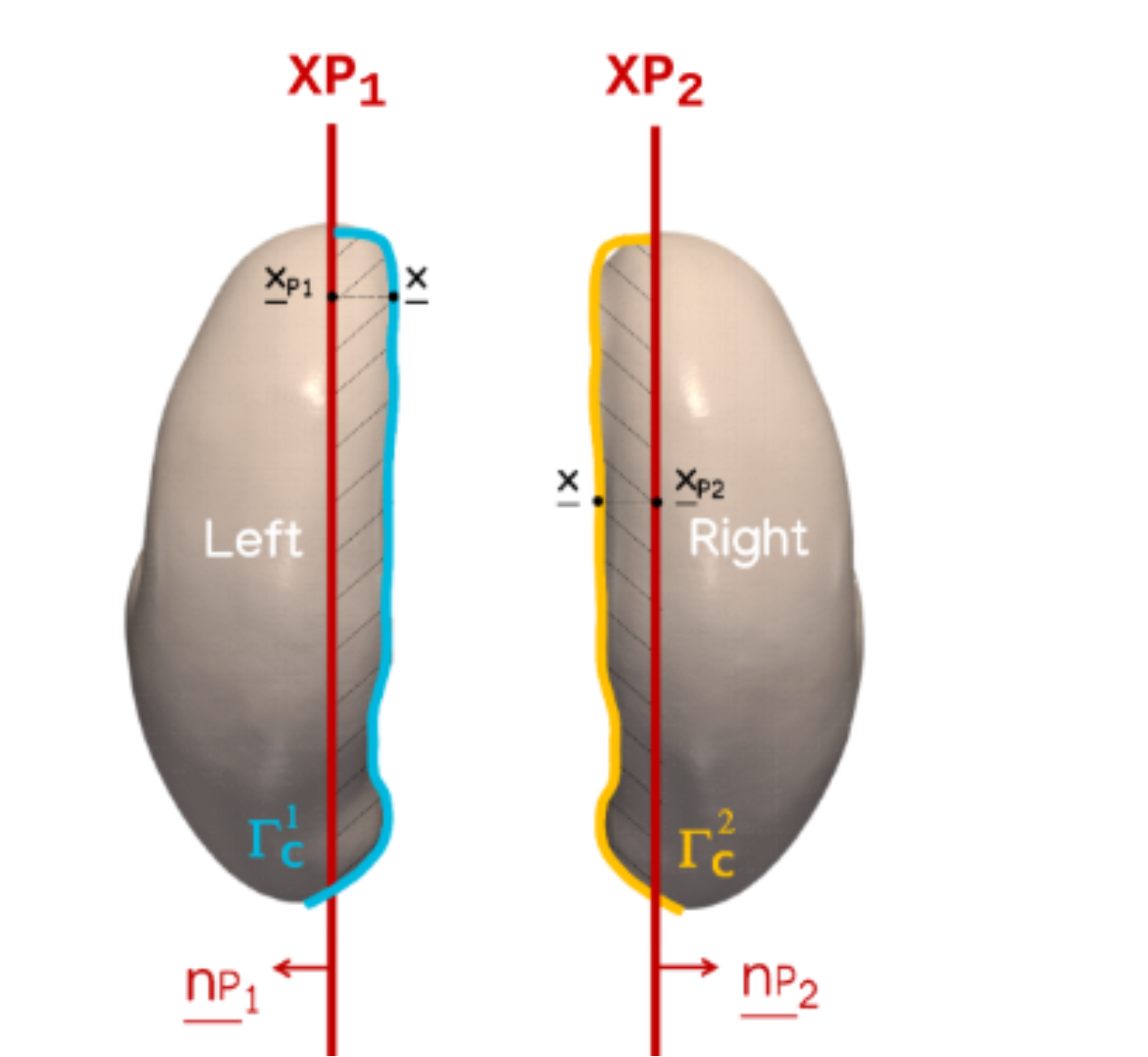}
  \includegraphics[width=4cm]{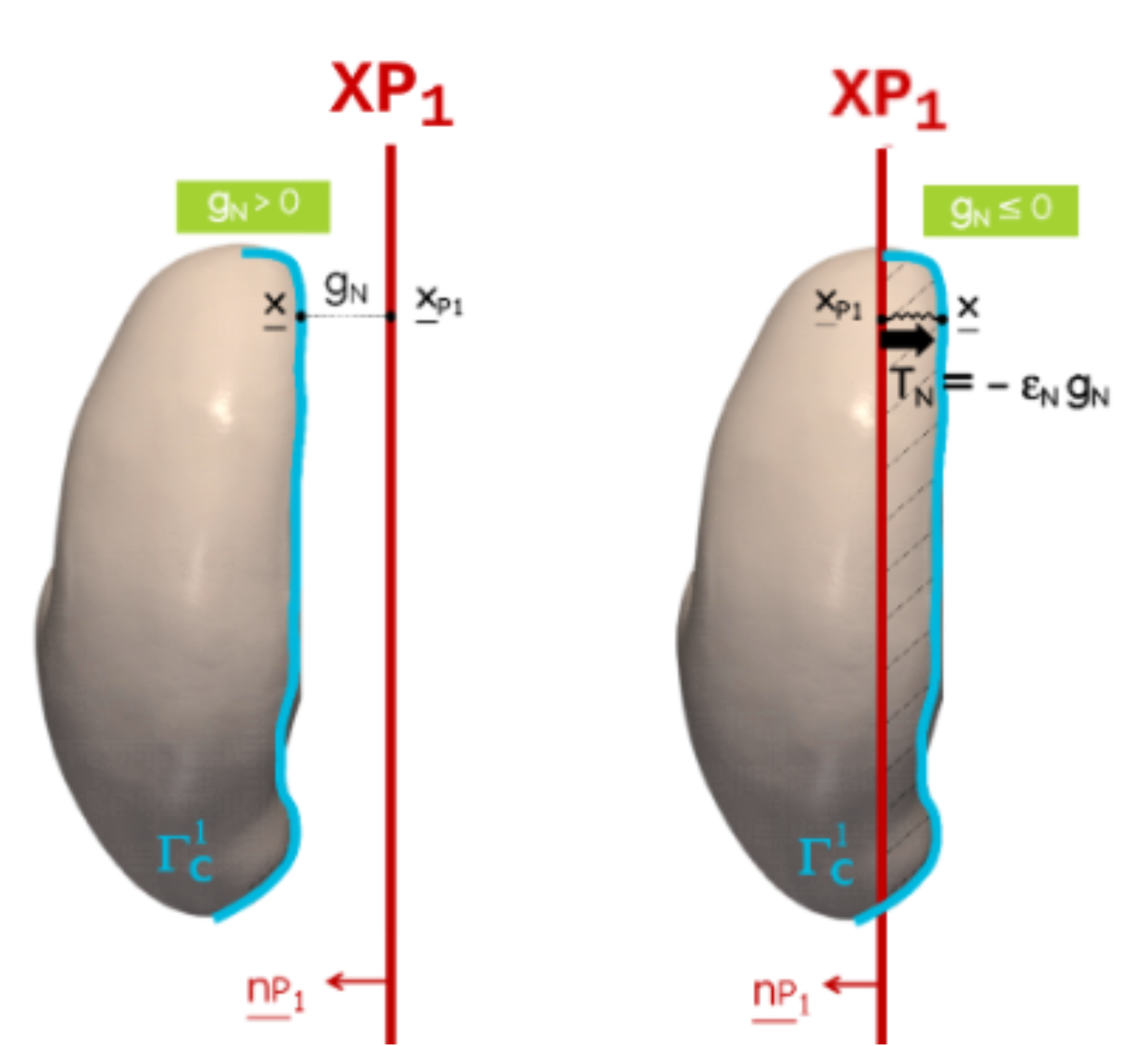}
  \caption{Penalizing interhemispheric contact. (a) and (b) The self-contact between the two hemispheres is modeled as unilateral contact between a deformable body ($\Gamma_\mathrm{c}^1$, $\Gamma_\mathrm{c}^2$) and its associated rigid fictive plane ($X_\mathrm{P_1}$, $X_\mathrm{P_2}$). (c) The penalty method is used to minimize the contact in the variational formulation; an example corresponding to the left hemisphere is provided. When the normal gap is positive, no penalty is applied. Conversely, when collision occur on the hemisphere, the normal gap $g_\mathrm{N}$ becomes negative and a normal traction $T_\mathrm{N}$ is applied on the contact boundary $\Gamma_\mathrm{c}^1$, proportional to the penetration magnitude, in order to minimize it}
  \label{fig:interhemispheric_contact_penalization} 
\end{figure}

In this work, we consider collisions between the two hemispheres (self-contact), through their contact boundaries $\Gamma_\mathrm{c}^1$ and $\Gamma_\mathrm{c}^2$,  can be modeled by an artificial contact occurring between each hemisphere against an intermediate \textit{fixed} rigid plane (i.e. $\mathbf{x}=X_\mathrm{P_1}$ refers to the equation of the left hemisphere plane; $\mathbf{x}=X_\mathrm{P_2}$ of the right hemisphere plane), as illustrated in Fig. \ref{fig:interhemispheric_contact_penalization} (a) and (b). We employ the theoretical developments discussed in \cite{VAYastrebov2013, JLengiewicz2011}, related to unilateral contact between a deformable body and a rigid plane. 
In contact mechanics, a first step consists in detecting the collisions; a second step consists in applying a method to "treat the contact constraints" \cite{JLengiewicz2011}, among which we choose the \textit{penalty method}, commonly used for large deformation problems. A third step would consist in smoothing the contact zones to make the treatment more accurate as the simulation of cortical folding progresses, but for the sake of simplification, we will not use any smoothing method.

\vskip0.2cm

\noindent Eq. \eqref{eqn:weakform_main} becomes:

\begin{align}\label{eqn:weakform_IPP_rigid_planes_3}
\begin{split}
     \int_{\Omega_\mathrm{t}}^{} \mathbf{P}:\partial \nabla{\mathbf{u}} \, dX \hspace{1mm} = \hspace{1mm} & \int_{\Gamma_\mathrm{c}^1}^{} T_\mathrm{N} \hspace{1mm} \mathbf{N}_\mathrm{t} \hspace{2mm} {\partial(\mathbf{u}_\mathrm{1} - \mathbf{u}_\mathrm{P1})} \, dS \hspace{1mm} + \hspace{1mm} \int_{\Gamma_\mathrm{c}^2}^{} T_\mathrm{N} \hspace{1mm} \mathbf{N}_\mathrm{t} \hspace{2mm} {\partial(\mathbf{u}_\mathrm{2} - \mathbf{u}_\mathrm{P2})} \, dS \hspace{1mm} 
\end{split}
\end{align}
\noindent where $T_\mathrm{N}$ the \textit{contact traction} \cite{JLengiewicz2011} or \textit{pressure} \cite{VAYastrebov2013}, represents the pressure exerted by any collision onto the brain geometry surface, that needs to be minimized; $\mathbf{u}_\mathrm{1}$ and $\mathbf{u}_\mathrm{2}$ the displacement fields in any point that belongs to the left and right hemisphere respectively; $\mathbf{u}_\mathrm{P1}$ and $\mathbf{u}_\mathrm{P2}$ the displacement of the left and right hemispheric planes described previously (null here). $T_\mathrm{N}$ is expressed in the \textit{Lagrangian} description and defined on the \textit{updated reference} configuration. 

\vskip0.2cm

The \textit{normal gap} $g_\mathrm{N}$ between one hemisphere node and its projection onto the associated rigid plane, as illustrated in Fig. \ref{fig:interhemispheric_contact_penalization} (c), is defined in the \textit{current} configuration (both for coordinates and normal vector) as in \cite{VAYastrebov2013, JLengiewicz2011}:

\begin{equation}\label{eqn:normal_gap}
    g_\mathrm{N}(\mathbf{x}) = (\mathbf{x} - \mathbf{x}_\mathrm{P}) \cdot \mathbf{n}_\mathrm{P}
\end{equation}
\noindent where $\mathbf{x}$ is the coordinates of any node at the contact surface of an hemisphere, $\mathbf{x}_\mathrm{P}$ is the coordinates of the associated fictive rigid plane and $\mathbf{n}_\mathrm{P}$ the outward normal at this plane (Fig. \ref{fig:interhemispheric_contact_penalization} (b)). 

\vskip0.4cm

\noindent The \textit{normal gap increment} reads:
\begin{equation}\label{eqn:normal_gap_variation}
    \delta g_\mathrm{N}(\mathbf{x}) = (\delta \mathbf{x} - \delta \mathbf{x}_\mathrm{P}) \cdot \mathbf{n}_\mathrm{P} 
\end{equation}

\noindent As we use the \textit{updated} Lagrangian description, we express $g_\mathrm{N}$ and $\delta g_\mathrm{N}$ in the \textit{updated reference} configuration.  

\vskip0.2cm

\noindent Equations \eqref{eqn:normal_gap} and \eqref{eqn:normal_gap_variation} become:
\begin{equation}\label{eqn:normal_gap_REFERENCE}
    g_\mathrm{N}(\mathbf{u}) = (\mathbf{X} + \mathbf{u} - (\mathbf{X}_\mathrm{P} + \mathbf{u}_\mathrm{P})) \cdot \mathbf{n}_\mathrm{P} = (\mathbf{X} + \mathbf{u} - \mathbf{X}_\mathrm{P}) \cdot \mathbf{n}_\mathrm{P}
\end{equation}

\begin{equation}\label{eqn:normal_gap_variation_REFERENCE}
    \delta g_\mathrm{N}(\mathbf{u}) = \delta (\mathbf{u} - \mathbf{u}_\mathrm{P}) \cdot \mathbf{n}_\mathrm{P} =  \delta \mathbf{u} \cdot \mathbf{n}_\mathrm{P}
\end{equation}

\vskip0.1cm

\noindent where $\mathbf{X}$ represents the coordinates of any node of the domain in the \textit{updated reference} configuration and $\mathbf{u}$ the displacement from \textit{updated reference} to \textit{current} configurations, defined by Eq. \eqref{eqn:u_x_X}. 

\vskip0.2cm

\noindent Following the work of \cite{VAYastrebov2013}, and assuming the contact between the two hemispheres is frictionless, the brain growth variational Eq. \eqref{eqn:weakform_IPP_rigid_planes_3} can then be rewritten as follows: 
\begin{equation}\label{eqn:brain_growth_variational_equation_penalty_method}
     \int_{\Omega_\mathrm{t}}^{} \mathbf{P}:\partial \nabla{\mathbf{u}} \, dX \hspace{1mm} = \hspace{1mm} \int_{\Gamma_\mathrm{c}^1 \cup \Gamma_\mathrm{c}^2}^{} T_\mathrm{N} \hspace{1mm} \delta g_\mathrm{N}(\mathbf{u}) \, dS \hspace{1mm}
\end{equation}

We place in frictionless contact, the \textit{Hertz–Signorini–Moreau}'s conditions non-penetration conditions must be satisfied \cite{VAYastrebov2013}. The penalty method assumes the pressure $T_\mathrm{N}$ exerted on the surface in contact is proportional to the penetration depth $g_\mathrm{N}$ according to Eq. \eqref{eqn:nominal_contact_traction}; it avoid each penetration by an elastic reaction force (Eq. \eqref{eqn:nominal_contact_traction}) that drives any colliding node to the opposite direction (Fig. \ref{fig:interhemispheric_contact_penalization} (c)).

\begin{equation}\label{eqn:nominal_contact_traction}
    T_\mathrm{N} = \epsilon_\mathrm{N} \langle - \hspace{1mm} g_\mathrm{N}\rangle
\end{equation}

\noindent where $\epsilon_\mathrm{N}$ is the \textit{penalty coefficient}, positive and chosen the same constant for every nodes belonging to the contact boundary whether on the left or right hemisphere. 

\vskip0.1cm

\noindent The \textit{mackauley brackets} $\langle \cdot\rangle$ are used as follows:
\begin{equation}\label{eqn:normal_gap_mackauley_brackets}
\langle - \hspace{1mm} g_\mathrm{N}\rangle \hspace{2mm} = \hspace{2mm} \left\{
                                           \begin{aligned}
                                            - \hspace{1mm} g_\mathrm{N} \hspace{2mm} &\text{if} \hspace{2mm} g_\mathrm{N} \leq 0 \hspace{2mm} \text{(contact)}\\
                                            0 \hspace{2mm} &\text{if} \hspace{2mm} g_\mathrm{N} > 0 \hspace{2mm} \text{(separation)}\\
                                           \end{aligned}
                                    \right.
\end{equation}

\vskip0.3cm

To avoid potential contact between $\Gamma_\mathrm{c}^1$ and $\Gamma_\mathrm{c}^2$, brain growth variational equation, Eq. \eqref{eqn:brain_growth_variational_equation_penalty_method} is penalized incorporating a contact pressure variational term as follows:

\vskip0.2cm

\begin{align}
    \begin{split}
     \int_{\Omega_\mathrm{t}}^{} \mathbf{P}:\partial \nabla{\mathbf{u}} \, dX \hspace{1mm} & = \int_{\Gamma_\mathrm{c}^1 \cup \Gamma_\mathrm{c}^2}^{} \epsilon_\mathrm{N} \hspace{1mm} \langle- \hspace{1mm} g_\mathrm{N}\rangle \hspace{1mm} \delta g_\mathrm{N} \, dS\\
     & = \hspace{1mm} \epsilon_\mathrm{N} \hspace{1mm} \int_{\Gamma_\mathrm{c}^1 \cup \Gamma_\mathrm{c}^2}^{} \hspace{1mm} \langle - \hspace{1mm} (\mathbf{X} + \mathbf{u} - \mathbf{X}_\mathrm{P}) \cdot \mathbf{n}_\mathrm{P} \rangle \hspace{1mm} \delta \mathbf{u} \cdot \mathbf{n}_\mathrm{P} \, dS \hspace{2mm} 
    \end{split}
\end{align}

\noindent The problem Eq. \eqref{eqn:weakform_all} then becomes:\\

\noindent Find the unknown displacement field $\mathbf{u}: (\Omega_\mathrm{t} \subset (\mathbb{R}^{+})^3$) $\rightarrow$ $(\mathbb{R}^{+})^3$, such that:

\begin{subequations}\label{eqn:brain_growth_all}
\begin{empheq}[left=\empheqlbrace]{align}
    &\begin{aligned}
        \int_{\Omega_\mathrm{t}} \mathbf{P}(\mathbf{u}):\partial \nabla{\mathbf{u}} \, dX & - \epsilon_\mathrm{N} \int_{\Gamma_\mathrm{c}^1} \langle (\mathbf{X} + \mathbf{u}) \cdot \mathbf{x} - X_\mathrm{P_1} \rangle \delta \mathbf{u} \cdot (-\mathbf{x}) \, dS \\
        & - \epsilon_\mathrm{N} \int_{\Gamma_\mathrm{c}^2} \langle - [(\mathbf{X} + \mathbf{u}) \cdot \mathbf{x} - X_\mathrm{P_2}] \rangle \delta \mathbf{u} \cdot \mathbf{x} \, dS = 0 
    \end{aligned} \label{eqn:variational_main} \\[10pt]
    &\int_{\Gamma_\mathrm{f}} \mathbf{P}_\mathrm{n}(\mathbf{u}) \cdot \partial\mathbf{u} \, dS = 0 \hspace{3.5cm} \text{(Neumann BCs)} \label{eqn:variational_neumann} \\[5pt]
    &\int_{\Gamma_\mathrm{u}} \mathbf{N}_\mathrm{t} \cdot \mathbf{P}(\mathbf{u}) \cdot \partial\mathbf{u} \, dS = 0 \hspace{3.3cm} \text{(Dirichlet BCs)} \label{eqn:variational_dirichlet}
\end{empheq}
\end{subequations}


\subsection{Open-source MRI-informed simulation framework}
\label{subsec_computational_model_simulation_framework}

\begin{figure}[htbp]
  \centering
  \hspace*{-0.5cm}
  \includegraphics[width=14.0cm]{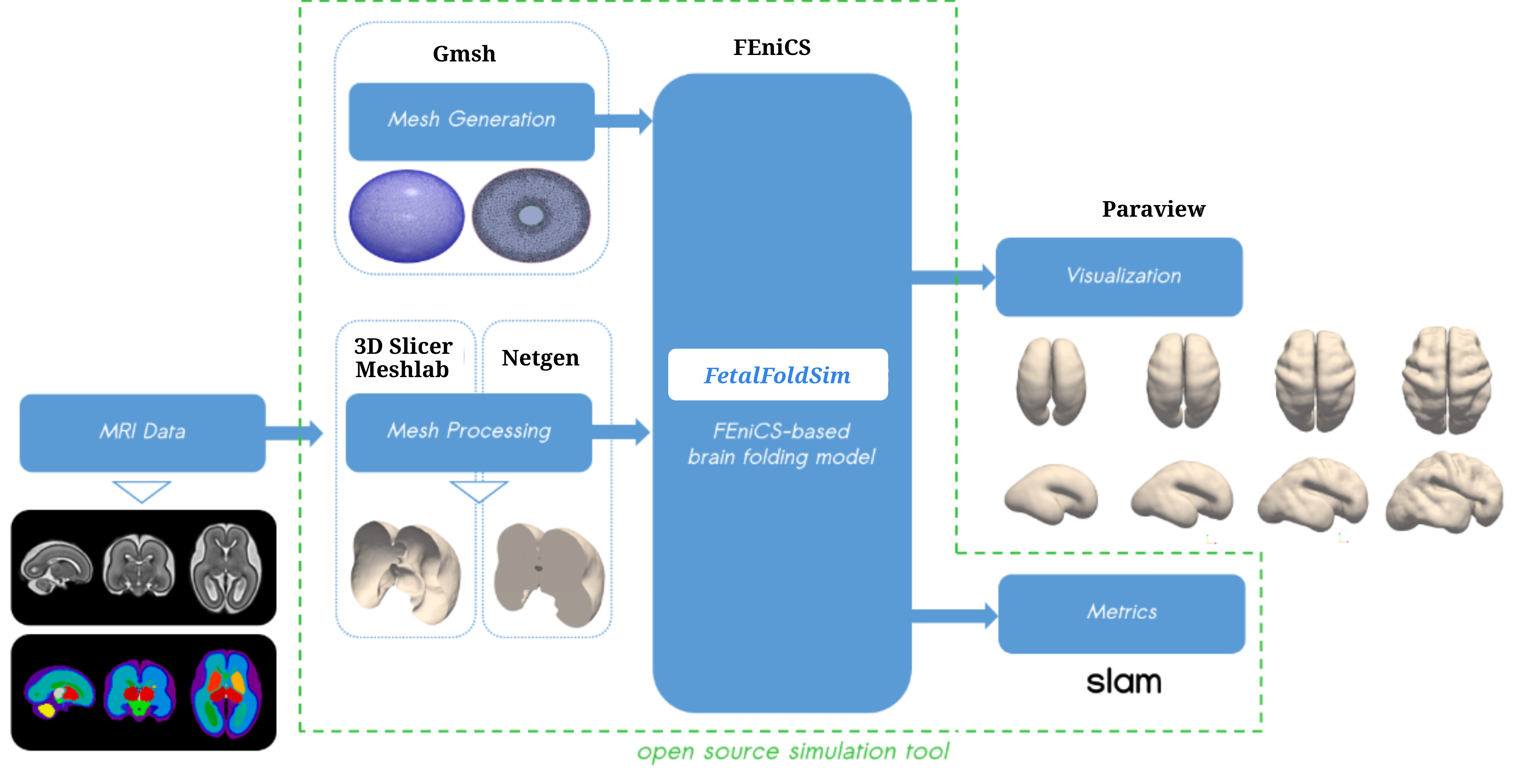}
  \caption{MRI-informed simulation framework around \textit{FetalFoldSim} (logos removed to respect copyright)}
  \label{fig:simulation_framework}
\end{figure}

\noindent A distinctive feature of our model, similarly to \cite{TTallinen2016, XWang2019, ZWang2021, MAlenya2022}, lies in its capacity to simulate cortical folding on the entire brain geometry. In \cite{MAlenya2022}, the authors introduced a comprehensive pipeline enabling to produce cortical folds from individual fetal MRI data, grounded in the biomechanical model of \cite{TTallinen2016}, and applicable to both healthy and pathological brains. Their work further proposed the computation of multiple quantitative metrics to compare simulated and real neurodevelopmental cortical surfaces. Building upon this approach that bridges biomechanical modeling and neuroimaging, this open-source simulation framework is illustrated in Fig. \ref{fig:simulation_framework}, centered on our finite element computational folding model, \textit{FetalFoldSim}\footnote{\url{https://github.com/gis-beachild/FetalFoldSim}}.
This framework aims to extend the modeling framework from raw fetal MRI input data to the prediction of whole-brain folds. It includes the use of other free software, \textit{3D Slicer}\footnote{\url{https://www.slicer.org/}} \cite{fedorov2012slicer}, \textit{Meshlab} 
and \textit{Netgen}
, to post-process medical imaging and generate input brain geometry; \textit{Gmsh}\footnote{\url{https://gmsh.info/}} \cite{geuzaine2009gmsh}, to generate more elementary geometries such as spheres or ellipsoids; \textit{Paraview} 
to visualize the simulation results; \textit{slam}\footnote{\url{https://github.com/brain-slam/slam}}, to compute geometrical and spectral evaluation metrics of the simulated cortical surfaces. 
It has been developed to be available to the community, modular, interpretable and adaptable.

\noindent Specifically, the framework uses fetal MRI data both to generate the input brain mesh and to obtain realistic estimates of key biomechanical parameters, including cortical growth rate. 
The simulations rely on biophysical parameters expressed in International System of Units (SI) and on a brain mesh in the RAS+ configuration.

\subsection{\textit{FetalFoldSim} computational model}\label{subsubsec_computational_model}

\subsubsection{\textit{FEniCS}-based model}

The \textit{FetalFoldSim} computational model of cortical folding is implemented using  \textit{FEniCS}\footnote{\url{https://en.wikipedia.org/wiki/FEniCS_Project}} \cite{MSAlnaes2015}, which enables to solve partial differential equations (PDEs) with the finite element method.\\
\textit{FEniCS}, a Python-based library, was selected for its open-source availability, owing to its large use in multiphysics simulations and its capability to semi-automate problem implementation. Solving a PDE first requires expressing its variational formulation. This expression allows discretization into a matrix system solvable by numerical methods through the computation of matrix coefficients. With \textit{FEniCS}, only the integrands of the variational form need to be specified according to the library syntax; the assembly of the system matrices is handled automatically, eliminating the need for explicit coefficient calculation. Solver parameters, particularly the linearization method and the linear solver, need nevertheless be defined explicitly. 
Furthermore, \textit{FEniCS} supports large deformations typical of biomedical tissues, such as those in cortical folding modeling, and has already been used to solve variational biomechanical problems involving the penalization of contact and, more particularly, self-contact \cite{CPatte2022, XChen2024, RTDjoumessi2025}.




\vskip0.2cm

\noindent 

\vskip0.2cm 

\subsubsection{Solver options}
In \textit{FetalFoldSim}, Eq. \eqref{eqn:variational_main} is linearized using the \textit{Newton-Raphson} method. 
Non-linear problems may exhibit difficult convergence, under-relaxation can be employed to prevent the iterations from diverging or oscillating around the solution. In practice, we use an under-relaxed scheme, with a relaxation parameter set to a value less than unity\footnote{\footnotesize\url{https://home.simula.no/~hpl/homepage/fenics-tutorial/release-1.0-nonabla/webm/nonlinear.html}}.
The direct solver \textit{mumps} has been chosen to straightforward allow parallel computation.   

\subsubsection{Delineation of the two brain layers}
The weighting function $gr_\mathrm{Cortex}$ is used to 
continuously delineate the cortical and subcortical layers, depending on the initial cortical thickness, as in \cite{TTallinen2016}. $gr_\mathrm{Cortex}$ is one in the cortical layer and tends to zero in the sublayers of the brain volume, with a smooth transition between the two layers.\\ 

\noindent In the \textit{FetalFoldSim} model, such delineation is first used to define locally the brain stiffness $\mu$ (Eq. \eqref{eqn:local_shear_modulus}) and bulk modulus $K$ and (Eq. \eqref{eqn:local_bulk_modulus}). In Eq. \eqref{eqn:local_shear_modulus} and \eqref{eqn:local_bulk_modulus}, $\mu_\mathrm{Core}$ and $K_\mathrm{Core}$ denote shear and bulk moduli of the sublayers, respectively, whereas $\mu_\mathrm{Cortex}$ and $K_\mathrm{Cortex}$ represent the shear and bulk moduli of the cortex. These quantities are prescribed input parameters of the model.

\begin{equation}\label{eqn:local_shear_modulus}
    \mu(\mathbf{X}) = \mu_\mathrm{Core} \cdot (1.0 - gr_\mathrm{Cortex}) + \mu_\mathrm{Cortex} \cdot gr_\mathrm{Cortex}
\end{equation}

\begin{equation}\label{eqn:local_bulk_modulus}
    K(\mathbf{X})  = K_\mathrm{Core} \cdot (1.0 - gr_\mathrm{Cortex}) + K_\mathrm{Cortex} \cdot gr_\mathrm{Cortex}
\end{equation}

\vskip0.2cm

\begin{figure}[htbp] 
  \centering
  \includegraphics[width=6.0cm]{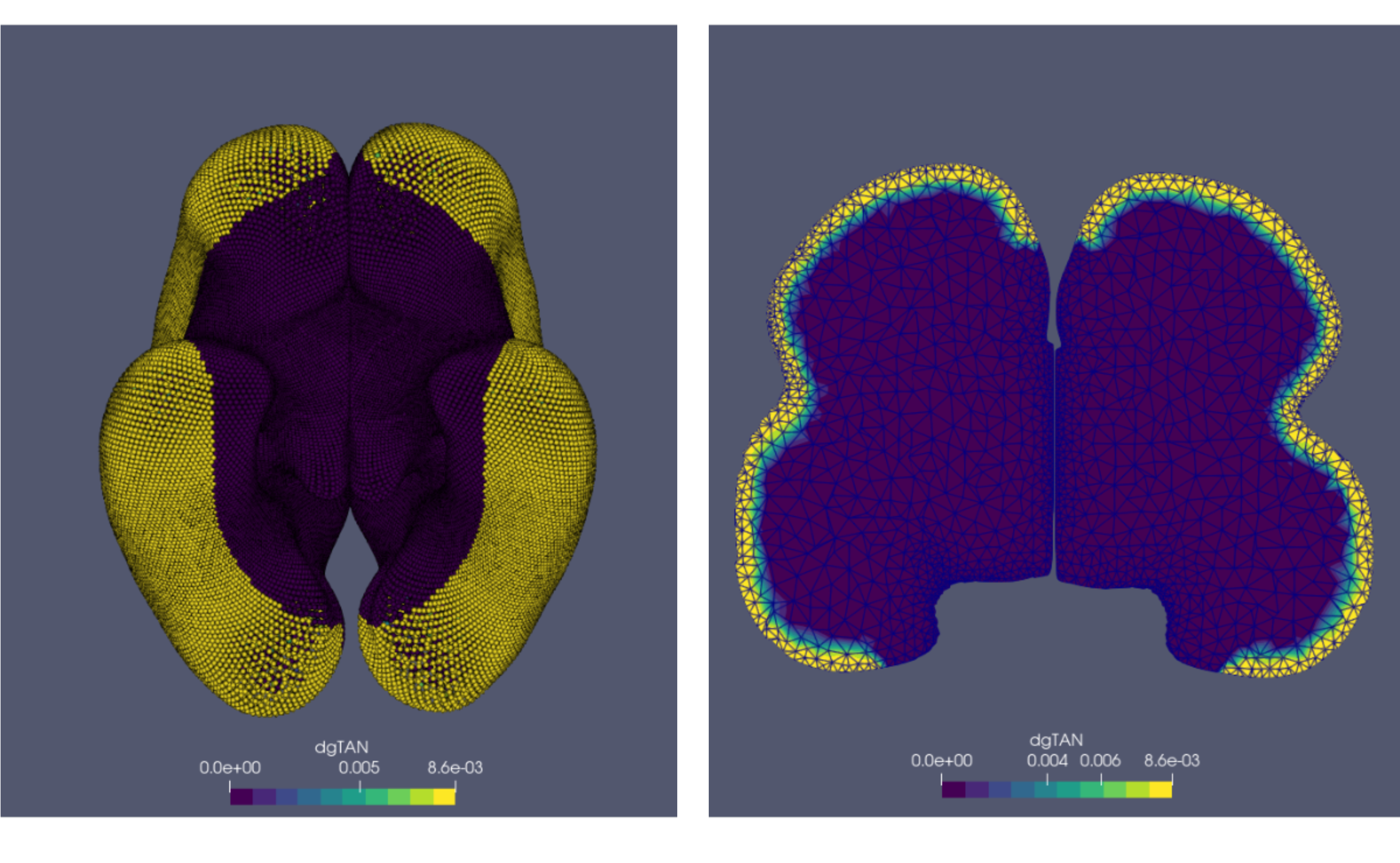}%
  \caption{Tangential growth stretch increment $\mathrm{d}g_{\mathrm{TAN}}$ in the whole brain}
\label{fig:dgTAN}
\end{figure}

\noindent The $gr_\mathrm{Cortex}$ function also allows to define the local tangential ($\mathrm{d}g_{\mathrm{TAN}}$) and radial ($\mathrm{d}g_{\mathrm{RAD}}$) growth stretch increments (Eq. \eqref{eqn:dgTAN_grCortex} and \eqref{eqn:dgRAD_grCortex}). $\alpha_{\mathrm{RAD}} $ denotes the radial growth rate in the sublayers; $\alpha_{\mathrm{TAN}} $ and $\alpha_{\mathrm{RAD}} $ are input parameters of the model. Through the $gr_\mathrm{Cortex}$ function, tangential growth is allocated to the cortical layer, while radial growth can be allocated to the sublayers of the brain volume (standing respectively for neuronal migration and differentiation). As mentioned in Section \ref{subsection_growth_law_cortex}, in this work, only cortex is growing, then, $\alpha_{\mathrm{RAD}} $ is set to $0$ to omit the sublayers growth. Fig. \ref{fig:dgTAN} illustrates the tangential growth stretch increment for all mesh nodes.

\begin{equation}\label{eqn:dgTAN_grCortex}
    \mathrm{d}g_{\mathrm{TAN}}(\mathbf{X})  = \alpha_{\mathrm{TAN}}  \cdot gr_\mathrm{Cortex} \cdot \mathrm{d}t 
\end{equation}

\begin{equation}\label{eqn:dgRAD_grCortex}
    \mathrm{d}g_{\mathrm{RAD}}(\mathbf{X})  = \alpha_{\mathrm{RAD}}  \cdot (1 - gr_\mathrm{Cortex}) \cdot \mathrm{d}t = 0 
\end{equation}
\subsubsection{Non-growing brain regions}
\begin{figure}[htbp] 
  \centering
  \includegraphics[width=3.0cm]{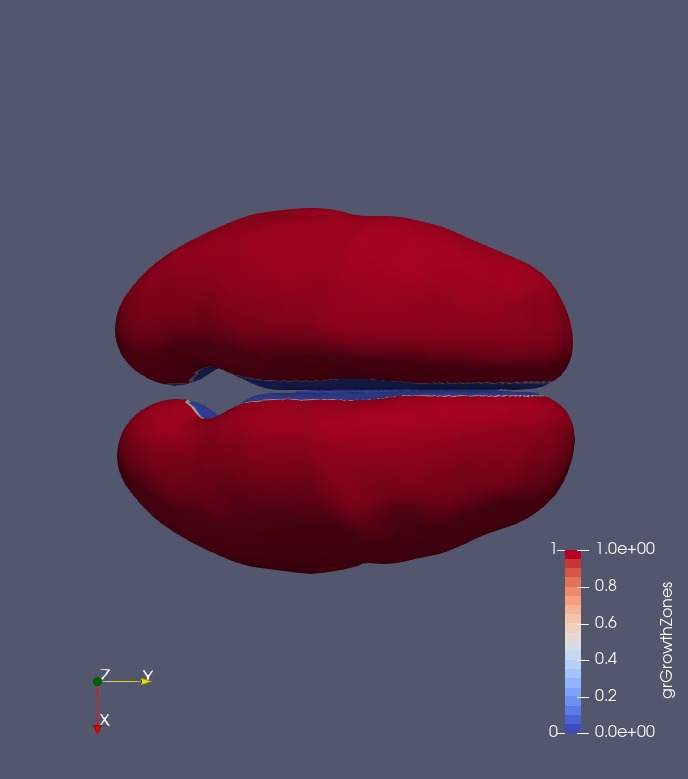}%
  \includegraphics[width=3.0cm]{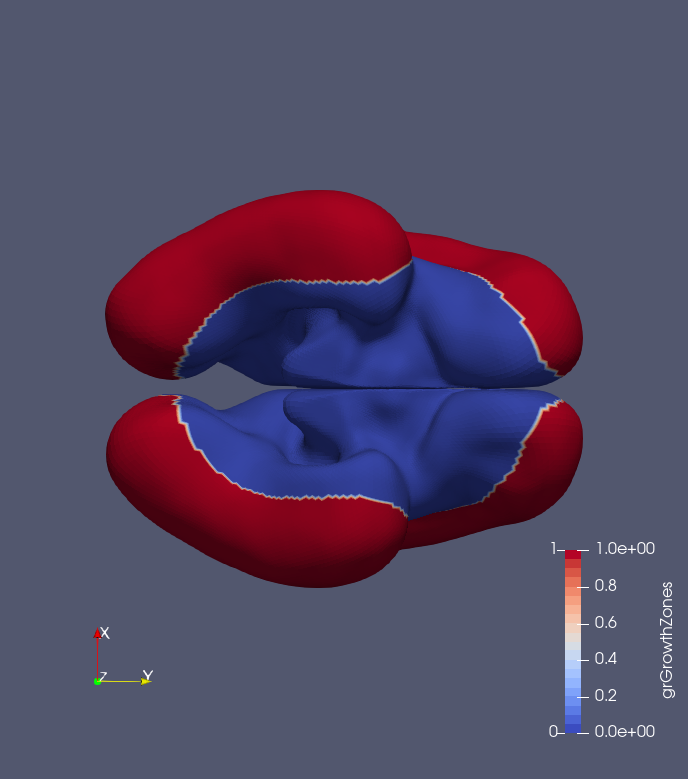}%
  \includegraphics[width=3.0cm]{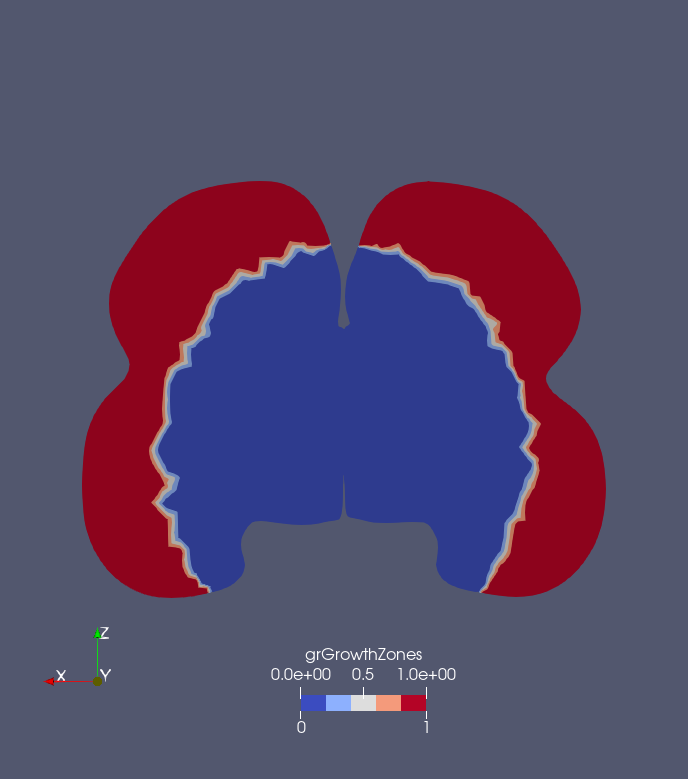}%
  \caption{Weighting function $gr_\mathrm{GrowthZones}$ that enables to define growth zones: equals 1 in growing regions (red); 0 in no-growth regions (blue)}
\label{fig:grGrowthZones}
\end{figure}

In \cite{TTallinen2016}, selected brain regions such as longitudinal fissure, ventricles and mammillary bodies were designated as non-growing regions in both the numerical model and the swelling brain gels experience.
Preventing their growth enables to avoid abnormal collision and non-physiological brain growth and deformations.
\vskip0.1cm
\noindent In \textit{FetalFoldSim}, the no-growth zones are defined based on an ellipsoid that intersects with the brain mesh, as proposed by \cite{TTallinen2016}, and defined by the growth weighting function $gr_\mathrm{GrowthZones}$ (Fig. \ref{fig:grGrowthZones}). 

\noindent In our computational model, to limit brain growth to the selected cortical regions described previously, the growth stretch increments for both cortex and the sublayers, obtained by Eq. \eqref{eqn:dgTAN_grCortex} and \eqref{eqn:dgRAD_grCortex}, are then multiplied by the weighting function $grGrowthZones$. The growth laws in the cortex and sublayers are then defined by Eq. \eqref{eqn:dgTAN_grCortex_grGrowthZones} and \eqref{eqn:dgRAD_grCortex_grGrowthZones}, which are the equations implemented in the \textit{FetalFoldSim} computational model.

\begin{equation}\label{eqn:dgTAN_grCortex_grGrowthZones}
    \mathrm{d}g_{\mathrm{TAN}}(\mathbf{X})  = \alpha_{\mathrm{TAN}}  \cdot gr_\mathrm{Cortex} \cdot gr_\mathrm{GrowthZones} \cdot \mathrm{d}t
\end{equation}

\begin{equation}\label{eqn:dgRAD_grCortex_grGrowthZones}
    \mathrm{d}g_{\mathrm{RAD}}(\mathbf{X})  = \alpha_{\mathrm{RAD}}  \cdot (1 - gr_\mathrm{Cortex}) \cdot gr_\mathrm{GrowthZones} \cdot \mathrm{d}t = 0
\end{equation}

\subsubsection{Adaptive growth tensor}
\begin{figure}[htbp]
    \centering
    \hspace*{-2cm}
    \includegraphics[width=16.0cm]{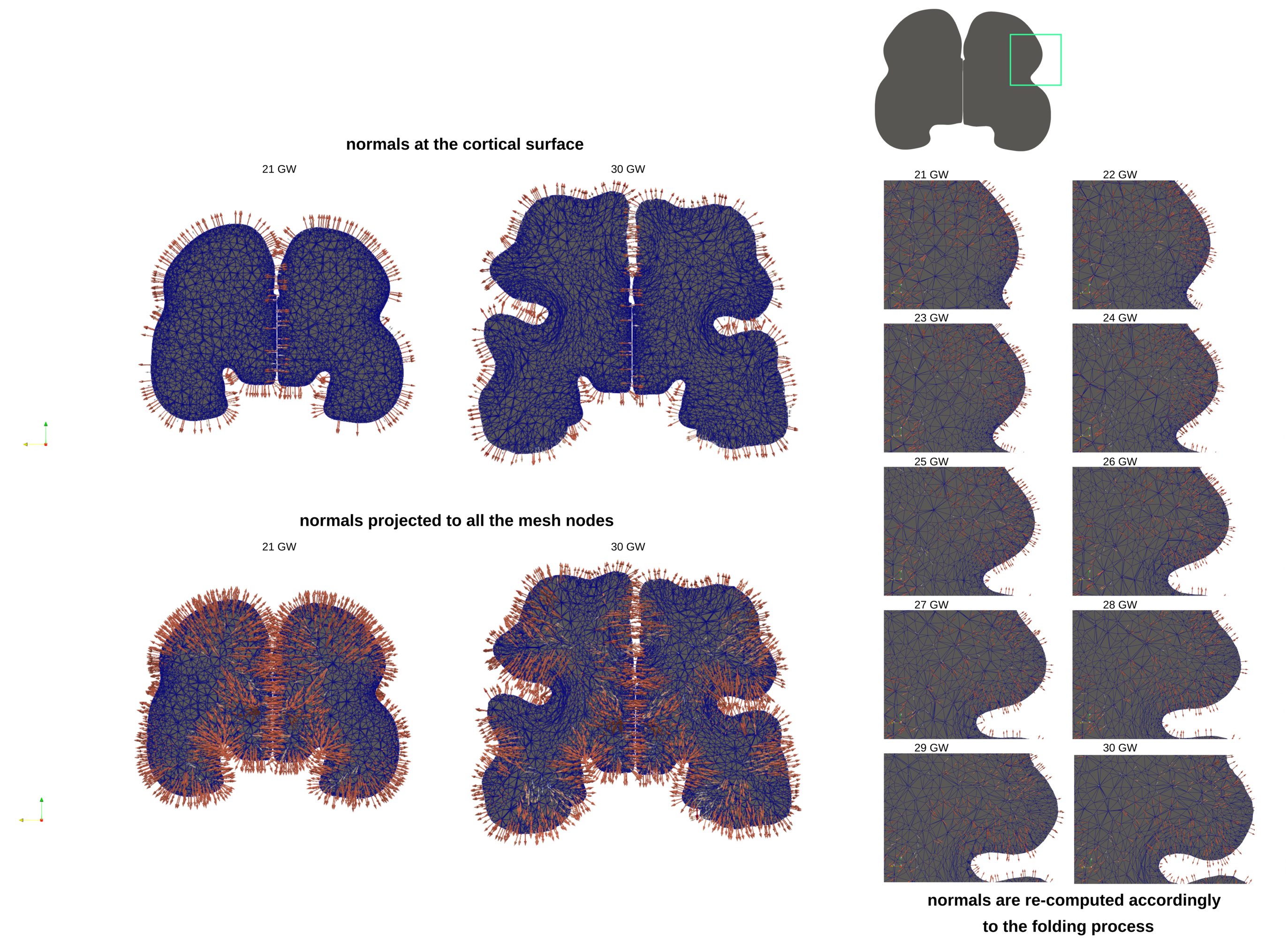}
    \caption{Adaptive mesh normals: surface normals of the brain mesh for a sample of surface nodes (top left image). For each volume node, the previous normal at the nearest neighboring surface node is projected; projected normals $N_t$ are visualized for a sample of volume nodes (bottom left). Normals are re-computed at each simulation step to adapt to the folding brain surface, as seen from 21 GW to 30 GW (right)}
\label{fig:mesh_nt_zoom_21_30GW}
\end{figure}

\noindent As the brain mesh undergoes folding during the simulation, the cortical surface normals undergo a progressive change in direction; the tangential direction of the growth should evolve accordingly \cite{ZWang2021}. To this end, at each time step of the simulation, the normals at the cortical surface (Fig. \ref{fig:mesh_nt_zoom_21_30GW}, top of left column) and the projected local normals $N_t$ at each vertex of the volume mesh (Fig. \ref{fig:mesh_nt_zoom_21_30GW}, bottom of left column) are re-computed. This update ensures an adaptive directional basis to re-compute the growth tensor (see Eq. \eqref{eqn:growth_tensor_Cortex}) 
(Fig. \ref{fig:mesh_nt_zoom_21_30GW}, right column). 

\subsubsection{Dirichlet boundary conditions}
\noindent To define the Dirichlet BCs, we propose to clamp the brain mesh regions which have no growth (blue regions in Fig. \ref{fig:grGrowthZones}), excluding the contact boundaries, defined in Section \ref{subsubsec_boundary_conditions} (yellow and blue surfaces in Fig. \ref{fig:brain932K_boundaries}) and the inner region between the two contact boundaries (purple). The resulting regions where Dirichlet BCs are applied are displayed in green in Fig. \ref{fig:brain932K_boundaries}.

\subsection{Model parameters}\label{subsec_parameters}

\subsubsection{Biophysical values for homogeneous parameters}\label{subsubsec_homogeneous_model_parameters}

\noindent Our cortical folding model presents six main input parameters that can impact the simulation results: 
\begin{itemize}
    \item the thickness of the brain cortex $H_0$; 
    \item the stiffnesses (shear moduli) of both the cortex and the sub-layers $\mu_\mathrm{Cortex}$ and $\mu_\mathrm{Core}$ and more importantly their ratio; 
\item the Poisson's ratio $\nu$; 
\item the growth ratio, in our case the tangential growth rate in the cortex layer $\alpha_{\mathrm{TAN}} $; 
\item the timestep $dt$. 
\end{itemize}
In this work, we assume all parameters are \textit{constant} in space and time, a strong assumption we will discuss in the last section.
\noindent Some parameters play a crucial role on defining the folding pattern. If a minimum intensity in the differential growth rate triggers primary and secondary folds \cite{SBudday2018}, other parameters like cortical thickness or the ratio of the shear modulus of the cortex to the shear modulus of the sub-layers influence the frequency and shape of the emerging folds \cite{XWang2019, XWang2021, MJRazavi2015, TTallinen2014}. 

\noindent The issue is to identify simulation parameters directly connected to patient data, that have realistic biophysical values and expressed in the international system (IS), interpretable both for scientists and clinicians. 

\noindent In this study, the parameter values listed in Table \ref{table:parameters} are used as reference values for the simulations. These parameters are taken from a literature review
\cite{LVasung2016,XXu2022,RdeRooij2018,MSZarzor2021,EGriffiths2023,SBudday2014,TTallinen2016},
and are expressed in SI units. The cortical growth rate is estimated from MRI data, as detailed in next Section.

\subsubsection{Estimation of the cortical growth rate from human healthy fetal cerebral MRI data} 
Cortical growth rate, a key model parameter \cite{MJRazavi2015bis, SBudday2018}, prove to be difficult to define and hard to compare from a computational model to another. 
\noindent We compute the cortical growth rate from the healthy fetal MRI data provided by the dHCP project, between 21 GW and 29 GW, to cover the gestational neurodevelopmental period during which primary folds emerge and develop \cite{HJYun2022}.

\vskip0.2cm 
The tangential cortical \textit{growth rate} \cite{SBudday2014, MAHolland2013}, noted as $\alpha_{\mathrm{TAN}}^\mathrm{Cortex}$, reflects the differential amount of growth between the cortex and the subcortical layers and is defined as follows: 

 \begin{equation}\label{cortical_tangential_growth_rate}
 (\theta_\mathrm{TAN}^\mathrm{Cortex})^{\mathrm{1/dim}} = 1 + \alpha_{\mathrm{TAN}} ^\mathrm{Cortex} \cdot \mathrm{d}t
 \end{equation}
 
\noindent 
where $\mathrm{dim}$ is the dimension of the material under growth  \cite{MAHolland2015, TTallinen2016}; and $\theta_\mathrm{TAN}^\mathrm{Cortex}$ is the \textit{growth multiplier} \cite{SBudday2014, MAHolland2013, SWang2021, TTallinen2016}.
We assume that the brain material under growth is the surface of the cerebral cortex, consequently, we consider dim=2.\\ $\theta_\mathrm{TAN}^\mathrm{Cortex}$ is expressed by the ratio between the cortical areas at two distinct gestational ages (e.g. Eq. \eqref{eqn:growth_multiplier}) . 

\begin{equation}\label{eqn:growth_multiplier}
    \theta_\mathrm{TAN}^\mathrm{Cortex} (21 GW \rightarrow 22 GW) = \displaystyle \frac{A_{22}^\mathrm{Cortex}}{A_{21}^\mathrm{Cortex}}
\end{equation}

\noindent Then the growth rate between two successive gestational ages is computed using Eq. \ref{eq:tangential_growth_rate_from_cortex_area}.

\begin{equation}\label{eq:tangential_growth_rate_from_cortex_area}
    \alpha_{\mathrm{TAN}} ^\mathrm{Cortex}(t_{GW} \rightarrow t_{GW} + 1) = \displaystyle \frac{(\theta_\mathrm{TAN}^\mathrm{Cortex})^{1/2} - 1}{\mathrm{d}t} = [({\displaystyle \frac{A_{\hspace{1mm}t_\mathrm{GW} + 1}^{\hspace{1mm} \mathrm{Cortex}}}{A_{\hspace{1mm}t_\mathrm{GW}}^{\hspace{1mm} \mathrm{Cortex}}})^{1/2}} - 1] \cdot \displaystyle \frac{1}{\Delta t_{\hspace{1mm} t_\mathrm{GW}\rightarrow t_\mathrm{GW} + 1}}
\end{equation}

\begin{figure}[htbp]

  \centering
\includegraphics[width=0.5\textwidth]{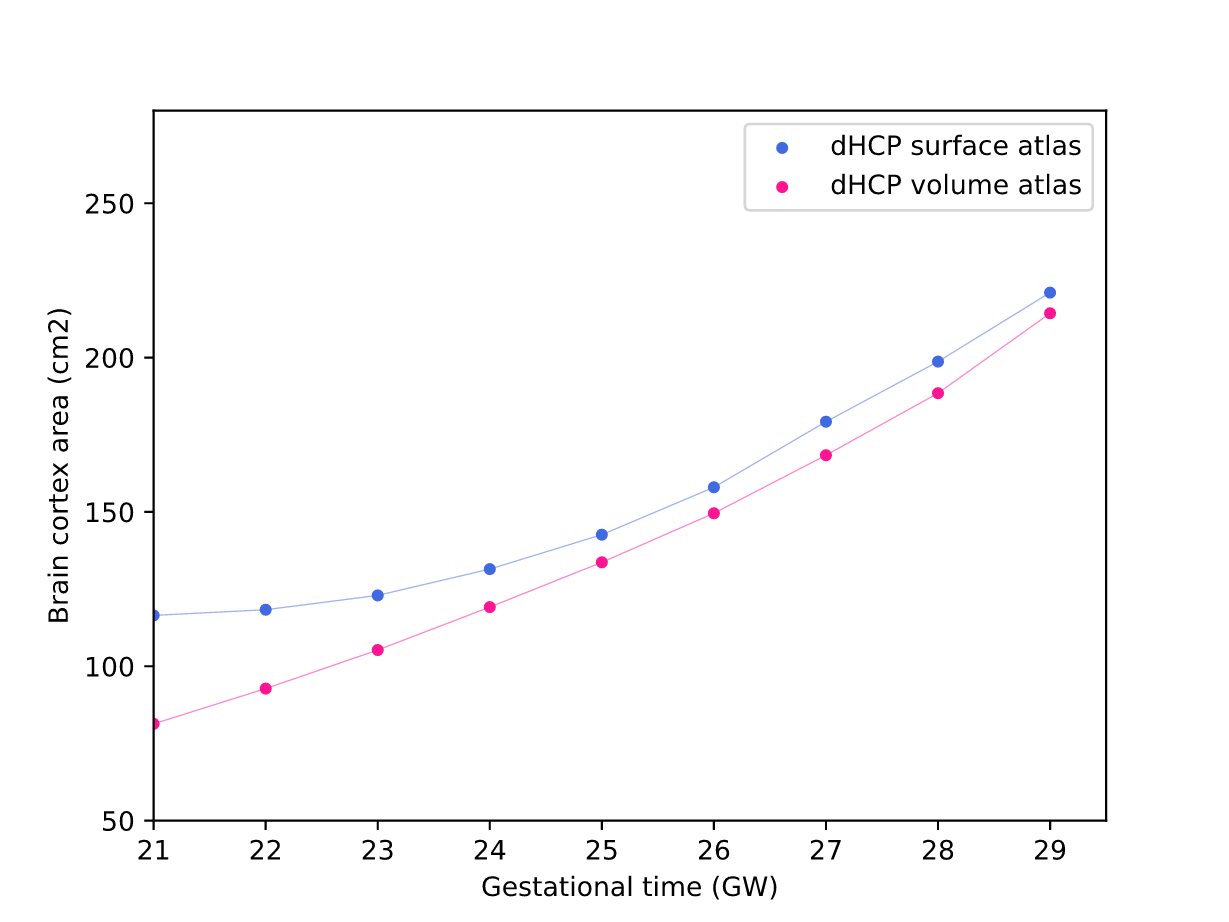} 
  
  \vspace{0.5cm}
  
\includegraphics[width=0.48\textwidth]{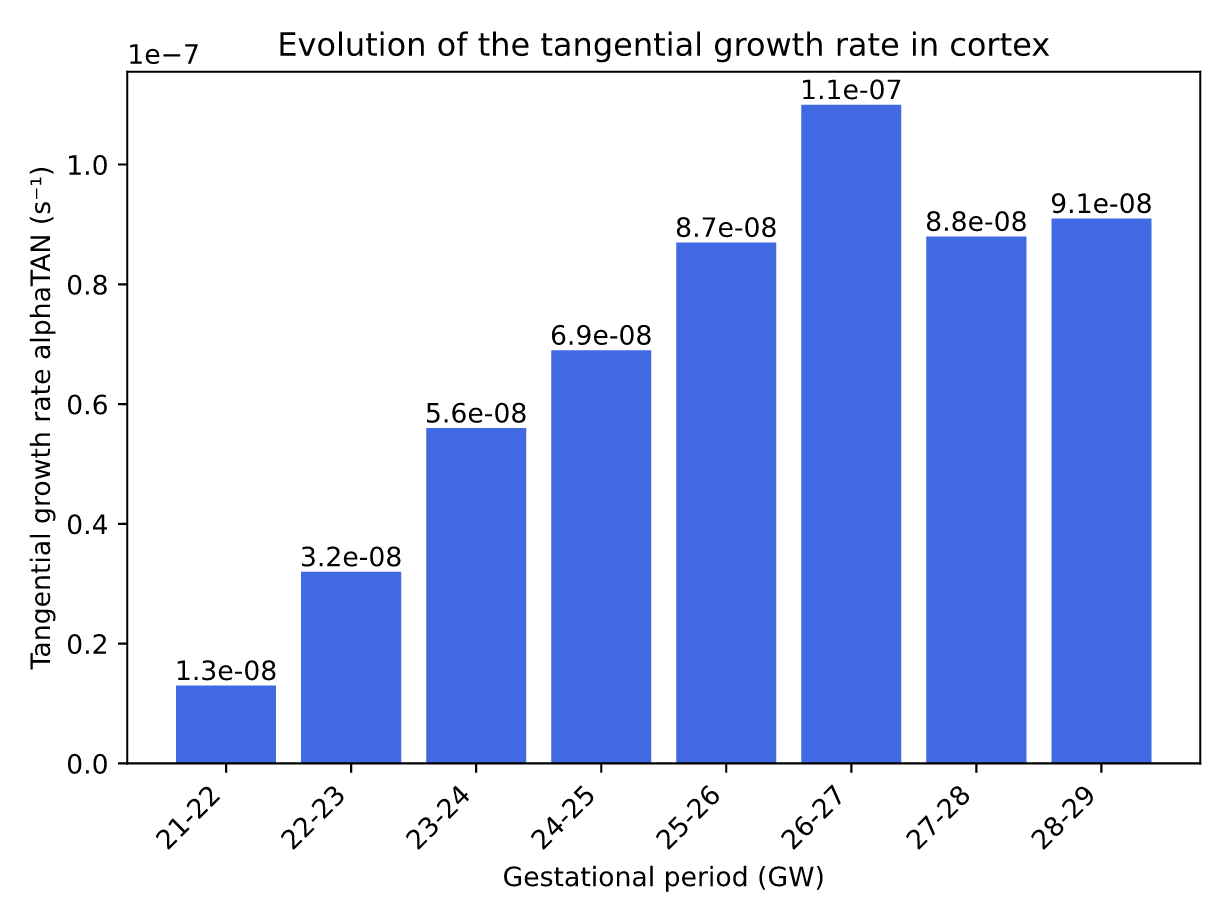} 
  \hfill
\includegraphics[width=0.48\textwidth]{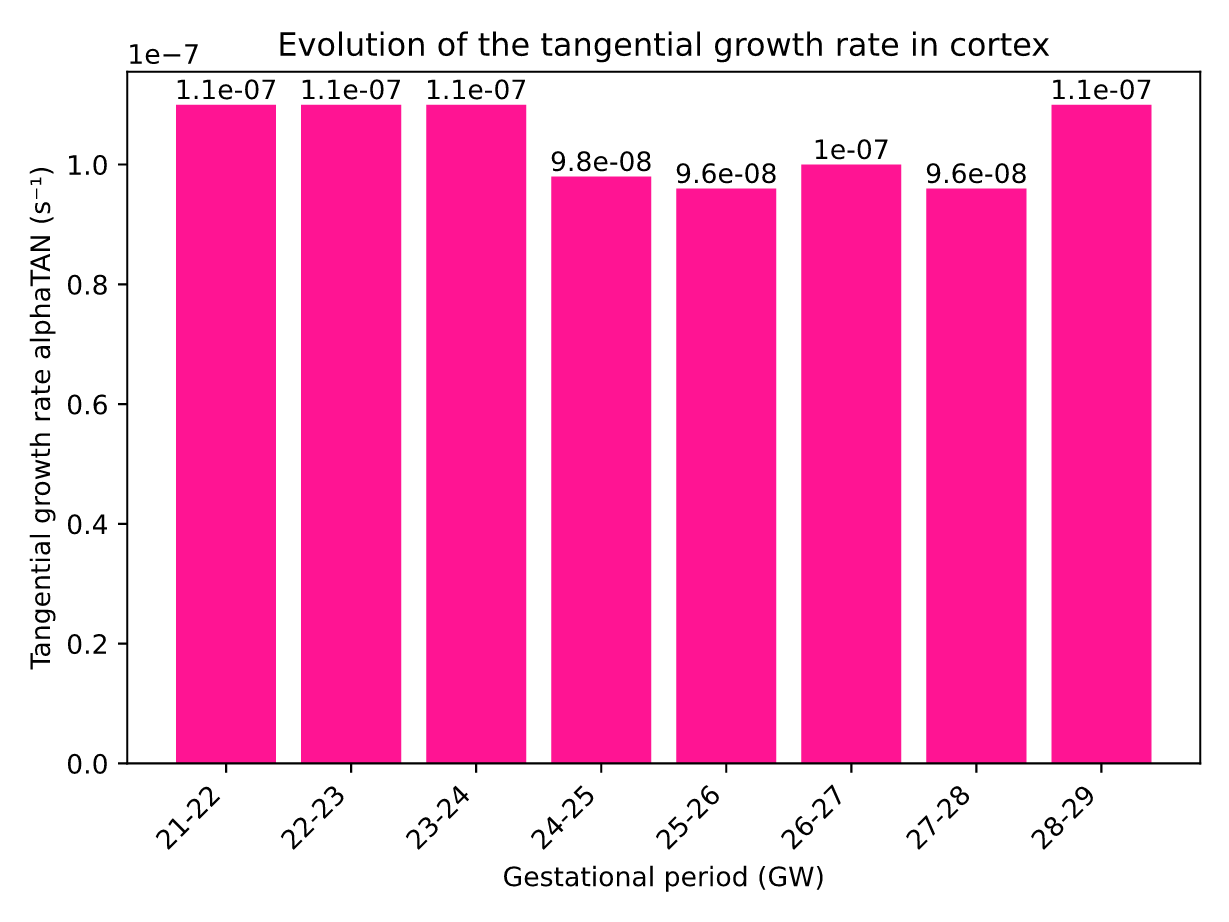} 
  
  \caption{Evolution of the cortex area for both the \textit{dHCP surface} and \textit{dHCP volume} meshes from 21 to 29 GW (a). Estimation of the tangential growth rates in the cortex layer from the \textit{dHCP surface} (b) and \textit{dHCP volume} (c) meshes for successive gestational periods (between 21 and 29 GW)}
\label{fig:tGW_to_cortex_areas_AND_alphaTANs_dHCPsurface_versus_dHCPvolume_21_29GW} 
\end{figure}

\vskip0.2cm 
dHCP atlas data are provided as surface meshes and volumetric data: the dHCP atlas fetal surface hemispheric meshes (\textit{dHCP surface} meshes) and the dHCP atlas fetal MR NIfTIs\footnote{\footnotesize\url{https://gin.g-node.org/kcl_cdb/fetal_brain_mri_atlas}} (see Fig. \ref{fig:dHCP_surface_data_21_36GW}) (\textit{dHCP volume} meshes). To find the more appropriate data type for growth modeling, we compared the two by computing the cortical areas from 21 to 29 GW. The evolution of cortical areas over gestational time is displayed in Fig. \ref{fig:tGW_to_cortex_areas_AND_alphaTANs_dHCPsurface_versus_dHCPvolume_21_29GW}.
The tangential growth rate in the cortex is deduced for gestational periods from $21GW\rightarrow22GW$ to $28GW\rightarrow29GW$ using Eq. \ref{eq:tangential_growth_rate_from_cortex_area}. The results are displayed in Fig. \ref{fig:tGW_to_cortex_areas_AND_alphaTANs_dHCPsurface_versus_dHCPvolume_21_29GW} (histograms). 

\vskip0.1cm
\noindent The evolution of the cortex area for the surface data demonstrates a greater degree of similarity to previously documented results in literature \cite{JDubois2019, XXu2022} than the curve for the volume data. The cortical tangential growth rate is therefore estimated at minimum $1.0 \times 10^{-7} s^{-1}$.

\subsection{Metrics}
\label{subsec_biometrics}

\noindent As we simulate cortical folding on a brain geometry derived from fetal neurodevelopmental MRI data, a comparative analysis between simulation outcomes and these real data is possible. Combining cortical folding simulations with MRI data thus facilitates the exploration on the influence of the model parameters on dedicated metrics that quantify the folding pattern of the cortical surface. Second, it opens the way to optimize the similarity between simulations and real data. 

\vskip0.1cm

Defining appropriate metrics to quantify fetal cortical folding is not trivial since the concept of “ground truth” is inherently undefined. First, each brain shape and folds are individual-specific; second; the reference frame for a same individual's brain evolves during development. Metric \textit{intrinsic} to the brain object (independent to the reference frame) are therefore preferred, since they can be computed for any brain geometry in any reference and their computation do not need registration of simulation or of real meshes at different gestational times. 

\vskip0.1cm

\textit{Global metrics} that quantify the folding pattern of the cortical surface, such as the \textit{brain volume} ($mL$), the \textit{area of the cortex surface}  ($cm^2$), and the \textit{index of gyrification}, referred as to GI, have been used to compare both simulation results and real data. In particular, GI is computed as the ratio of the area of the cortex surface mesh to the area of its smooth ”convex hull” and as already proposed by \cite{SBudday2014, TTallinen2016, XWang2021}. It is frequently used as a reference metric in quantifying the degree of cortical folding. However, as brain development is a phenomenon that evolves following a spatio-temporal gradient, being heterogeneous, global metrics do not adequately account for the local dynamics of cortical folding. Therefore, \textit{local metrics} are preferred in this work. Among them, we present and compute two types of local metrics: the \textit{shape} and \textit{curvedness indices} \cite{JJKoenderink1992, HHHu2013, JLefevre2015, NDemirci2024}, that quantify the local curvature of the cortical surface; the \textit{SpAnGy} spectral analysis method \cite{DGermanaud2012} to analyze the local degree of folding of the cortical surface. 

\subsubsection{Shape and curvedness indices}
\noindent The shape index characterizes the local shape of a surface. It is defined by Eq. \eqref{eqn:shape_index} \cite{JJKoenderink1992} where $\kappa_1$ and $\kappa_2$ are the two principal curvatures. It ranges from $-1$ and $1$: $SI \in [-1, -0.5]$: sulcal pit (sulcus); $SI \in [-0.5, 0]$: sulcal saddle; $[0, 0.5]$: gyral saddle and $[0.5, 1]$: gyral node (gyrus). The shape index thus provides information on the type of fold. Additionally, it is independent of the brain mesh size \cite{JLefevre2015, NDemirci2024}.

\vskip0.2cm

\noindent The curvedness index (CVD), defined by Eq. \eqref{eqn:curvedness_index} \cite{JJKoenderink1992}, is a positive measure that represent the power of folding \cite{JLefevre2015}, i.e. the sharpness of the fold. In other words, the more pronounced the fold, the higher the curvedness index.

\begin{equation}\label{eqn:shape_index}
    SI = \dfrac{2}{\pi} \mathrm{arctan}(\dfrac{\kappa_1 + \kappa_2}{\kappa_2 - \kappa_1}), \hspace{2mm} \kappa_1 \geq \kappa_2
\end{equation}

\vskip0.1cm

\begin{equation}\label{eqn:curvedness_index}
    CVD = \sqrt{\dfrac{\kappa_1^2 + \kappa_2^2}{2}}
\end{equation}

\subsubsection{SpAnGy}\label{SpAnGy}
A method for spectral analysis of gyrification using frequency bands (\textit{SpAnGy}) was proposed by \cite{DGermanaud2012}. \cite{JDubois2019} employed this method to characterize the evolution of folding in premature infants during brain development from 27 GW. \cite{MAlenya2022} used \textit{SpAnGy} to characterize cortical curvature in both simulations and real fetal brain data, to validate simulations. 
\noindent In this study, we propose to examine the contribution of the \textit{SpAnGy} method, with a particular focus on its spatial frequency bands $B4$, $B5$ and $B6$, to assess the ability of the \textit{FetalFoldSim} model to reproduce the emergence of the primary folds \textit{antenatally}. To do so, we compute the band spectrum with the \textit{SpAnGy} code (package \textit{slam})\footnote{\url{https://github.com/gauzias/slam/blob/master/slam/spangy.py}} to compare the folding dynamics of the simulations with those shown by the dHCP data presented in Section \ref{subsec_mri_atlas_data}.
\vskip0.2cm

\textit{SpAnGy} decomposes a folding proxy (e.g. curvature, shape index) defined on the cerebral cortex surface mesh on the basis of eigenmodes of the Laplace-Beltrami operator, and merges the resulting coefficients into spatial frequency bands. The spectral density and relative spectral density (to the total spectral density) are computed for each frequency band. Depending on the local value of the proxy, the cortical surface is associated with a dominant band of spatial frequencies. As the folded surface is composed by negative and positive curvature zones, the associated frequency bands extend from $B_{-6}$ to $B_6$, with $B_{-i}$ corresponding to the zones where the curvature is negative, and $B_{i}$ equivalent zones where the curvature is positive. 
\vskip0.2cm

The spatial frequency limits of each band have been chosen so that the transition from a band $B_n$ to the higher frequency band $B_{n+1}$ mimics the emergence of new folds of half the wavelength. The wavelength of each eigenmode is defined from the corresponding eigenvalue. Starting from the main wavelength $WL_0$, the minimum wavelength of the band $B_n$ is deduced \cite{DGermanaud2012} (Eq. \ref{eq:WL_n}). 
\vskip0.2cm

Weyl's law (Eq. \ref{eq:Weyl_number}) \cite{ZGao2014} gives the order $N(\lambda)$ of the eigenvalue $\lambda$ for large values, in the continuous case. By applying Eq. \ref{eq:lambda_max}, the maximum eigenvalue for the $B_6$ band can be determined, allowing us to estimate the approximate number of modes that need to be computed.

\noindent
\begin{minipage}{0.26\textwidth}
\begin{equation}
  WL_n = \dfrac{WL_0}{2^{\mathrm{n}}} 
  \label{eq:WL_n}
\end{equation}
\end{minipage}
\hfill
\begin{minipage}{0.28\textwidth}
\begin{equation}
  \lambda_{max} = \dfrac{2\pi}{{WL_6}^2} 
  \label{eq:lambda_max}
\end{equation}
\end{minipage}
\hfill
\begin{minipage}{0.36\textwidth}
\begin{equation}
  N(\lambda) \mathop{\sim}\limits_{+\infty} \dfrac{Area(\Omega) \cdot \lambda}{4\pi}
  \label{eq:Weyl_number}
\end{equation}
\end{minipage}

\vskip0.3cm

The following data are computed from the surface mesh to be characterized: the spectral density for each band $B_4$, $B_5$ and $B_6$ of spatial frequencies associated with the fold; the relative spectral density (to the total spectral density) for each band $B_4$, $B_5$ and $B_6$; the eigenvalues of the spectrum of the Laplace-Beltrami operator and the modal composition of each band ($B_{-6}$ to $B_6$).
\vskip0.1cm

The canonical definition of the \textit{SpAnGy} band power spectrum considers the mean curvature as the folding proxy to be decomposed, and the association between B4, B5, and B6 with the primary, secondary, and tertiary folds of \cite{DGermanaud2012, JDubois2019} has been demonstrated based on this proxy. Nevertheless, we propose here to use the \textit{shape index} to represent the folding pattern, as it is, unlike mean curvature, independent of the brain mesh size \cite{NDemirci2024} and may enable a more robust comparison of the local states of curvature for the real data, and all the simulations. 

\vskip0.3cm

\section{Simulating primary cortical folding in healthy conditions}\label{sec3} 

\noindent We simulate cortical folding between 22 and 29 GW with the \textit{FetalFoldSim} model from a whole brain 3D mesh at 21 gestational weeks. The mesh is generated from the post-processed hemispheric surface atlas data from the dHCP project (see Section \ref{subsec_mri_atlas_data}). The model parameters are the reference ones in Table \ref{table:parameters}.

\subsection{Qualitative assessment of the \textit{FetalFoldSim} simulation results} 

\noindent Fig. \ref{fig:brain_growth_simulation_REF} compares simulation predictions with the brain meshes built from the MRI atlas data of human healthy fetal brains presented in Section \ref{subsec_mri_atlas_data} (dHCP surface), for gestational times from 22 to 29 GW.

\noindent The simulations reveal a progressive folding of the cortical surface as the brain develops. Interestingly, some sulci that are characteristic of normal fetal neuro-development are visible (refer to \cite{TTallinen2014, PAHabas2012, HdeVareilles2023} for the early folds location). Indeed, the central right and superior temporal sulci (CS, STS) develop from the initial simulation time. Additionally, the pre- and post-central sulci (preCS, postCS), as well as the superior and inferior frontal sulci (SFS, IFS), are clearly discernible at 28 GW simulated.
\noindent In contrast to the folding simulations proposed by \cite{TTallinen2016}, the location of the folds appears to be more realistic. However, the development of fold seems to be less pronounced than in \cite{MAlenya2022,TTallinen2016}, which may be attributable to the fact that collisions are not handled globally, potentially leading to an anticipated crash of the simulation (around 28/30 GW). Like in the simulation of \cite{TTallinen2016}, a prevalence of primary folds is observable. This feature is coherent with what we know of the appearance of secondary and tertiary folds after 28 GW.

\subsection{Assessment of the computational model physical validity}

\subsubsection{Computational time and energy cost}
\noindent Simulating cortical folding between 21 and 33 GW required $\sim$ 28 hours on Intel Xeon 4 CPUs.
This corresponds to approximately 16 hours and 20 minutes for the period between 21 and 28 gestational weeks, or 2 hours and 20 minutes per gestational week. By extrapolation, simulating cortical folding over the entire third trimester (21 to 40 gestational weeks) would require an estimated 44 hours and 20 minutes.
\noindent According to the carbon emissions calculator \textit{Green Algorithms calculator} \footnote{\url{https://calculator.green-algorithms.org/}}, running a simulation for one gestational week using \textit{FetalFoldSim} on Intel Xeon 4 CPUs consumes approximately 200 Wh of energy and results in an estimated emission of 10 g $CO_2$e. For a simulation of the entire process of \textit{in utero} cortical folding, this would correspond to an energy consumption of 3.72 kWh and emissions of 191 g $CO_2$e.

\vskip0.2cm

\subsubsection{Physically acceptable displacement field}
\begin{figure}[htbp]
\vspace{4pt}

\centering
\footnotesize

\begin{minipage}[t]{\textwidth}
\centering

\begin{minipage}[t]{0.55\textwidth}  
\centering  

{\footnotesize
\begin{tabular}{@{}lllllll@{}}
\toprule
Parameter & $H_0$ & $\mu_\mathrm{Cortex}$ & $\mu_\mathrm{Core}$  & $\nu$ & $\alpha_{\mathrm{TAN}} $ & $\alpha_{\mathrm{RAD}} $\\
\midrule
Value    & 1.5 & 1500 & 300 & 0.45 & $2.0 \hspace{1mm} 10^{-7}$ & 0.0\\
         & mm & Pa & Pa & - & $s^{-1}$ & $s^{-1}$ \\
\end{tabular}

\vspace{2pt}
\begin{tabular}{@{}lllllll@{}}
\toprule
Parameter & $\epsilon_\mathrm{N}$ & $T_0$ & $T_\mathrm{max}$ & dt\\
\midrule
Value    & $5.0 \hspace{1mm} 10^5$ & 21 & 36 & 43200 \\
         & $kg.m^{-2}.s^{-2}$ & GW & GW & s \\
\end{tabular}
}

\parbox{0.60\textwidth}{%
    \captionof{table}{First table: Biomechanical parameters (thickness, stiffnesses of Cortex and Core, Poisson's ratio, tangential and radial growth rates. Second table: numerical parameters (penalty coefficient for collisions, time instants for the simulation, time step)}
    \label{table:parameters}
}


\end{minipage}%
\begin{minipage}[c]{0.35\textwidth}  
\includegraphics[width=\linewidth]{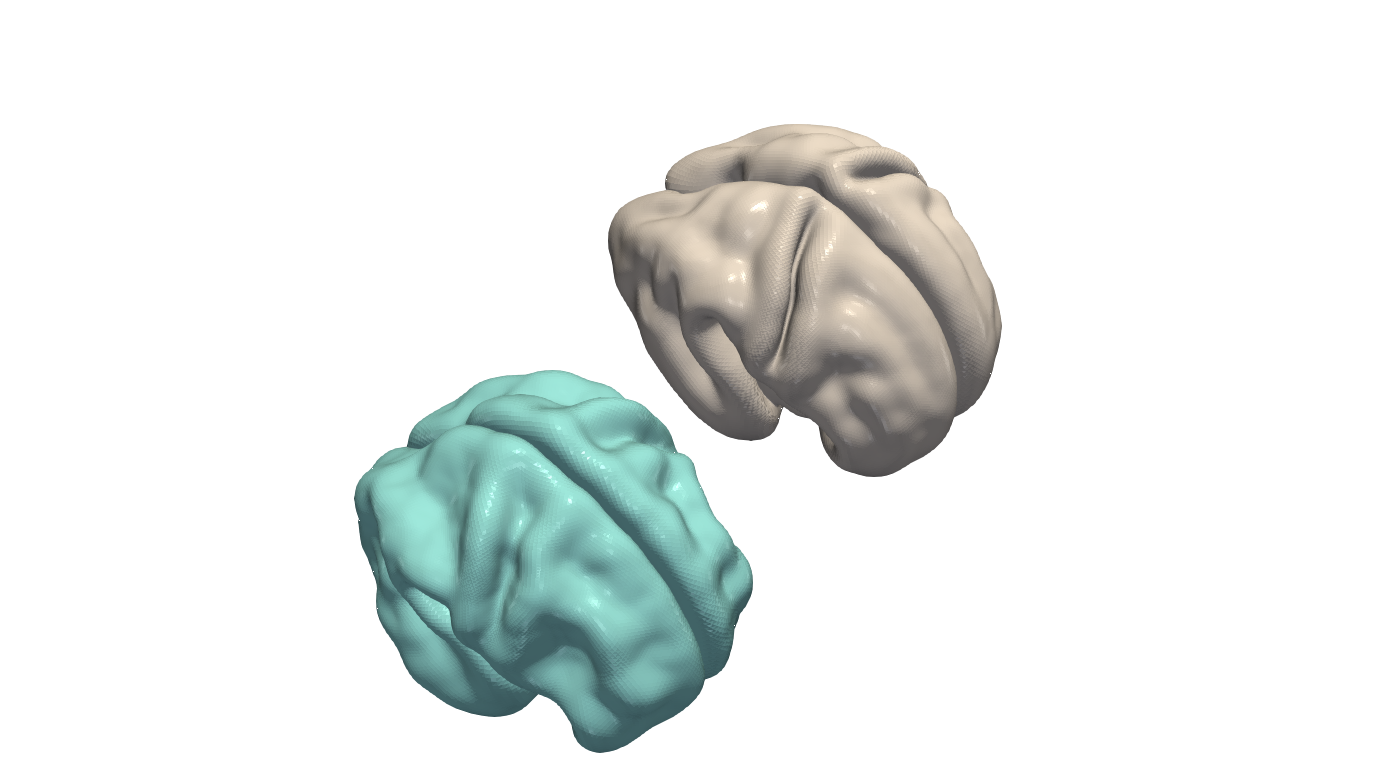}
\captionof{figure}{Folding brain simulated with \textit{FetalFoldSim} on initial brain geometry from Section \ref{subsec_mri_atlas_data}, Fig. \ref{fig:3D_input_brain_geometry_21GW_from_dHCP_Hemispheric_meshes}, with the parameters in Table~\ref{table:parameters} (blue mesh) compared to \textit{dHCP surface} real brain (beige mesh) at 28 GW}  
\label{fig:brain_growth_simulation_REF_28GW} 
\end{minipage}
\end{minipage}

\vspace{4pt}

\centering
\includegraphics[width=0.85\textwidth]{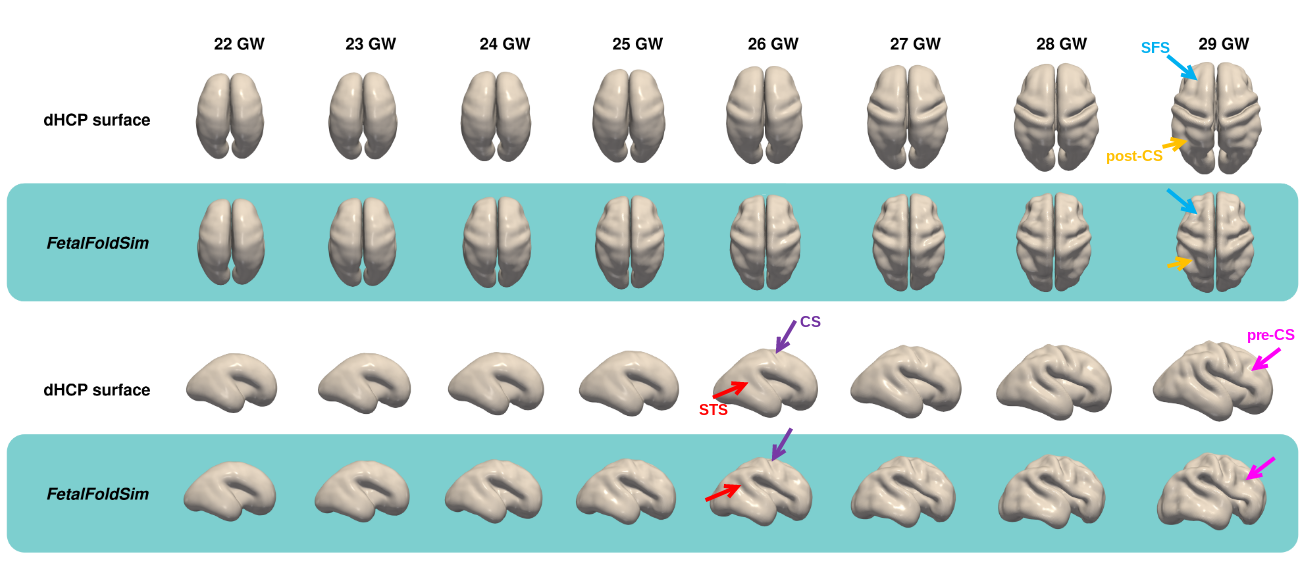}
\captionof{figure}{Simulation of a folding brain with \textit{FetalFoldSim} (blue lines), compared to the surface meshes post-processed from the hemispheric surface atlas meshes from the dHCP real brain data (white lines), for each gestational age from 22 to 29 GW. 
The input brain geometry (21 GW) used for the simulation is the one presented in Section \ref{subsec_mri_atlas_data}, Fig. \ref{fig:3D_input_brain_geometry_21GW_from_dHCP_Hemispheric_meshes}.
Simulation parameters are given in Table~\ref{table:parameters}. The simulations result have been smoothed in \textit{Paraview} with a coefficient of 500} 
\label{fig:brain_growth_simulation_REF}

\vspace{3pt}

\begin{minipage}[t]{\textwidth}
\hspace{0.08\textwidth} 
\begin{minipage}[t]{0.53\textwidth}
\vspace{0pt}
\centering
\includegraphics[width=\linewidth]{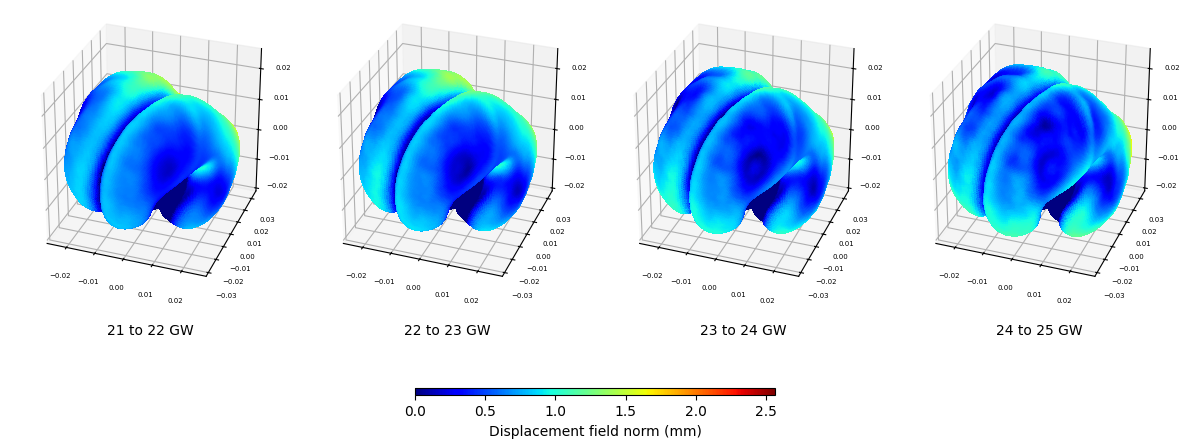}
\captionof{figure}{Norm of the displacement field (mm) at the brain mesh surface by gestational period (colormap built from the norm of the field value between 21 and 25 GW)}
\label{fig:braingowthFEniCS_displacement_field_21_to_25GW}
\end{minipage}
\hspace{0.04\textwidth}
\begin{minipage}[t]{0.14\textwidth}
\vspace{0pt}
\centering
\includegraphics[width=\linewidth]{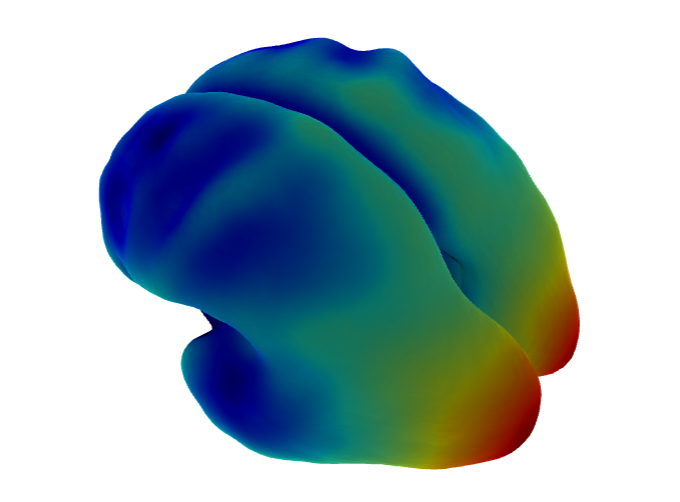}
\captionof{figure}{Maximum of the displacement field (mm) between 24 and 25 GW}
\label{fig:braingowthFEniCS_displacement_field_24_to_25GW}
\end{minipage}
\end{minipage}

\vspace{6pt}

{\footnotesize
\begin{tabular}{@{}lllll@{}}
\toprule
\textbf{- max $||u||$ -} & \textit{21-22 GW} & \textit{22-23 GW} & \textit{23-24GW} & \textit{24-25GW} \\
\midrule
\textit{FetalFoldSim} & 2.51 & 2.22 & 2.37 & 2.57 \\
\midrule
\cite{ZWang2021} & 1.24 & 2.29 & 3.65 & 5.79 \\
\end{tabular}
}\\
{\footnotesize
\begin{tabular}{@{}lllll@{}}
\toprule
\textbf{- mean $||u||$ - } & \textit{21-22 GW} & \textit{22-23 GW} & \textit{23-24GW} & \textit{24-25GW} \\
\midrule
\textit{FetalFoldSim} & 0.42 & 0.4 & 0.38 & 0.4 \\
\midrule
\cite{ZWang2021} & 0.45 & 0.88 & 1.91 & 3.03 \\
\end{tabular}
}
\captionof{table}{Max/mean values of the displacement norm $||u||$ (mm) by gestational period: \textit{FetalFoldSim} simulations versus real brain data \cite{ZWang2021}}
\label{table:max_mean_norm_displacement_field_braingrowthFEniCS_versus_ZWang2021}

\end{figure}

\noindent In \cite{ZWang2021}, the authors map the displacement field of the cortical surface using the spatiotemporal MRI atlas of fetal brain data between 21 and 25 gestational weeks from \cite{AGholipour2017}. 
They have shown a progressive increase in the mean and max values of the norm of the displacement field for each gestational period, as shown in Table \ref{table:max_mean_norm_displacement_field_braingrowthFEniCS_versus_ZWang2021}. Our simulation shows a comparable range in the magnitude of the two values over the two first weeks. However, conversely to \cite{ZWang2021}, we have constant norm values. Moreover, \cite{ZWang2021} showed an homogeneous displacement field for 21-22GW and 22-23GW gestational periods on the whole brain surface, while heterogeneous displacement field for 23-24GW and 24-25GW gestational periods, our model capture an heterogeneous field for all these periods. Fig. \ref{fig:braingowthFEniCS_displacement_field_21_to_25GW} shows a displacement field which is not characterized by a increasing intensity gradient like in real data highlighted in \cite{ZWang2021}, nevertheless, the location of the max displacement seem similar to the one from these real data
, maybe except that our \textit{FetalFoldSim} model shows a max displacement in the occipital lobe, as illustrated in Fig. \ref{fig:braingowthFEniCS_displacement_field_24_to_25GW}. This is likely attributable to the fact that growth is modeled as a linear process in our model (constant growth rate), whereas, in reality, it exhibits inherently non-linear characteristics. 

\newgeometry{margin=4pt}
\restoregeometry


\section{Spectral optimization of the similarity of simulated folding pattern to healthy neurodevelopmental data} 
\label{sec:spectral_opt}
\subsection{Cortical thickness values approximating the degree of folding observed in the real brain data}\label{SpAnGy_H0} 

\subsubsection{Experimental conditions}
\noindent We want to assess the influence of the one of the key model parameter, the cortical thickness \cite{XWang2019}, on the folding pattern and to estimate the optimal value for which primary folds are the most similar to the ones observed in real healthy neurodevelopmental data. To do so, we make cortical thickness vary to 0.8, 1.5, 2.25, and 6 millimeters, respectively, and simulate brain growth with the \textit{FetalFoldSim} model. The cortical thickness $H_0$ is set constant in space and time in the whole brain mesh regions, as in \cite{TTallinen2016}. The values of the other biophysical parameters are provided in Table \ref{table:parameters_EXPE2}. 

\vskip0.2cm

\begin{minipage}[t]{\textwidth}
\centering  
{\footnotesize
\begin{tabular}{@{}lllllll@{}}
\toprule
Parameter & $\mu_\mathrm{Cortex}$ & $\mu_\mathrm{Core}$  & $\nu$ & $\alpha_{\mathrm{TAN}} $ & $\alpha_{\mathrm{RAD}} $\\
\midrule
Value     & 3000 & 300 & 0.45 & $2.0 \hspace{1mm} 10^{-7}$ & 0.0\\
          & Pa & Pa & - & $s^{-1}$ & $s^{-1}$ \\
\end{tabular}

\vspace{2pt}

\begin{tabular}{@{}lllllll@{}}
\toprule
Parameter & $\epsilon_\mathrm{N}$ & $T_0$ & $T_{max}$ & dt\\
\midrule
Value    & $5.0 \hspace{1mm} 10^5$ & 21 & 42 & 86400 (43200s for $H_0=1.5$ mm) \\
         & $kg.m^{-2}.s^{-2}$ & GW & GW & s \\
\end{tabular}
}
\captionof{table}{Fixed biomechanical and numerical parameters to investigate the influence of cortical thickness on the folding pattern}
\label{table:parameters_EXPE2}
\end{minipage}

\vskip0.2cm

\subsubsection{Qualitative analysis of the simulated folding patterns}
\begin{figure}[htbp]
  \centering
  
\roundedcolorbox{white}{white}{10pt}{%
    \parbox{\textwidth}{%
      \centering
      \footnotesize

      \makebox[0.18\textwidth][c]{dHCP surface}%
      \hspace{0.02\textwidth}%
      \makebox[0.18\textwidth][c]{$H_0 = 0.8 \hspace{1mm} mm$}%
      \hspace{0.02\textwidth}%
      \makebox[0.18\textwidth][c]{$H_0 = 1.5\hspace{1mm} mm$}
      \hspace{0.02\textwidth}%
      \makebox[0.18\textwidth][c]{$H_0 = 2.25\hspace{1mm} mm$}
      \hspace{0.02\textwidth}%
      \makebox[0.18\textwidth][c]{$H_0 = 6.0\hspace{1mm} mm$}
      
      \vspace{0.5cm}

      \includegraphics[width=0.18\textwidth]{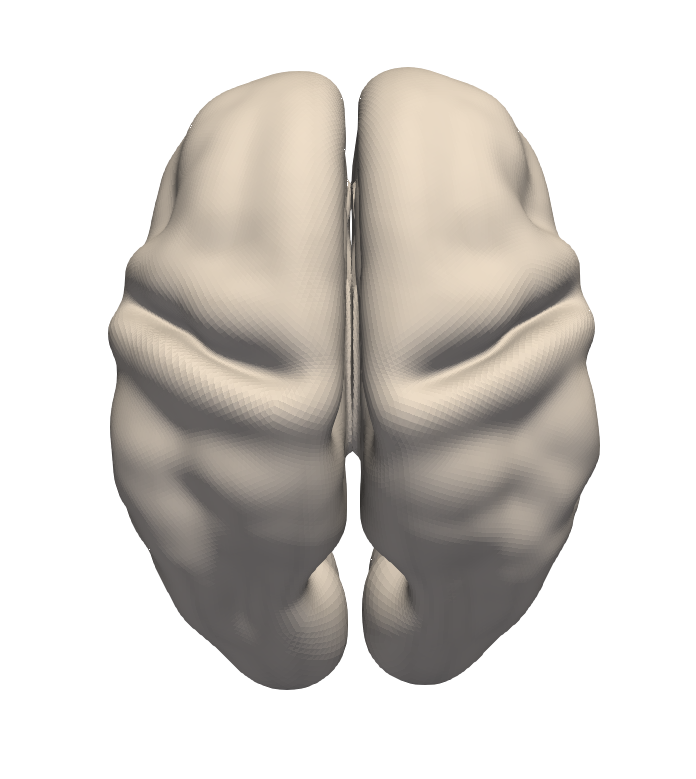}%
      \hspace{0.02\textwidth}%
      \includegraphics[width=0.18\textwidth]{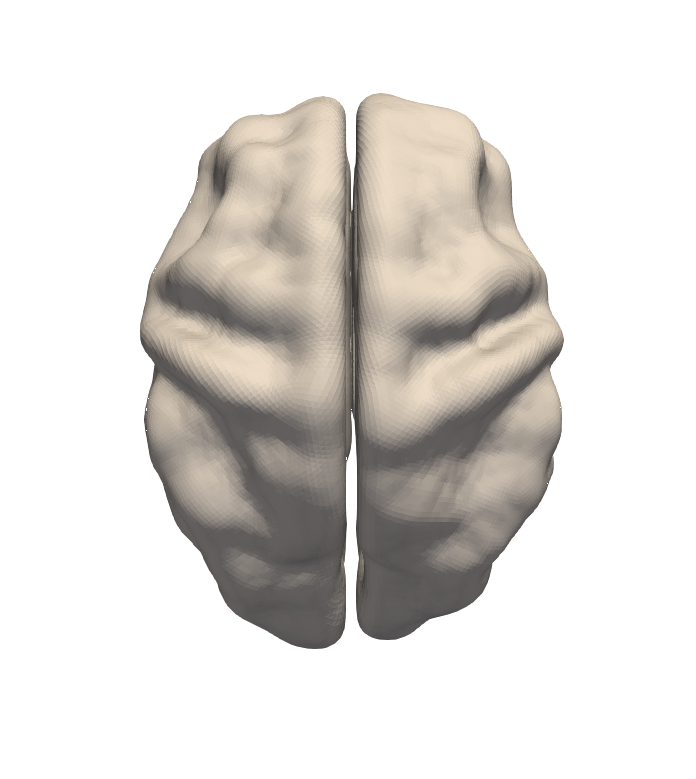}%
      \hspace{0.02\textwidth}%
      \includegraphics[width=0.18\textwidth]{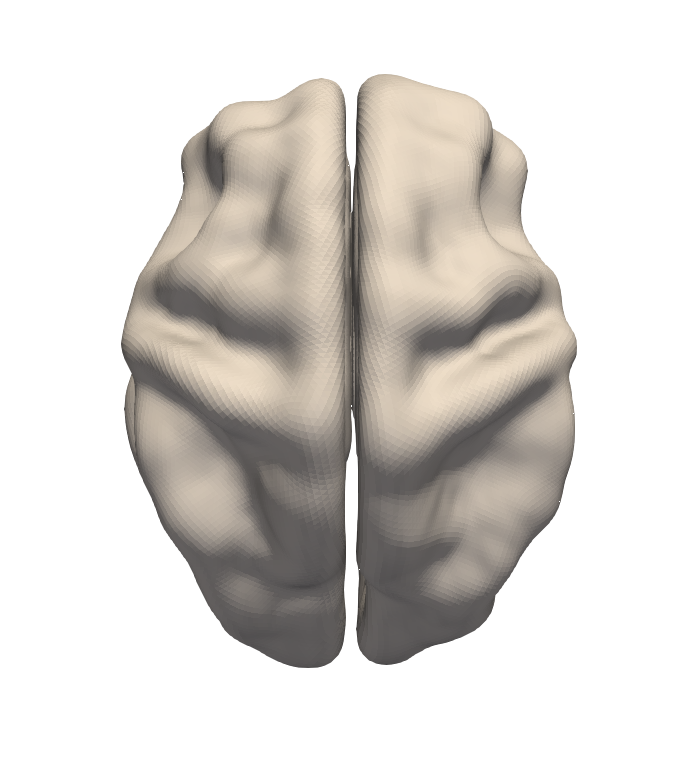}
      \hspace{0.02\textwidth}%
      \includegraphics[width=0.18\textwidth]{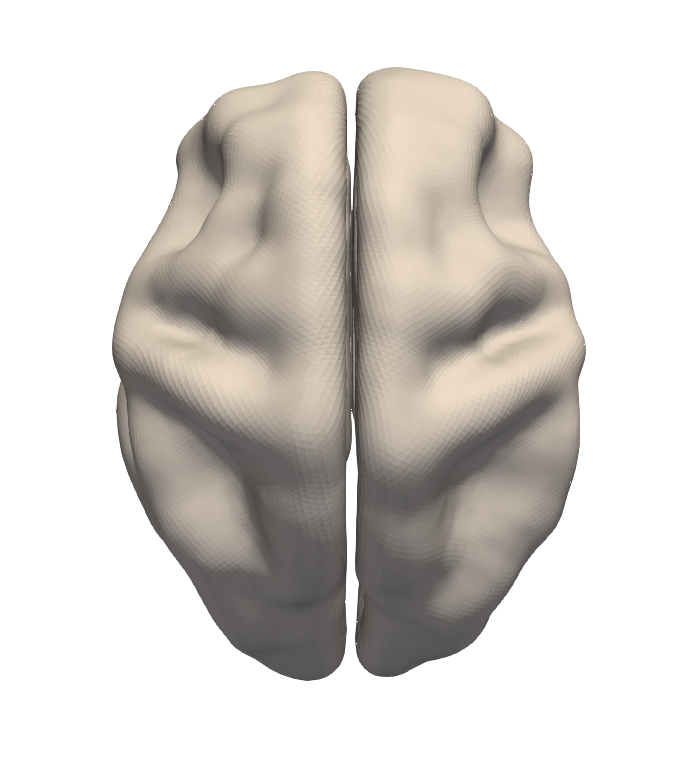}
      \hspace{0.02\textwidth}%
      \includegraphics[width=0.18\textwidth]{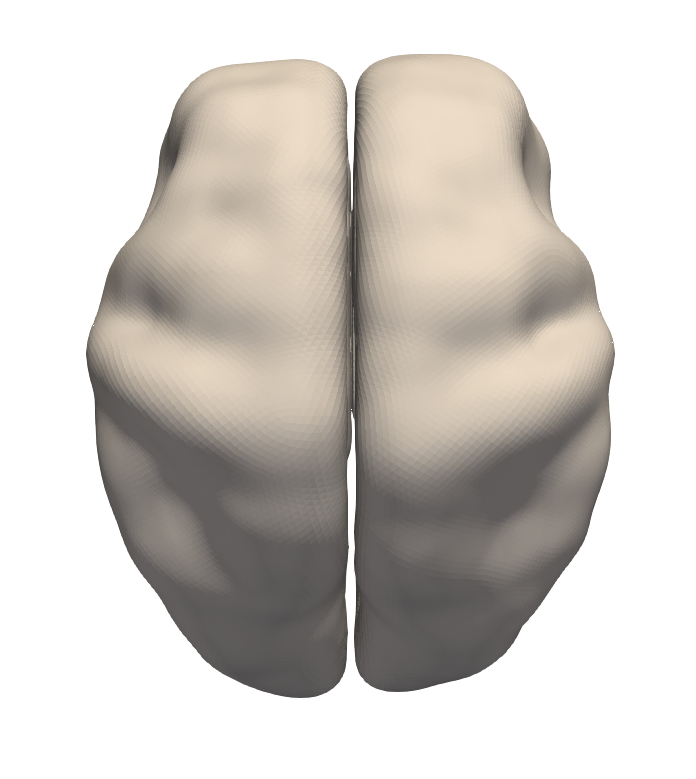}
    }%
  }

  \vspace{0.03cm}

\roundedcolorbox{white}{white}{10pt}{%
    \parbox{\textwidth}{%
      \centering

      \includegraphics[width=0.18\textwidth]{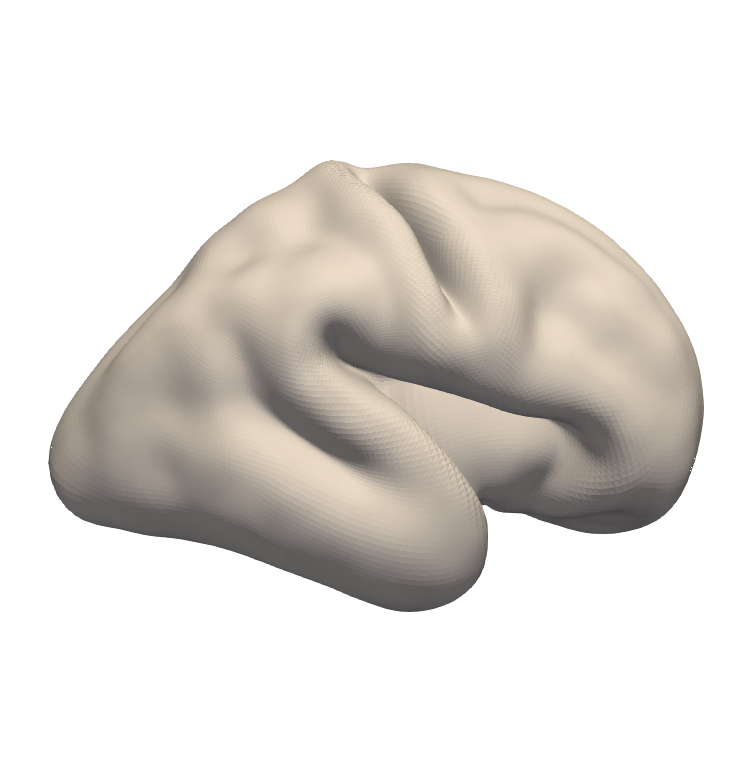}%
      \hspace{0.02\textwidth}%
      \includegraphics[width=0.18\textwidth]{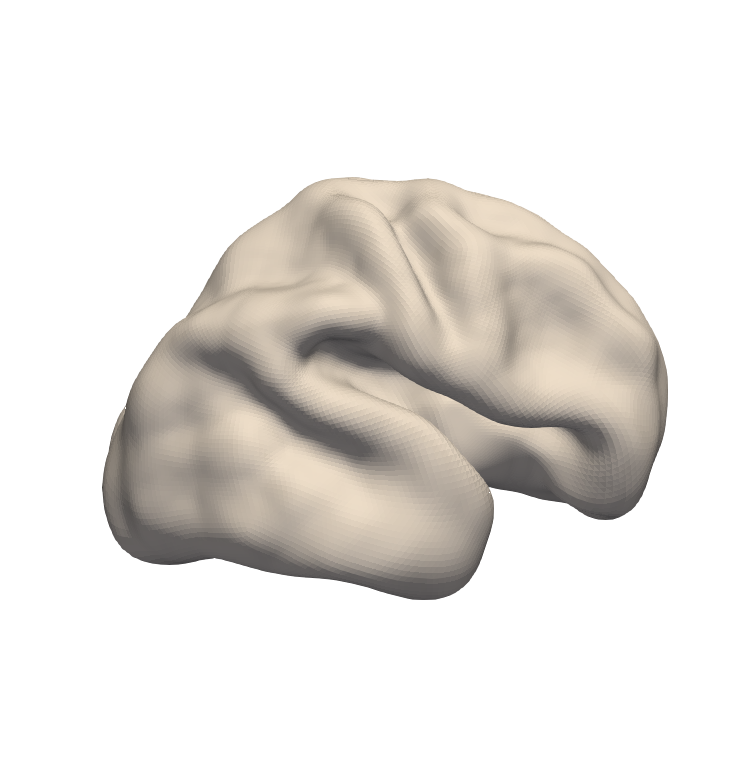}%
      \hspace{0.02\textwidth}%
      \includegraphics[width=0.18\textwidth]{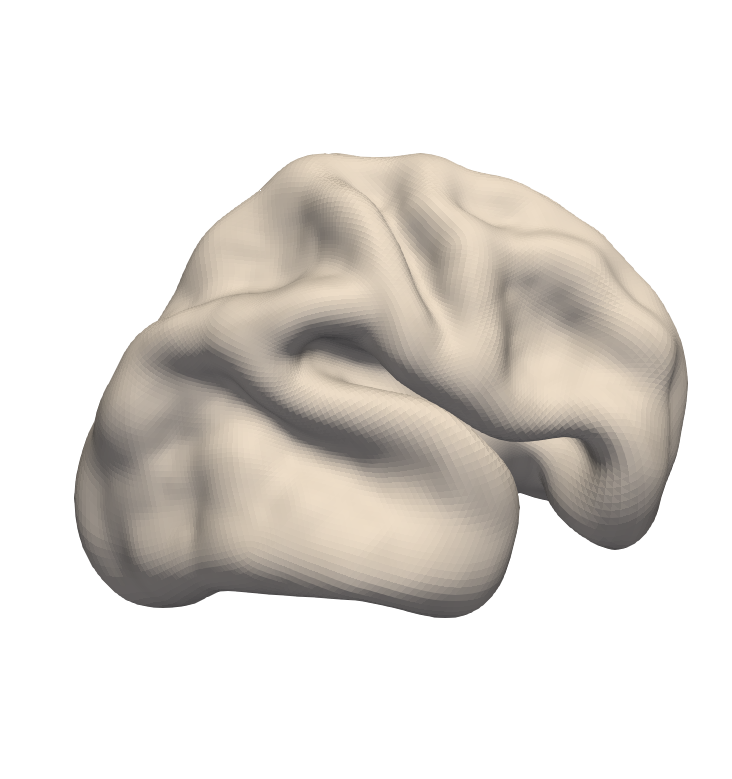}
      \hspace{0.02\textwidth}%
      \includegraphics[width=0.18\textwidth]{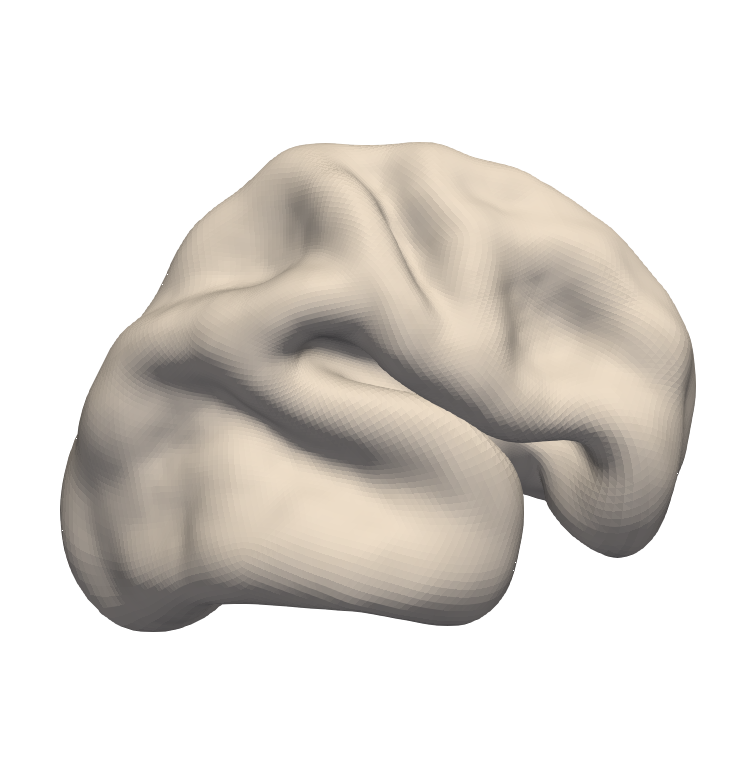}
      \hspace{0.02\textwidth}%
      \includegraphics[width=0.18\textwidth]{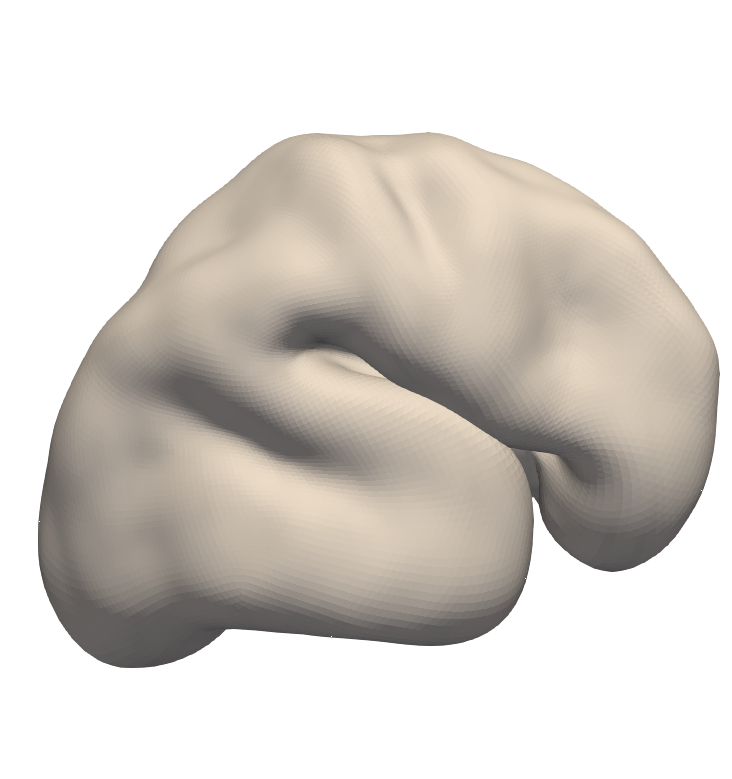}
    }%
  }

  \vspace{0.05cm}
  \includegraphics[width=\textwidth, height=0.18\textheight]{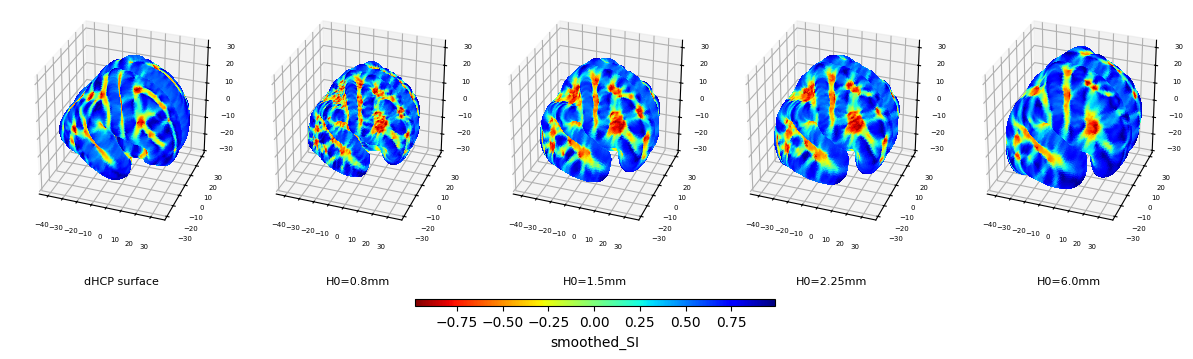}
  
  \caption{Influence of the cortical thickness on the folding pattern, at 28GW. (a) and (b) 
  Simulations of cortical folding with different values of cortical thickness, compared to real data (dHCP surface). Input brain geometry (21 GW) used for the simulation: as presented in Section \ref{subsec_mri_atlas_data}, Fig. \ref{fig:3D_input_brain_geometry_21GW_from_dHCP_Hemispheric_meshes}. Other simulation parameters are from Table \ref{table:parameters_EXPE2}. Meshes are smoothed in \textit{Paraview} (smoothing coefficient = 200). (c) Smoothed shape index map on both the simulated folded brains with $H_0$ = 0.8; 1.5; 2.25 and 6.0 mm and the \textit{dHCP surface} brain mesh at 28 GW (common scale)}
  \label{fig:brain_growth_sensitivity_analysis_H0}
\end{figure}

\noindent Fig. \ref{fig:brain_growth_sensitivity_analysis_H0} (top) shows the results of the \textit{FetalFoldSim} simulations with different values of cortical thickness at the maximum common gestational age at which the simulations all converge without any collisions, i.e., 28GW, compared to the \textit{dHCP surface} atlas brain mesh at 28GW. 

\noindent In Fig. \ref{fig:brain_growth_sensitivity_analysis_H0} (top), we can observe primary folds emerge on the surface of the simulated brain and have similar locations than in the real brain meshes: central, superior temporal, pre-central, superior frontal, post-central and intraparietal sulci. 
\noindent However, certain discrepancies can be observed. First, for all the cortical thickness values, the central sulci in both left and right hemispheres do not stop close to the longitudinal fissure, as in the real data, this phenomenon is particularly noticeable at 28 GW. This could be linked with the contact penalty method 
(the force applied to correct the collision is defined in the orthogonal direction; other type of method to correct collisions could be more precise; the fixed penalty coefficient may be too high) and associated boundaries maybe too large, implemented in the computational model.\\ 
Secondly, a collapse of the simulated brain is observed in the frontal and occipital lobes. This result is in agreement with the observations reported by \cite{TTallinen2016}. 
\noindent Thirdly, for most of cortical thickness values, 1.5, 2.25 and 6 mm, the lateral lobes join, thereby enveloping the sylvian fissure, which is not the case till 28 GW in real data. This can be due to unadapted growth around this region, or also 
a too high stiffness modeling of the cortex in this area.

\subsubsection{SpAnGy frequency band analysis}

\begin{figure}
  \centering 
  \hspace*{\fill}%
  \includegraphics[width=0.7\textwidth]{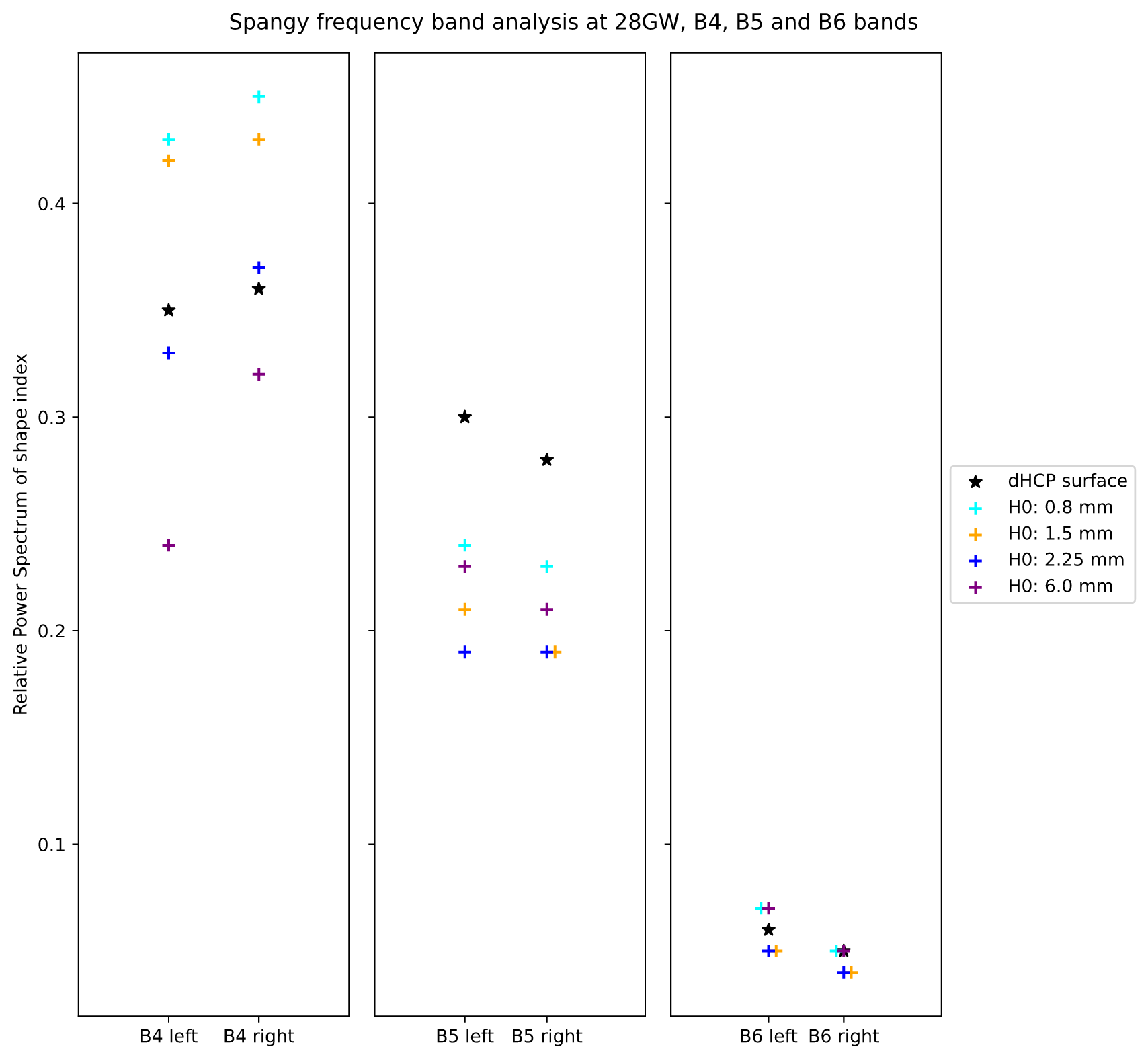}
  \hspace*{\fill}
  \caption{\textit{SpAnGy} relative power spectrum of shape index (power spectrum normalized to total power spectrum) - frequency bands B4, B5 and B6 - for both \textit{FetalFoldSim} simulations with distinct values of $H_0$ and \textit{dHCP surface} mesh, at 28GW. For each mesh and band, left and right hemispheres are analyzed separately} 
\label{fig:Spangy_relative_power_spectrum_simulationsH0_left_right_hemisphere} 
\end{figure}


\vskip0.2cm

\noindent We hereby propose to analyze the surface spectrum of the meshes obtained by the \textit{FetalFoldSim} simulations with the different cortical thickness values at simulated 28 GW, using the \textit{SpAnGy} method described in Section \ref{SpAnGy}. The spectral band analysis will provide the spectral power and relative spectral power for bands B4, B5, and B6, which are associated with primary, secondary, and tertiary folds, respectively \cite{DGermanaud2012, JDubois2019}. These powers will be compared to the one from the \textit{dHCP surface} real mesh at 28 GW. 
\noindent All the brain meshes exhibit a distinct separation between the two hemispheres, which comes from the way the {dHCP surface} data has been post-processed. In order to carry out the \textit{SpAnGy} analysis, the brain meshes are preliminarily divided into two sub-meshes, corresponding to the two hemispheres and \textit{SpAnGy} is run on each hemisphere distinctly. In the event that the initial mesh is connex, this step is not necessary. 
\noindent To obtain six spectral bands from B0 to B6 (common to gyri and sulci at this level), the shape index is decomposed on the basis of the geometry's eigenvectors (see the definition of \textit{SpAnGy} in Section \ref{SpAnGy}). In order to do so, it is first necessary to determine the number of eigenmodes (eigenvectors and eigenvalues) that enable to obtain the spectral band B6. For that purpose, we employ the theoretical Weyl number formula (Eq. \ref{eq:Weyl_number}, Section \ref{SpAnGy}). For each hemisphere, the $N_{Weyl}$ value is calculated for all geometries that are compared, and the maximum value is retained (it equals 1970 for the right hemisphere and 1833 for the left hemisphere).

\noindent Fig. \ref{fig:Spangy_relative_power_spectrum_simulationsH0_left_right_hemisphere} enables to compare the relative power spectrum of shape index for B4, B5 and B6 bands of the real fetal atlas brain at 28 GW (black star) with the simulations with different values of cortical thickness.
We hypothesize that B4, B5, and B6 also represent three successive degrees of folding, characterized by their respective wavelength divided by two (see Section \ref{SpAnGy}), while using the shape index as folding proxy and not the mean curvature (See Section \ref{SpAnGy}). However, as we simulate folding till 28 GW, secondary and tertiary folds are not supposed to have emerged yet, and bands B5/B6 should be interpreted carefully. For this reason, we might keep an interpretation of the \textit{SpAnGy} B4 bands and associated primary folds.

\noindent Both for the real data and for the simulation results, spatial variation in the nature of the folds is characterized by a higher B4 spectral density in the right hemisphere than in the left. This tendency is evocative of the asymmetry observed in the early cortical folding, wherein primary folds emerge earlier on the right hemisphere compared to the left \cite{PAHabas2012}. The asymmetry seem to be reversed for spectral densities B5 and B6, which reflect a higher-frequency folding pattern for both the real data and the simulations. 
\noindent Then, folds whose pattern is characterized by the B4 frequency band (primary folds) are predominant at 28 GW for both real data and simulations.; those associated with the B6 band (tertiary folds) are very weakly represented. 
\vskip0.1cm
\noindent Finally, a cortical thickness of 2.25 millimeters simulates the primary folds (B4) that correspond to the ones observed in the real data; while cortical thickness of 0.8 millimeters simulates the secondary folds (B5) that correspond to the ones observed in the real data. Nevertheless, \textit{FetalFoldSim} simulation produces a folding pattern which is less complex, since the relative power spectrum of B5 is higher for real data than for any simulation result.

\subsection{Stiffness ratio values approximating the degree of folding observed in the real brain data}\label{SpAnGy_H0}

\begin{minipage}[t]{\textwidth}
\centering  
{\footnotesize
\begin{tabular}{@{}lllllll@{}}
\toprule
Parameter & $H_0$ & $\mu_\mathrm{Core}$  & $\nu$ & $\alpha_{\mathrm{TAN}} $ & $\alpha_{\mathrm{RAD}} $\\
\midrule
Value    & 1.5 & 300 & 0.45 & $2.0 \hspace{1mm} 10^{-7}$ & 0.0\\
         & mm & Pa & - & $s^{-1}$ & $s^{-1}$ \\
\end{tabular}

\vspace{2pt}

\begin{tabular}{@{}lllllll@{}}
\toprule
Parameter & $\epsilon_\mathrm{N}$ & $T_0$ & $T_{max}$ & dt\\
\midrule
Value    & $5.0 \hspace{1mm} 10^5$ & 21 & 42 & 43200s\\
         & $kg.m^{-2}.s^{-2}$ & GW & GW & s \\
\end{tabular}
}
\captionof{table}{Fixed biomechanical and numerical parameters to investigate the influence of the cortical stiffness (stiffness ratio) on the folding pattern}
\label{table:parameters_EXPE3}
\end{minipage}

\clearpage
\thispagestyle{empty} 
\begin{figure}[p]
  \centering

  \footnotesize

  \begin{minipage}{\textwidth}
  \centering 
  
  \makebox[0.21\textwidth][c]{dHCP 29GW}%
  \hspace{0.01\textwidth}%
  \makebox[0.18\textwidth][c]{$300 \hspace{1mm} Pa$}%
  \hspace{0.01\textwidth}%
  \makebox[0.18\textwidth][c]{$600 \hspace{1mm} Pa$}
  \hspace{0.01\textwidth}%
  \makebox[0.18\textwidth][c]{$900 \hspace{1mm} Pa$} 

  \raisebox{-4.5mm}{\includegraphics[width=0.21\textwidth]{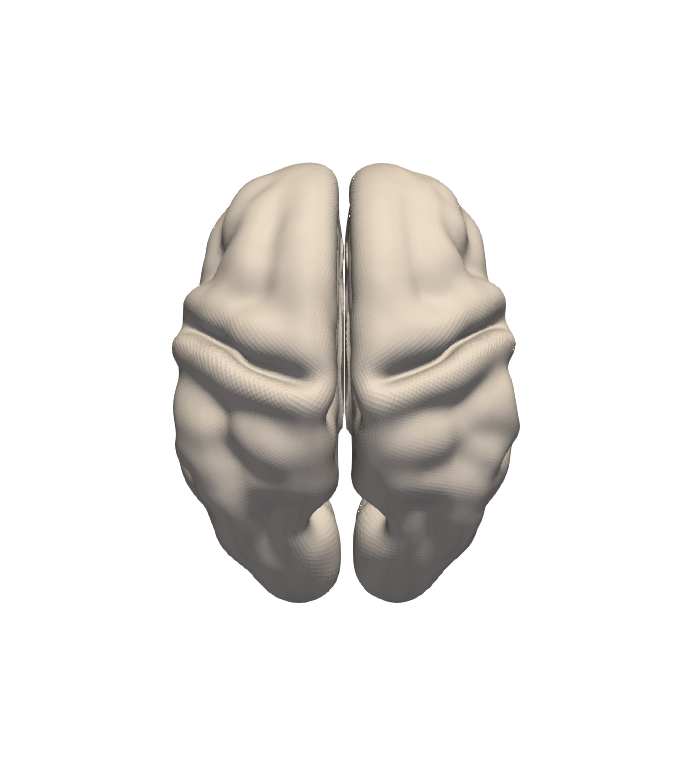}}
  \hspace{0.01\textwidth}%
  \includegraphics[width=0.18\textwidth]{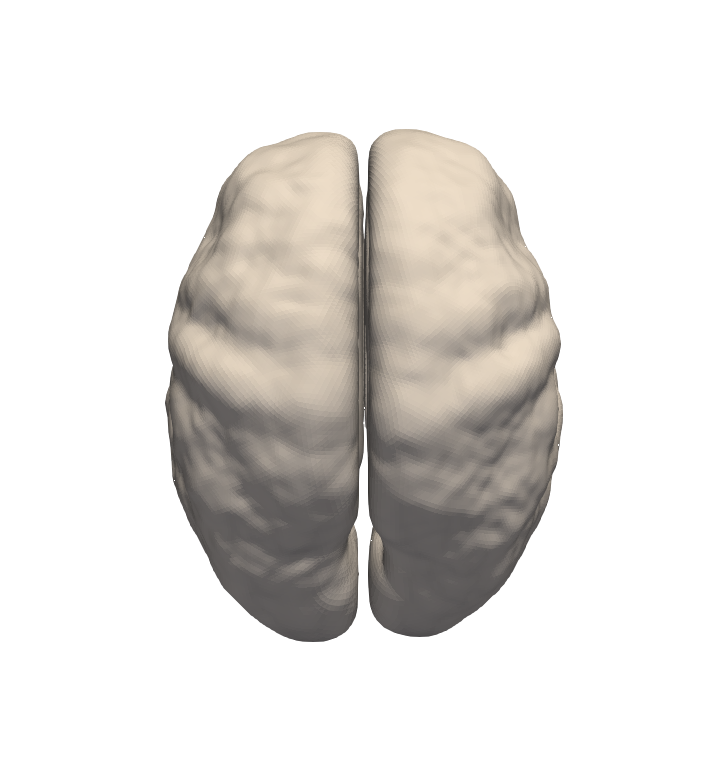}%
  \hspace{0.01\textwidth}%
  \includegraphics[width=0.18\textwidth]{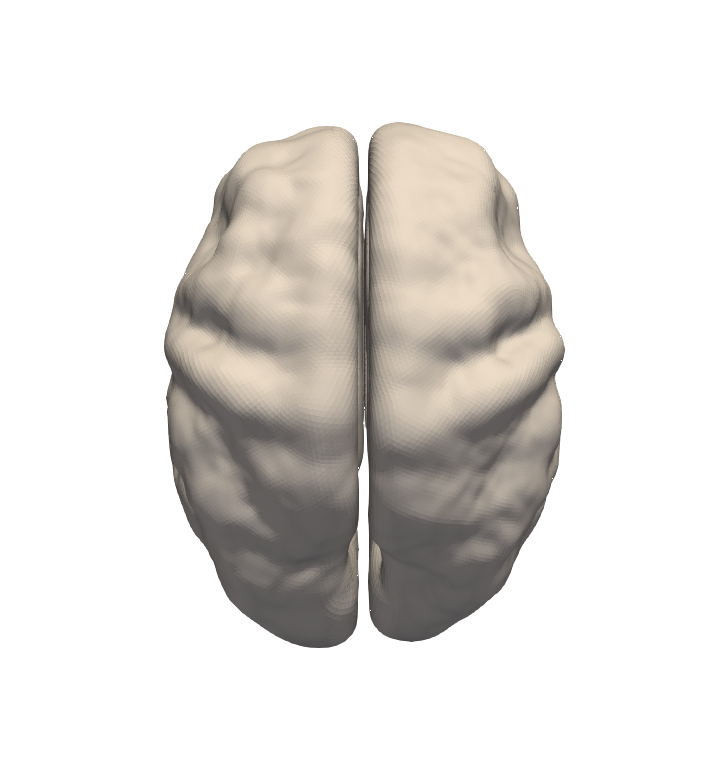}
  \hspace{0.01\textwidth}%
  \includegraphics[width=0.18\textwidth]{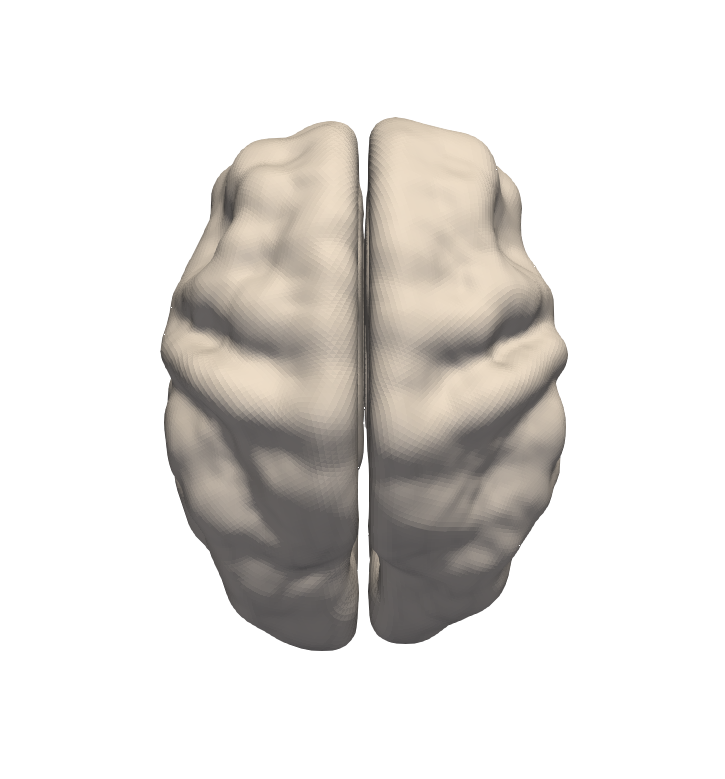}

  \vspace{-0.8cm}

  \raisebox{-4.5mm}{\includegraphics[width=0.21\textwidth]{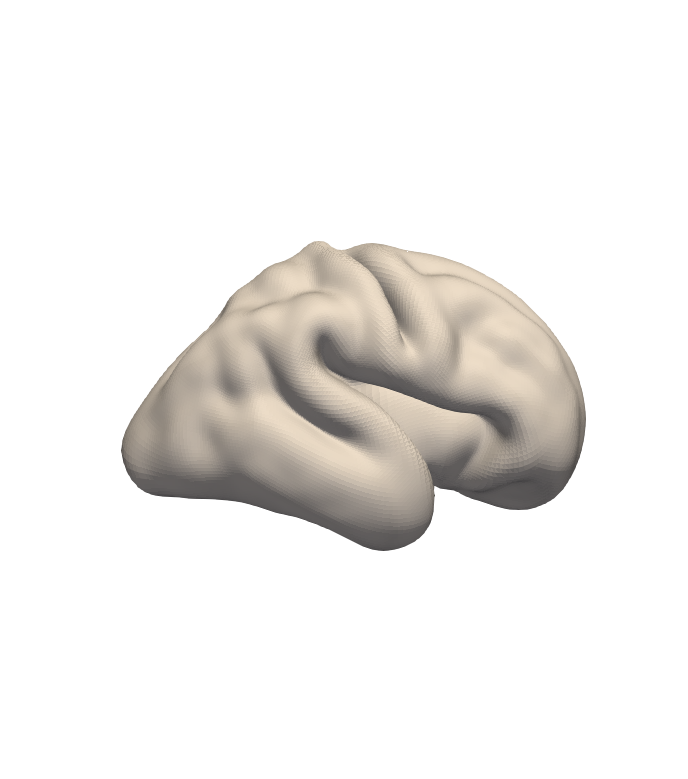}}
  \includegraphics[width=0.18\textwidth]{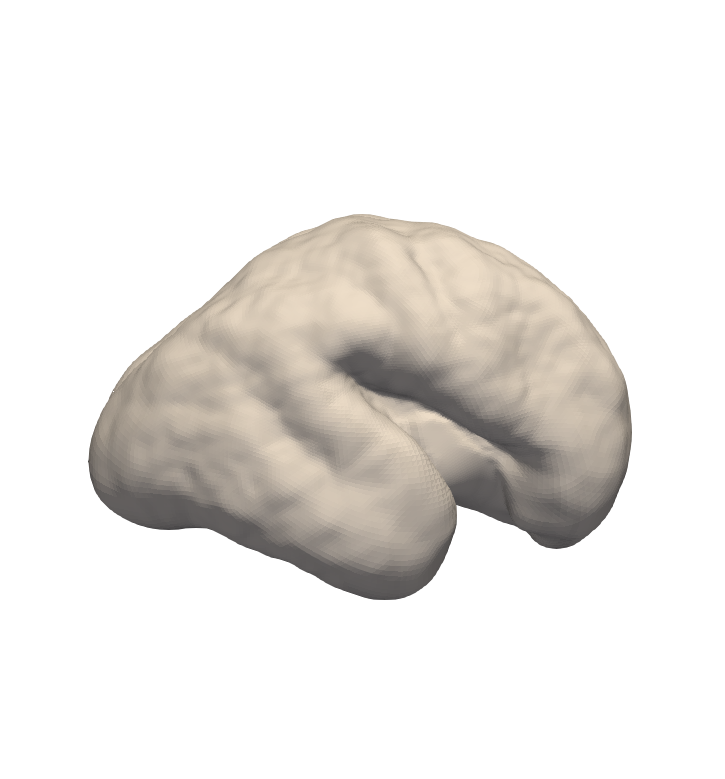}%
  \hspace{0.01\textwidth}%
  \includegraphics[width=0.18\textwidth]{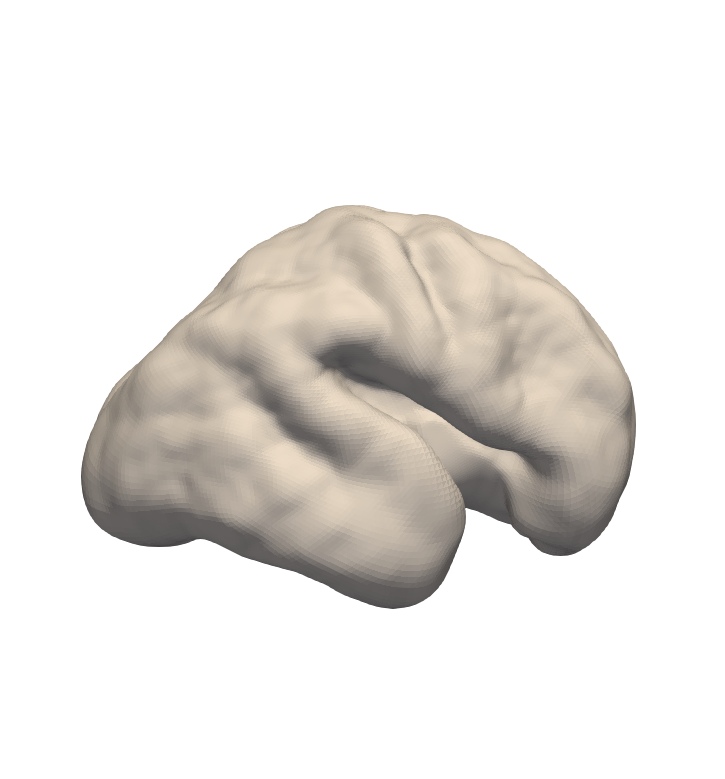}
  \hspace{0.01\textwidth}%
  \includegraphics[width=0.18\textwidth]{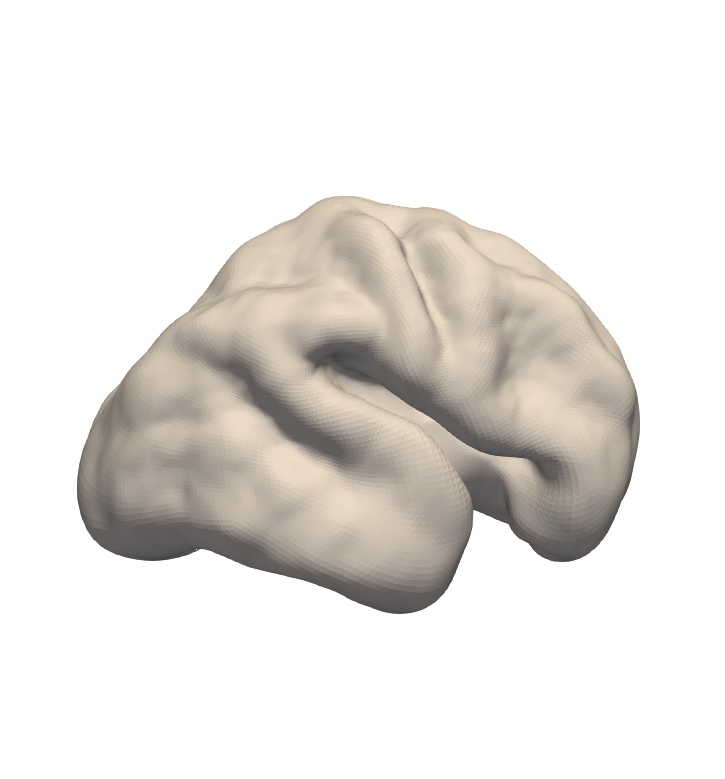}
 
  \makebox[0.18\textwidth][c]{$1200 \hspace{1mm} Pa$}%
  \hspace{0.01\textwidth}%
  \makebox[0.18\textwidth][c]{$1500 \hspace{1mm} Pa$}%
  \hspace{0.01\textwidth}%
  \makebox[0.18\textwidth][c]{$3000 \hspace{1mm} Pa$}%
  \hspace{0.01\textwidth}%
  \makebox[0.18\textwidth][c]{$4500 \hspace{1mm} Pa$}

  \includegraphics[width=0.18\textwidth]{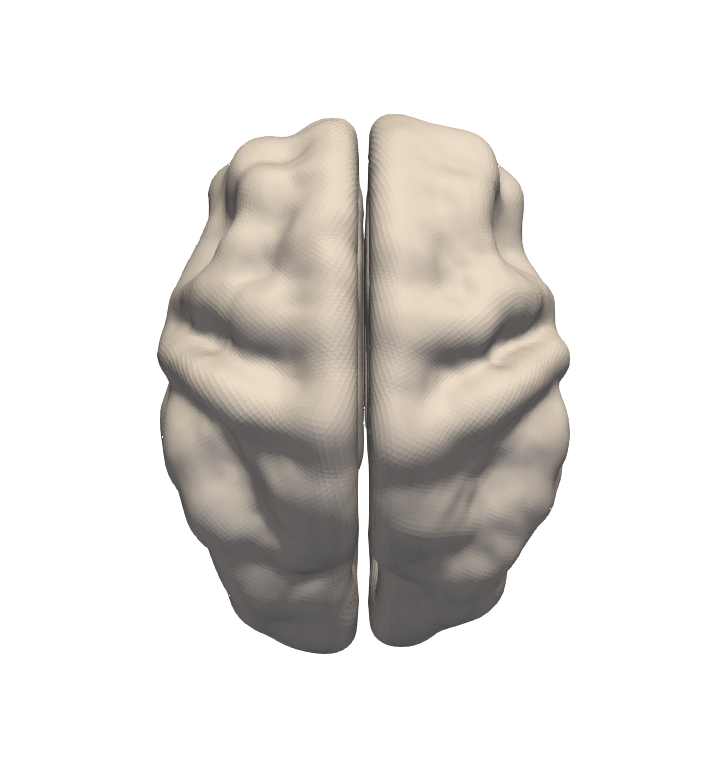}
  \hspace{0.01\textwidth}%
  \includegraphics[width=0.18\textwidth]{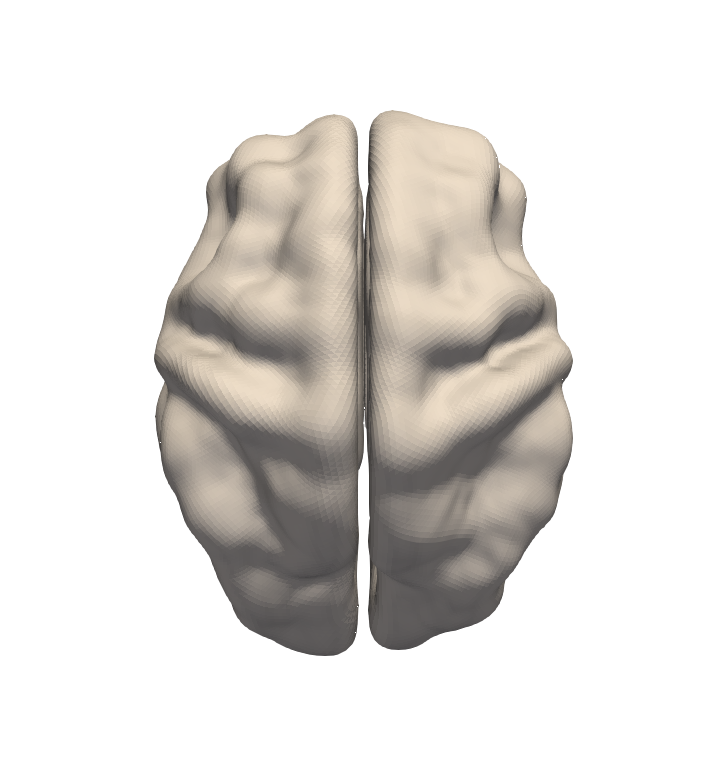}
  \hspace{0.01\textwidth}%
  \includegraphics[width=0.18\textwidth]{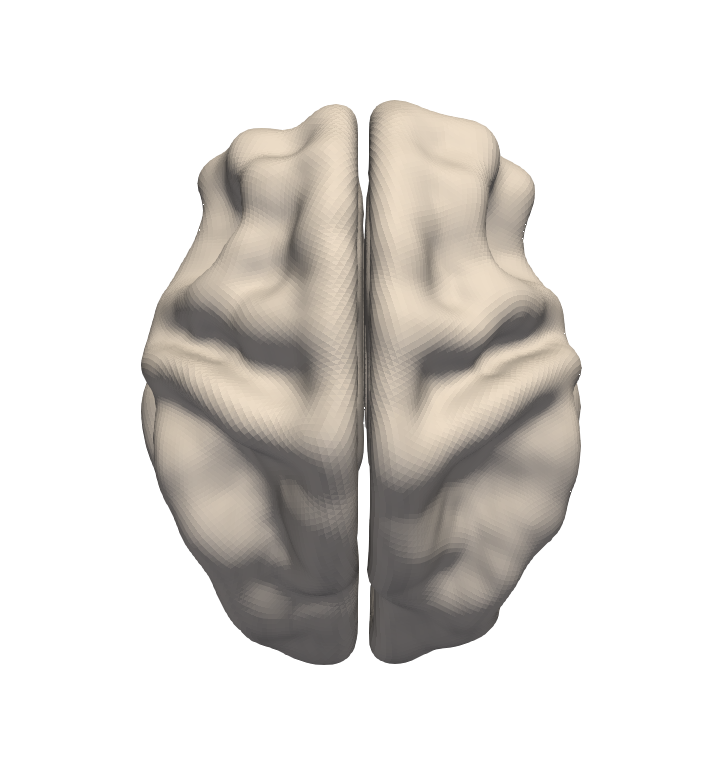}
  \hspace{0.01\textwidth}%
  \includegraphics[width=0.18\textwidth]{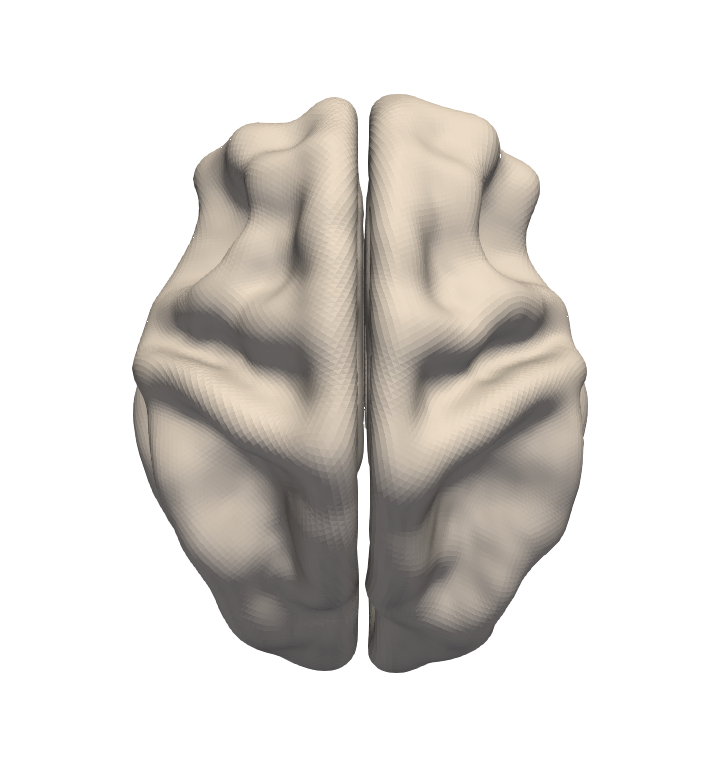}

  \includegraphics[width=0.18\textwidth]{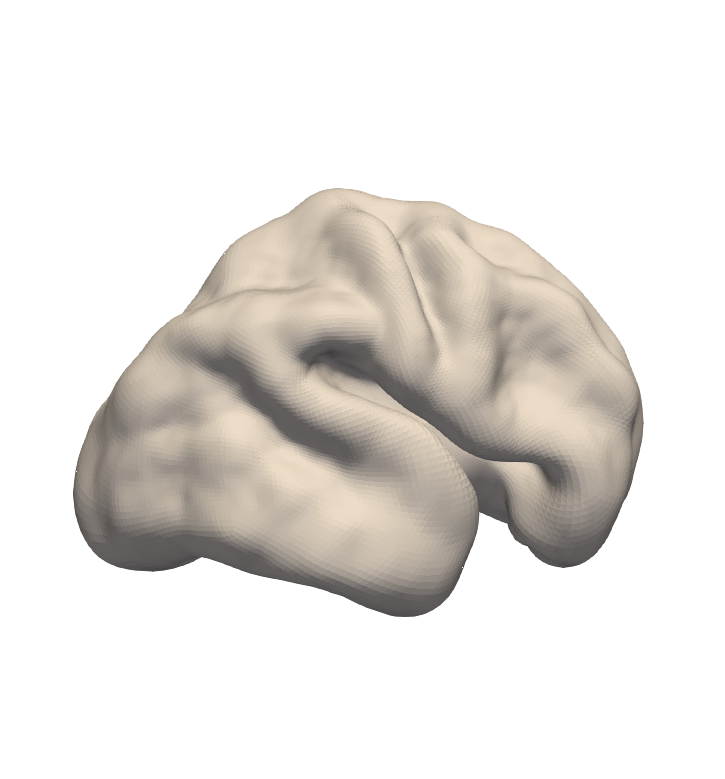}
  \hspace{0.01\textwidth}%
  \includegraphics[width=0.18\textwidth]{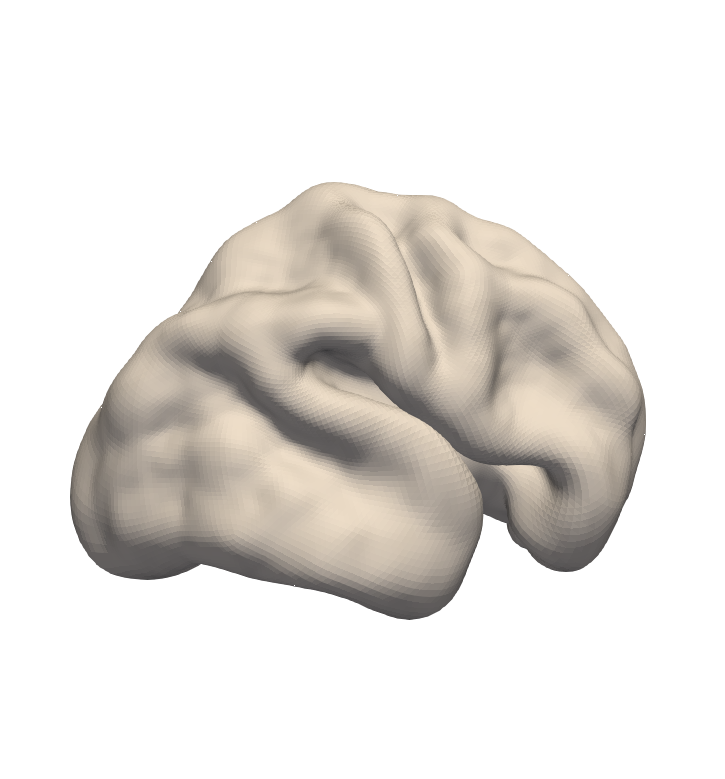}
  \hspace{0.01\textwidth}%
  \includegraphics[width=0.18\textwidth]{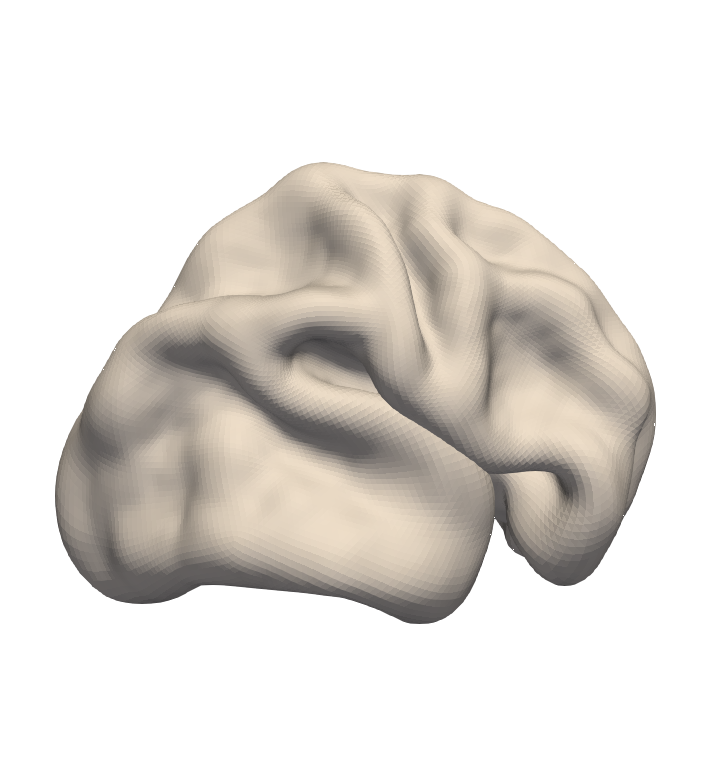}
  \hspace{0.01\textwidth}%
  \includegraphics[width=0.18\textwidth]{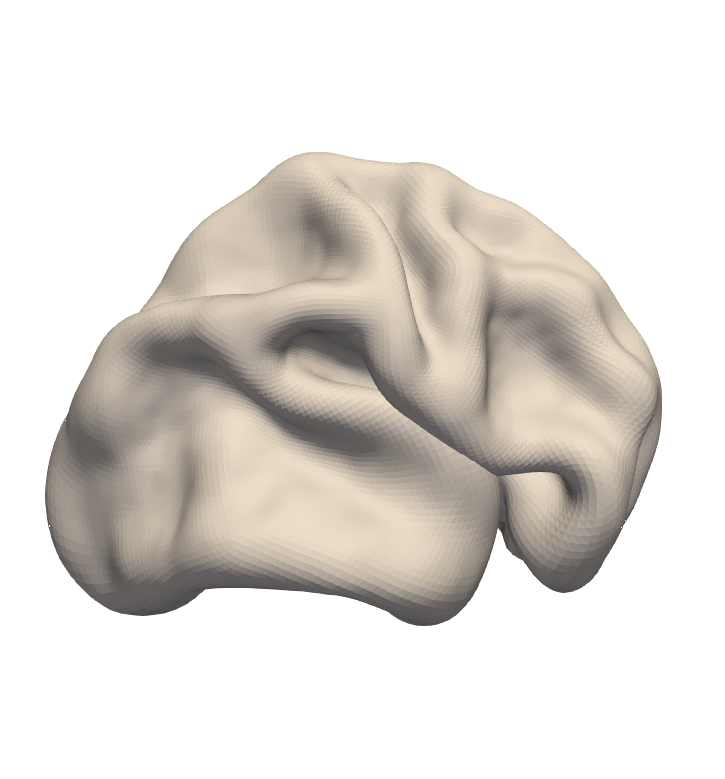}

  \subcaption{} 
        
  \end{minipage}

  \vspace{0.4cm}

  \begin{minipage}{\textwidth}
  \centering 

  \includegraphics[width=1.05\textwidth, height=0.42\textheight]{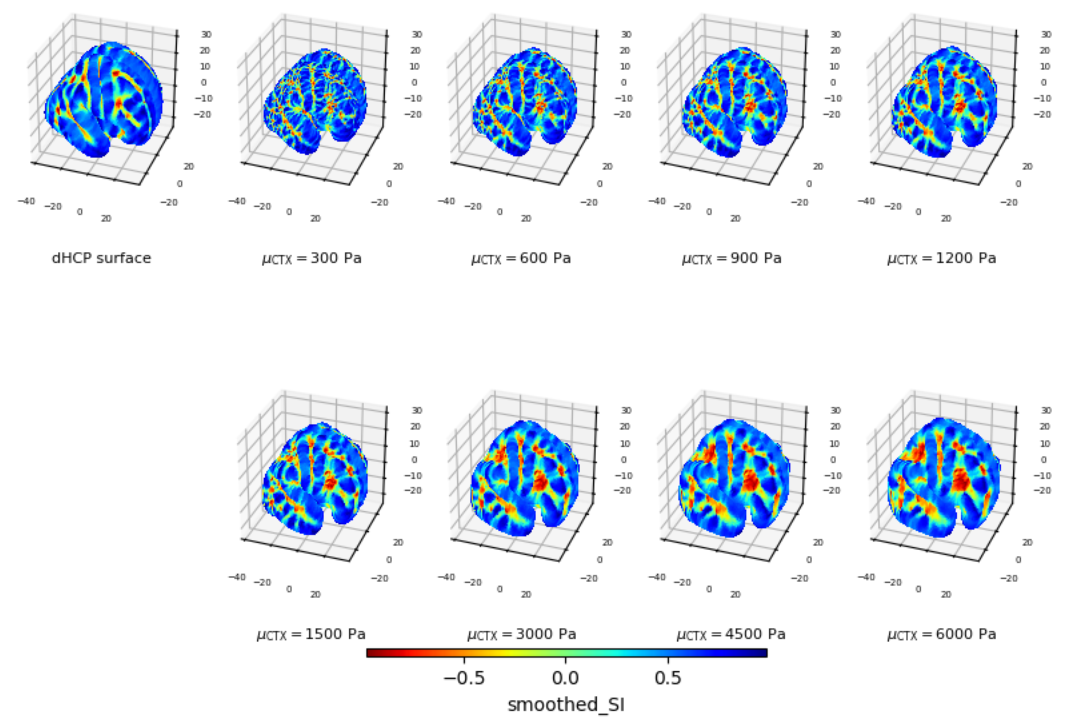}

  \subcaption{} 
  
  \end{minipage}

  \captionsetup{width=1.2\textwidth}
  \caption{Influence of the stiffness ratio $\mu_\mathrm{Cortex} / \mu_\mathrm{Core}$ on the folding pattern, at 29GW. (a) Simulations of cortical folding with different values of cortical stiffnesses (ratios from 1 to 15), compared to real data (\textit{dHCP surface}). Input brain geometry (21 GW) used for the simulation: as presented in Section \ref{subsec_mri_atlas_data}, Fig. \ref{fig:3D_input_brain_geometry_21GW_from_dHCP_Hemispheric_meshes}. Other simulation parameters are from Table \ref{table:parameters_EXPE2}. Meshes are smoothed in \textit{Paraview} (smoothing coefficient = 300). (b) Smoothed shape index map on both simulated folded brains with $\mu_\mathrm{Cortex}$ from 300 to 6000 Pa (ratios from 1 to 20)
  and the \textit{dHCP surface} brain mesh at 28 GW}
  
  \label{fig:brain_growth_sensitivity_analysis_muCortex_qualitative}
\end{figure}
\clearpage

\begin{figure}
  \centering 
  \hspace*{\fill}%
  \includegraphics[width=0.7\textwidth]{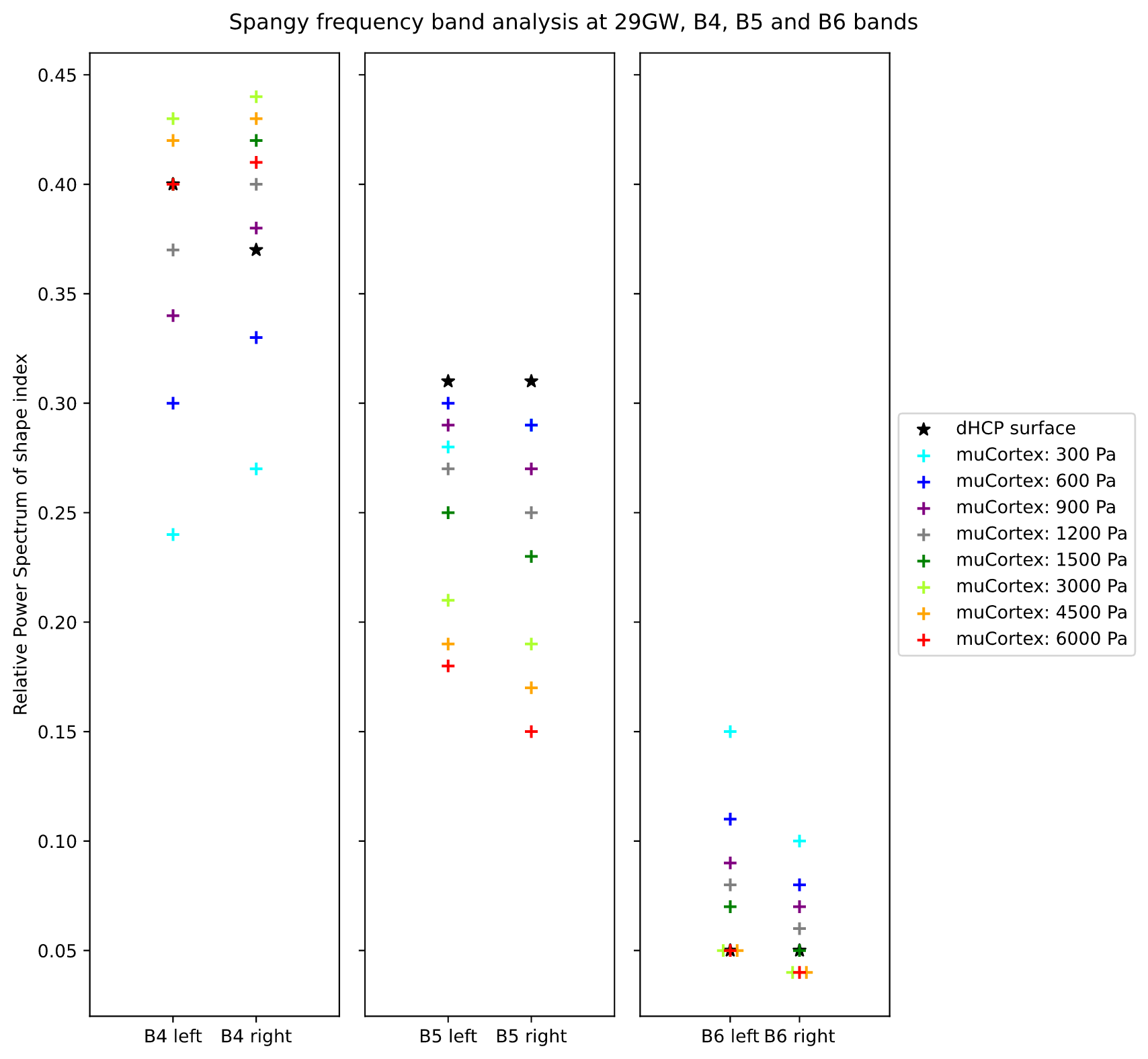}
  \hspace*{\fill}
  \caption{\textit{SpAnGy} relative power spectrum of shape index (power spectrum normalized to total power spectrum) - frequency bands B4, B5 and B6 - for both \textit{FetalFoldSim} simulations with distinct values of $\mu_\mathrm{Cortex}$ and \textit{dHCP surface} mesh, at 29GW. For each mesh and band, left and right hemispheres are analyzed separately} 
  
\label{fig:Spangy_relative_power_spectrum_simulations_muCortex_left_right_hemisphere} 
\end{figure}


\section{Discussion}\label{sec12}

Computational models to simulate cortical folding enable enhanced investigation on biomechanical factors influencing both healthy and pathological \textit{in utero} fold development. A key challenge lies in developing realistic models that can be validated against fetal brain MRI data and be interpretable.
In this work, we have developed a novel open-source model of cortical folding implemented with \textit{FEniCS}, namely the \textit{FetalFoldSim} computational model, on realistic brain geometry. In particular, we have reformulated the biomechanical folding model of \cite{TTallinen2016} using a finite element framework. We have introduced penalization of the collisions between the two hemispheres to be able to model folding on the whole brain geometry. \textit{FetalFoldSim} is available to the community, modular, interpretable, and has been developped to be extensible. 
Additionally, we have proposed a MRI-informed simulation framework that enable to simulate cortical folding from raw MRI fetal data. This framework uses MRI data to generate the input mesh, quantify simulation results and inform parameters. It also particularly includes metrics to characterize the influence of parameters on the folding pattern (e.g. on the emergence of primary, secondary and tertiary folds). The metrics are used for the first time to optimize the similarity of the model to healthy neurodevelopmental data. Using MRI fetal atlas data from the dHCP project, our model succeeded in reproducing the development of some primary folds, from 21 to 28 GW, on the surface of the cortex during fetal neurodevelopment.\\

\noindent However, several limitations should be highlighted for potential improvement.
First of all, our model succeeds in modeling folding up to approximately 28 GW. Beyond this, collisions more global than interhemispheric ones occur. But, we know that secondary and tertiary folds appear from this age. It is therefore necessary to develop a method to manage global collisions and then to question the hypotheses underlying the emergence of these higher-degree folds. 

Second, it seems essential to study the incorporation of heterogeneities in the biomechanical parameters of the model, while we have assumed them to be constant in time and space in this work. Such study may focus primarily on three model's key parameters that strongly influence the folding pattern: cortical thickness, the ratio of cortex stiffness to sub-layer stiffness, and the cortical growth rate. 

 \noindent Indeed, \cite{NDemirciandMAHolland2022} have shown that cortical thickness varies regionally in a normally developed brain and is related to local curvature, therefore to the type of fold. Furthermore, \cite{SWang2021} have shown that realistic distinct cortical thicknesses in the gyri versus the sulci appear in condition that the stiffness ratio close to 1 produces. However, our model cannot use such a low ratio for now, limiting itself to 5 for numerical reasons. Work remains to be done to understand how to make this ratio and/or the resulting gyri/sulci more realistic.
 
 \noindent \cite{Sbudday2015bis} have elegantly shown through a 2D gyrification model, that if the cortical growth ratio 
 is gradually increased over time, the cortex sequentially crosses three critical thresholds, each of which induces a transition in the folding mode, leading to the emergence of primary, secondary, and subsequently tertiary folds. According to \cite{PVBayly2013}, the emergence of secondary folds is produced by the introduction of heterogeneous growth perturbations, either within the cortical sub-layers or in the cortex itself. \cite{MJRazavi2015bis} showed that incorporating a distinct and higher growth rate in some cortical regions induced a reproducible and consistent folding pattern whatever the growth ratio between cortex and sub-layers. Conversely, the folding pattern was not consistent with homogeneous cortical growth.
 
\noindent The growth model we currently use assumes that only the cortex grows. This reflects the maturation of neurons that have already migrated into the cortical plate, but at the onset of this second phase, the distribution of neurons is already heterogeneous, resulting from a first phase of migration of neurons through the sublayers to the cortex. It would be appropriate in a second step to incorporate a model of growth specific to the sublayers and to investigate its influence on the biomechanical constraint in the cortex preceding cortical growth. 

Thirdly, it should be required to test simulating cortical folding with the improved \textit{FetalFoldSim} on individual MRI fetal data instead of meshes obtained from an atlas which is smoother and less realistic.

Finally, the \textit{FetalFoldSim} computational model may be extended towards a multi-scale approach, that \textit{FEniCS} enables to deal with. 
\textit{FEniCS} also facilitates the development of multi-phase coupled models (see \cite{TLavigne2023}, \cite{SUrcun2023}). When modeling bio-materials, characterized by various compartments (fluid, solid, cells), like brain, it is necessary to transfer the equations from Eulerian to a Lagrangian framework. \textit{FEniCS} allows precisely to use both a mixed-element function space (e.g., coupled unknowns like displacement [solid] and pressure [fluid]), and a combined Eulerian-Lagrange description (ALE).





\paragraph{Acknowledgements} 

The research leading to these results has received funding from ANR (AI4CHILD ANR-19-CHIA-0015-01).




\paragraph{Credit authorship contribution statement}

Conceptualization (AK, JL, FR),
Data curation (AK, FR),
Formal analysis (AK),
Funding acquisition (FR),
Investigation (AK),
Methodology (AK, TL, SU, JL, FR),
Project administration (AK, JL, FR),
Resources (FR),
Software (AK, TL, SU, JL, FR),
Supervision (JL, FR),
Validation (AK),
Visualization (AK),
Writing – original draft (AK),
Writing – review \& editing (all authors)

\paragraph{Competing Interests}
The authors declare that they have no competing interest.

\paragraph{Declaration of generative AI and AI-assisted technologies in the manuscript preparation process}
During the preparation of this work the author(s) used Perplexity and deepL in order to improve some sentences. After using this tool/service, the author(s) reviewed and edited the content as needed and take(s) full responsibility for the content of the published article.









\bibliographystyle{unsrt}
\bibliography{sn-bibliography.bib}

@article{cignoni2008meshlab,
  author    = {Cignoni, Paolo and Callieri, Marco and Corsini, Massimiliano and Dellepiane, Matteo and Ganovelli, Fabio and Ranzuglia, Guido},
  title     = {MeshLab: an Open-Source Mesh Processing Tool},
  journal = {Eurographics Italian Chapter Conference},
  editor    = {Vittorio Scarano and Rosario De Chiara and Ugo Erra},
  publisher = {The Eurographics Association},
  year      = {2008},
  pages     = {129--136}
}

@article{ahrens2005paraview,
  author    = {Ahrens, James and Geveci, Berk and Law, Charles},
  title     = {ParaView: An End-User Tool for Large Data Visualization},
  journal = {Visualization Handbook},
  publisher = {Elsevier},
  year      = {2005}
}

@article{schoberl1997netgen,
  author  = {Sch{\"o}berl, Joachim},
  title   = {NETGEN An advancing front 2D/3D-mesh generator based on abstract rules},
  journal = {Computing and Visualization in Science},
  year    = {1997},
  volume  = {1},
  pages   = {41--52},
  doi     = {10.1007/s007910050004}
}

@article{MSAlnaes2015,
  title     = {The {FEniCS} Project Version 1.5},
  author    = {Alnaes, M. S. and Blechta, J. and Hake, J. and Johansson, A. and Kehlet, B. and Logg, A. and Richardson, C. and Ring, J. and Rognes, M. E. and Wells, G. N.}, 
  journal   = {Archive of Numerical Software},
  year      = {2015},
  volume    = {3},
  doi       = {10.11588/ans.2015.100.20553},
}

@article{fedorov2012slicer,
  author  = {Fedorov, A. and Beichel, R. and Kalpathy-Cramer, J. and Finet, J. and Fillion-Robin, J.-C. and Pujol, S. and Bauer, C. and Jennings, D. and Fennessy, F. and Sonka, M. and Buatti, J. and Aylward, S. and Miller, J. V. and Pieper, S. and Kikinis, R.},
  title   = {3D Slicer as an Image Computing Platform for the Quantitative Imaging Network},
  journal = {Magnetic Resonance Imaging},
  year    = {2012},
  volume  = {30},
  number  = {9},
  pages   = {1323--1341},
  month   = {November},
  doi     = {10.1016/j.mri.2012.05.001},
  pmid    = {22770690}
}

@article{geuzaine2009gmsh,
  author  = {Geuzaine, C. and Remacle, J.-F.},
  title   = {Gmsh: a three-dimensional finite element mesh generator with built-in pre- and post-processing facilities},
  journal = {International Journal for Numerical Methods in Engineering},
  year    = {2009},
  volume  = {79},
  number  = {11},
  pages   = {1309--1331},
  doi     = {10.1002/nme.2579}
}

@article{NDemirciandMAHolland2022,
  title={Cortical thickness systematically varies with curvature and depth in healthy human brains},
  author={Demirci, N. and Holland, M. A.},
  journal={Human Brain Mapping},
  volume={43},
  number={6},
  pages={2064--2084},
  year={2022}
}

@article{HJYun2022,
  title={Quantification of sulcal emergence timing and its variability in early fetal life: hemispheric asymmetry and sex difference},
  author={Yun, H. J. and Lee, H. J. and Lee, J. Y. and Tarui, T. and Rollins, C. K. and Ortinau, C. M. and Feldman, H. A. and Grant P. E. and Im, K.},
  journal={NeuroImage},
  volume={263},
  pages={119629},
  year={2022}
}

@article{HdeVareilles2023,
  title={Development of cortical folds in the human brain: An attempt to review biological hypotheses, early neuroimaging investigations and functional correlates},
  author={{de Vareilles}, H. and Rivière, D. and Mangin, J. F. and Dubois, J.},
  journal={Developmental cognitive neuroscience},
  volume={61},
  pages={101249},
  year={2023},
  url={https://doi.org/10.1016/j.dcn.2023.101249}
}

@article{ERodriguez1994,
  title={Stress-dependent finite growth in soft elastic tissues},
  author={Rodriguez, E. K. and Hoger, A. and McCulloch, A. D.},
  journal={Journal of biomechanics},
  volume={27},
  number={4},
  pages={455-467},
  year={1994}
}

@article{MBenAmar2005,
  title={Growth and instability in elastic tissues},
  author={Ben Amar, M. and Goriely, A.},
  journal={Journal of the Mechanics and Physics of Solids},
  volume={53},
  number={10},
  pages={2284-2319},
  year={2005}
}

@article{AGoriely2007,
  title={On the definition and modeling of incremental, cumulative, and continuous growth laws in morphoelasticity},
  author={Goriely, A. and Ben Amar, M.},
  journal={Biomechanics and modeling in mechanobiology},
  volume={6},
  number={5},
  pages={289--296},
  year={2007}
}

@article{JDervaux2008,
  title={Morphogenesis of growing soft tissues},
  author={Dervaux, J. and Ben Amar, M.},
  journal={Physical review letters},
  volume={101},
  number={6},
  pages={068101},
  year={2008}
}

@article{MJRazavi2015bis,
  title={Cortical folding pattern and its consistency induced by biological growth},
  author={Razavi, M. J. and Zhang, T. and Liu, T. and Wang, X.},
  journal={Scientific reports},
  volume={5},
  number={1},
  pages={14477},
  year={2015}
}

@article{Sbudday2015bis,
  title={Secondary instabilities modulate cortical complexity in the mammalian brain},
  author={Budday, S. and Steinmann, P. and Kuhl, E.},
  journal={Philosophical Magazine},
  volume={95},
  number={28--30},
  pages={3244--3256},
  year={2015}
}

@article{SBudday2015,
  title={Physical biology of human brain development},
  author={Budday, S. and Steinmann, P. and Kuhl, E.},
  journal={Frontiers in cellular neuroscience},
  volume={9},
  pages={257},
  year={2015}
}

@article{MAHolland2013,
  title={On the mechanics of thin films and growing surfaces},
  author={Holland, M. A. and Kosmata, T. and Goriely, A. and Kuhl, E.},
  journal={Mathematics and Mechanics of Solids},
  volume={18},
  number={6},
  pages={561--575},
  year={2013}
}

@article{SBudday2014,
  title={A mechanical model predicts morphological abnormalities in the developing human brain},
  author={Budday, S. and Raybaud, C. and Kuhl, E.},
  journal={Scientific reports},
  volume={4},
  number={1},
  pages={5644},
  year={2014}
}

@article{TTallinen2016,
  title={On the growth and form of cortical convolutions},
  author={Tallinen, T. and Chung, J. Y. and Rousseau, F. and Girard, N. and Lefèvre, J. and Mahadevan, L.},
  journal={Nature Physics},
  volume={12},
  number={6},
  pages={588--593},
  year={2016}
}

@article{MAlenya2022,
  title={Computational pipeline for the generation and validation of patient-specific mechanical models of brain development},
  author={Alenyà, M. and Wang, X. and Lefèvre, J. and Auzias, G. and Fouquet, B. and Eixarch, E. and Rousseau, F. and Camara, O.},
  journal={Brain Multiphysics},
  volume={3},
  pages={100045},
  year={2022}
}

@article{SiBudday2015,
  title={Size and curvature regulate pattern selection in the mammalian brain},
  author={Budday, S. and Steinmann, P. and Goriely, A. and Kuhl, E. },
  journal={Extreme Mechanics Letters},
  volume={4},
  pages={193--198},
  year={2015}
}

@article{AJavili2013,
  title={A novel strategy to identify the critical conditions for growth-induced instabilities},
  author={Javili, A. and Steinmann, P. and Kuhl, E.},
  journal={Journal of the mechanical behavior of biomedical materials},
  volume={29},
  pages={20--32},
  year={2014}
}

@article{VernerGarikipati2018,
  title={A computational study of the mechanisms of growth-driven folding patterns on shells, with application to the developing brain},
  author={Verner, S. N. and Garikipati, K. A.},
  journal={Extreme Mechanics Letters},
  volume={18},
  pages={58--69},
  year={2018}
}

@article{TTallinen2014,
  title={Gyrification from constrained cortical expansion},
  author={Tallinen, T. and Chung, J. Y. and Biggins, J. S. and Mahadevan, L.},
  journal={Proceedings of the National Academy of Sciences},
  volume={111},
  number={35},
  pages={12667--12672},
  year={2014}
}

@article{XXu2022,
  title={Spatiotemporal atlas of the fetal brain depicts cortical developmental gradient},
  author={Xu, X. and Sun, C. and Sun, J. and Shi, W. and Shen, Y. and Zhao, R. and Wanrong, L. and Mingyang, L. and Wang, G. and Wu, D.},
  journal={Journal of Neuroscience},
  volume={42},
  number={50},
  pages={9435--9449},
  year={2022}
}

@article{LVasung2016,
  title={Quantitative and qualitative analysis of transient fetal compartments during prenatal human brain development},
  author={Vasung, L. and Lepage, C. and Radoš, M. and Pletikos, M. and Goldman, J. S. and Richiardi, J. and Raguz, M. and Fischi-Gomez, E. and Karama, S. and Huppi, P. S. and Evans, A. C. and Kostovic, I.},
  journal={Frontiers in neuroanatomy},
  volume={10},
  pages={11},
  year={2016}
}

@article{EGriffiths2023,
  title={On the importance of using region-dependent material parameters for full-scale human brain simulations},
  author={Griffiths, E. and Hinrichsen, J. and Reiter, N. and Budday, S.},
  journal={European Journal of Mechanics-A/Solids},
  volume={99},
  pages={104910},
  year={2023}
}

@article{XWang2021,
  title={The influence of biophysical parameters in a biomechanical model of cortical folding patterns},
  author={Wang, X. and Lefevre, J. and Bohi, A. and Harrach, M. A. and Dinomais, M. and Rousseau, F.},
  journal={Scientific Reports},
  volume={11},
  number = {1},
  pages={7686},
  year={2021}
}

@article{PVBayly2013,
  title={A cortical folding model incorporating stress-dependent growth explains gyral wavelengths and stress patterns in the developing brain},
  author={Bayly, P. V. and Okamoto, R. J. and Xu, G. and Shi, Y. and Taber, L. A.},
  journal={Physical biology},
  volume={10},
  number = {1},
  pages={016005},
  year={2013}
}

@article{SBudday2018,
  title={On the influence of inhomogeneous stiffness and growth on mechanical instabilities in the developing brain},
  author={Budday, S. and Steinmann, P.},
  journal={International Journal of Solids and Structures},
  volume={132},
  pages={31--41},
  year={2018}
}

@article{MAHolland2015,
  title={Emerging brain morphologies from axonal elongation},
  author={Holland, M. A. and Miller, K. E. and Kuhl, E.},
  journal={Annals of biomedical engineering},
  volume={43},
  pages={1640--1653},
  year={2015}
}

@article{SWang2021,
  title={Numerical investigation of biomechanically coupled growth in cortical folding},
  author={Wang, S. and Demirci, N. and Holland, M. A. },
  journal={Biomechanics and Modeling in Mechanobiology},
  volume={20},
  number={2},
  pages={555--567},
  year={2021}
}

@article{MSZarzor2021,
  title={A two-field computational model couples cellular brain development with cortical folding},
  author={Zarzor, M. S. and Kaessmair, S. and Steinmann, P. and Blümcke, I. and Budday, S.},
  journal={Brain Multiphysics},
  volume={2},
  pages={100025},
  year={2021}
}

@article{MSZarzor2023,
  title={Exploring the role of the outer subventricular zone during cortical folding through a physics-based model},
  author={Zarzor, M. S. and Blumcke, I. and Budday, S.},
  journal={elife},
  volume={12},
  pages={e82925},
  year={2023}
}

@article{RdeRooij2018,
  title={A physical multifield model predicts the development of volume and structure in the human brain},
  author={{de Rooij}, R. and Kuhl, E.},
  journal={Journal of the Mechanics and Physics of Solids},
  volume={112},
  pages={563--576},
  year={2018}
}

@article{TLavigne2023,
  title={Single and bi-compartment poro-elastic model of perfused biological soft tissues: FEniCSx implementation and tutorial},
  author={Lavigne, T. and Urcun, S. and Rohan, P. Y. and Sciumè, G. and Baroli, D. and Bordas, S. P.},
  journal={Journal of the Mechanical Behavior of Biomedical Materials},
  volume={143},
  pages={105902},
  year={2023}
}

@article{VFernandez2016,
  title={Cerebral cortex expansion and folding: what have we learned?},
  author={Fernández, V. and Llinares‐Benadero, C. and Borrell, V.},
  journal={The EMBO journal},
  volume={35},
  number={10},
  pages={1021-1044},
  year={2016}
}

@article{SNVerner2018,
  title={Computational study of the mechanisms of growth-driven folding patterns on shells, with application to the developing brain},
  author={Verner, S. N. and Garikipati, K. A.},
  journal={Extreme Mechanics Letters},
  volume={18},
  pages={58--69},
  year={2018}
}

@article{ZWang2021,
  title={An inverse modelling study on the local volume changes during early morphoelastic growth of the fetal human brain},
  author={Wang, Z. and Martin, B. and Weickenmeier, J. and Garikipati, K.},
  journal={Brain multiphysics},
  volume={2},
  pages={100023},
  year={2021}
}

@book{VAYastrebov2013,
  title     = "Numerical methods in contact mechanics",
  author    = "Yastrebov, V. A.",
  year      = 2013,
  pages={151--155},
  publisher = "John Wiley \& Sons"
}

@article{JLengiewicz2011,
  title={Automation of finite element formulations for large deformation contact problems},
  author={Lengiewicz, J. and Korelc, J. and Stupkiewicz, S.},
  journal={International Journal for Numerical Methods in Engineering},
  volume={85},
  number={10},
  pages={1252--1279},
  year={2011}
}

@article{RTDjoumessi2025,
  title={A self-contact electromechanical framework for intestinal motility},
  author={Djoumessi, R. T. and Lenarda, P. and Gizzi, A. and Paggi, M.},
  journal={arXiv preprint},
  number={2502.05285},
  year={2025}
}

@article{XChen2024,
  title={Modelling midline shift and ventricle collapse in cerebral oedema following acute ischaemic stroke},
  author={Chen, X. and Józsa, T. I. and Cardim, D. and Robba, C. and Czosnyka, M. and Payne, S. J.},
  journal={PLOS Computational Biology},
  volume={20},
  number={5},
  pages={e1012145},
  year={2024}
}

@article{CPatte2022,
  title={A quasi-static poromechanical model of the lungs},
  author={Patte, C. and Genet, M. and Chapelle, D.},
  journal={Biomechanics and Modeling in Mechanobiology},
  volume={21},
  number={2},
  pages={527--551},
  year={2022}
}

@article{MJRazavi2015,
  title={Role of mechanical factors in cortical folding development},
  author={Razavi, M. J. and Zhang, T. and Li, X. and Liu, T. and Wang, X.},
  journal={Physical Review E},
  volume={92},
  number={3},
  pages={032701},
  year={2015}
}

@article{RToro2005,
  title={A morphogenetic model for the development of cortical convolutions},
  author={Toro, R. and Burnod, Y.},
  journal={Cerebral cortex},
  volume={15},
  number={12},
  pages={1900--1913},
  year={2005}
}

@article{SUrcun2023,
  title={Non-operable glioblastoma: proposition of patient-specific forecasting by image-informed poromechanical model},
  author={Urcun, S. and Baroli, D. and Rohan, P. Y. and Skalli, W. and Lubrano, V. and Bordas, S. P. and Sciume, G.},
  journal={Brain Multiphysics},
  volume={4},
  pages={100067},
  year={2023}
}

@article{DGermanaud2012,
  title={Larger is twistier : spectral analysis of gyrification (SPANGY) applied to adult brain size polymorphism},
  author={Germanaud, D. and Lefèvre, J. and Toro, R. and Fischer, C. and Dubois, J. and Hertz-Pannier, L. and Mangin, J-F.},
  journal={NeuroImage},
  volume={63},
  number={3},
  pages={1257--1272},
  year={2012}
}

@article{ZGao2014,
  title={A compact shape descriptor for triangular surface meshes},
  author={Gao, Z. and Yu, Z. and Pang, X.},
  journal={Computer-Aided Design},
  volume={53},
  pages={62--69},
  year={2014}
}

@article{JDubois2019,
  title={The dynamics of cortical folding waves and prematurity-related deviations revealed by spatial and spectral analysis of gyrification},
  author={Dubois, J. and Lefèvre, J. and Angleys, H. and Leroy, F. and Fischer, C. and Lebenberg, J. and Dehaene-Lambertz, G. and Borradori-Tolsa, C. and Lazeyras, F. and Hertz-Pannier, L. and Mangin, J-F. and Hüppi, P. S. and Germanaud, D.},
  journal={NeuroImage},
  volume={185},
  pages={934--946},
  year={2019}
}

@article{XWang2019,
  title        = {On early brain folding patterns using biomechanical growth modeling},
  author       = {Wang, X. and Bohi, A. and Al Harrach, M. and Dinomais, M. and Lefèvre, J. and Rousseau, F.},
  year         = 2019,
  journal    = {41st Annual International Conference of the IEEE Engineering in Medicine and Biology Society (EMBC). IEEE},
  pages        = {146--149},
}

@article{NDemirci2024,
  title={Scaling patterns of cortical folding and thickness in early human brain development in comparison with primates},
  author={Demirci, N. and Holland, M. A.},
  journal={Cerebral Cortex},
  volume={34},
  number={2},
  pages={bhad462},
  year={2024}
}

@article{JJKoenderink1992,
  title={Surface shape and curvature scales},
  author={Koenderink, J. J. and Van Doorn, A. J.},
  journal={Image and vision computing},
  volume={10},
  number={8},
  pages={557--564},
  year={1992}
}

@article{HHHu2013,
  title={Shape and curvedness analysis of brain morphology using human fetal magnetic resonance images in utero},
  author={Hu, H-H. and Chen, H. Y. and Hung, C. I. and Guo, W. Y. and Wu, Y. T.},
  journal={Brain Structure and Function},
  volume={218},
  pages={1451--1462},
  year={2013}
}

@article{JLefevre2015,
  title={Are developmental trajectories of cortical folding comparable between cross-sectional datasets of fetuses and preterm newborns?},
  author={Lefèvre, J. and Germanaud, D. and Dubois, J. and Rousseau, F. and de Macedo Santos, I. and Angleys, H. and Mangin, J-F. and Hüppi, P. S. and Girard, N. and de Guio, F.},
  journal={Cerebral cortex},
  volume={26},
  pages={3023--3035},
  year={2015}
}

@article{AGholipour2017,
  title={A normative spatiotemporal MRI atlas of the fetal brain for automatic segmentation and analysis of early brain growth},
  author={Gholipour, A. and Rollins, C. K. and Velasco-Annis, C. and Ouaalam, A. and Akhondi-Asl, A. and Afacan, O. and Ortinau, C. M. and Clancy, S. and Limperopoulos, C. and Yang, E. and Estroff, J. A. and Warfield, S. K.},
  journal={Scientific reports},
  volume={7},
  number={1},
  pages={476},
  year={2017}
}

@article{PAHabas2012,
  title={Early folding patterns and asymmetries of the normal human brain detected from in utero MRI},
  author={Habas, P. A. and Scott, J. A. and Roosta, A. and Rajagopalan, V. and Kim, K. and Rousseau, F. and Barkovich, A. J. and Glenn, O. A. and Studholme, C.},
  journal={Cerebral cortex},
  volume={22},
  number={1},
  pages={13--15},
  year={2012}
}

\end{document}